\documentclass[%
twocolumn,
superscriptaddress,
floatfix,
amsmath,amssymb,
aps,
pre,
a4paper,
]{revtex4-2}

\usepackage{booktabs}

\usepackage{lipsum}
\usepackage{comment}

\usepackage[hidelinks,unicode=true]{hyperref}
\hypersetup{colorlinks=true,
  linkcolor=blue,
  urlcolor=blue,
  citecolor=blue,
  pdfhighlight=/N
}

\usepackage{graphicx}% Include figure files
\usepackage[caption=false]{subfig}

\usepackage{dcolumn}% Align table columns on decimal point
\usepackage{bm}% bold math
\usepackage{xcolor}

\usepackage{graphicx}% Include figure files
\usepackage{dcolumn}% Align table columns on decimal point
\usepackage{bm}% bold math
\graphicspath{{fig_main/}, {images_SM/}}

\begin{document}

\title{Leveraging spurious Omori-Utsu relation in the nearest-neighbor
  declustering method}

\author{Andreu Puy}
\affiliation{Departament de Física, Universitat Politècnica de Catalunya, Campus Nord B4, 08034 Barcelona, Spain}

\author{Jordi Baró}
\affiliation{
    Departament de Fisica de la Materia Condensada, Universitat de Barcelona, 08028 Barcelona, Spain}
\affiliation{
    Universitat de Barcelona Institute of Complex Systems (UBICS), Universitat de Barcelona, 08028 Barcelona, Spain}

\author{J\"orn Davidsen}
\affiliation{
 Complexity Science Group, Department of Physics and Astronomy,
 University of Calgary, Calgary, Alberta T2N 1N4, Canada
}
\affiliation{Hotchkiss Brain Institute, University of Calgary, Calgary, Alberta T2N 4N1, Canada}

\author{Romualdo Pastor-Satorras}
\email{romualdo.pastor@upc.edu}
\affiliation{Departament de Física, Universitat Politècnica de Catalunya, Campus Nord B4, 08034 Barcelona, Spain}


\date{\today}% It is always \today, today,

\begin{abstract}
  Static and dynamic stress changes in the Earth's crust induced by an
  earthquake typically trigger other earthquakes. Identifying such aftershocks
  is an important step in seismic hazard assessment but has remained
  challenging, especially in cases involving natural fluid migration or
  anthropogenic fluid injections, which can occur with varying time scales
  and/or episodically, leading to strong temporal variations in earthquake
  occurrences. Here, we demonstrate analytically and numerically that earthquake
  catalogs without triggering can lead to spurious Omori-Utsu and productivity
  relations for the commonly used nearest-neighbor declustering method. However,
  we show that the robustness of the Omori-Utsu exponent on newly introduced
  parameters of the method allows one to determine whether indeed aftershocks
  are present. For a natural swarm catalog, which lacks a clear distinction
  between triggered and background events, we find that the catalog is dominated
  by aftershocks.
\end{abstract}

\maketitle

%%%%%%%%%%%%%%%%%%%%%%%%%%%%%%%%%%%%%%%%%%%%%%%
%
%  BODY TEXT
%
%%%%%%%%%%%%%%%%%%%%%%%%%%%%%%%%%%%%%%%%%%%%%%%

%%% Suggested section heads:
% \section{Introduction}
%
% The main text should start with an introduction. Except for short
% manuscripts (such as comments and replies), the text should be divided
% into sections, each with its own heading.

% Headings should be sentence fragments and do not begin with a
% lowercase letter or number. Examples of good headings are:

% \section{Materials and Methods}
% Here is text on Materials and Methods.
%
% \subsection{A descriptive heading about methods}
% More about Methods.
%
% \section{Data} (Or section title might be a descriptive heading about data)
%
% \section{Results} (Or section title might be a descriptive heading about the
% results)
%
% \section{Conclusions}

\section{Introduction}

The study of temporal or spatio-temporal correlations is essential to assess
event probabilities beyond Poisson rates in the context of natural
hazards~\cite{corralScalingCorrelationsDynamics2008,
  mollerStructuredSpatioTemporalShotNoise2010,
  dalelaneRobustEstimatorIntensity2013,
  morinaProbabilityEstimationCarringtonlike2019}, structural
hazards~\cite{locknerRoleAcousticEmission1993,
  baroStatisticalSimilarityCompression2013} and
beyond~\cite{chavez-demoulinEstimatingValueriskPoint2005,filimonovApparentCriticalityCalibration2015,herreraPointProcessModels2018}. In
the case of earthquakes, spatio-temporal clustering in instrumental catalogs
escapes the description of seismic activity in terms of spatially heterogeneous
Poisson processes often considered in hazard measures
\cite{ogataSpaceTimePointProcessModels1998,vere-jonesStatisticalSeismology2005}
and improvements in short and mid-term forecasts can be achieved by introducing
well parameterized clustering into the
model~\cite{scholzMechanicsEarthquakesFaulting2019,
  zaliapinPerspectivesClusteringDeclustering2022}.  Declustering methods are
used to identify and parameterize clusters and categorize events among
them. Most remarkably, the same methods can be used to identify and characterize
the underlying generating mechanisms, which is well-known to be a delicate
task~\cite{zaliapinPerspectivesClusteringDeclustering2022}. Clusters might be
caused by episodic changes in the driving forces, such as volcanic, geothermal,
or anthropogenic activity, or by mechanisms of event-event triggering, i.e.,
when the stress changes induced by an earthquake trigger a subsequent earthquake
or aftershock. Aftershocks manifest most prominently as a rise in activity
directly after and in the vicinity of large earthquakes. Aftershock activity is
observed to scale exponentially with the size of the mainshock and decay in time with a power
law, phenomena described by the productivity and Omori-Utsu relations,
respectively~\cite{omoriAftershocksEarthquakes1894,
  utsuCentenaryOmoriFormula1995,shcherbakov2005aftershock,wetzlerRegionalStressDrop2016,davidsenSelfsimilarAftershockRates2016}.

Based on the original approach by~\cite{baiesiScalefreeNetworksEarthquakes2004,
  baiesiComplexNetworksEarthquakes2005}, the nearest-neighbor (NN) declustering
method introduced by~\cite{zaliapinClusteringAnalysisSeismicity2008} is
currently employed in statistical seismology of both
natural~\cite{zaliapinEarthquakeClustersSouthern2013,
  guTriggeringCascadesStatistical2013, pageTuringStyleTests2018,
  fieldImprovementsThirdUniform2021} and anthropogenic
origin~\cite{schoenballSystematicAssessmentSpatiotemporal2017,
  maghsoudiIntereventTriggeringMicroseismicity2018,
  martinez2018comparative,karimiSeparatingPrimarySecondary2023}, rock
mechanics~\cite{davidsenTriggeringProcessesRock2017,
  davidsenWhatControlsPresence2021}, as well as in other systems exhibiting
Omori-like avalanche behavior, such as structural
transitions~\cite{baroAvalancheCorrelationsMartensitic2014}, crackling
noise~\cite{zapperiCracklingNoiseStatistical2022,Laurson2009} and turning
avalanches in schooling fish~\cite{puySignaturesCriticalityTurning2024}. It is
largely non-parametric, robust against catalog imperfections, and has been
reported to provide good accuracy in identifying clustered regions while
preserving heterogeneous features and non-stationary background
seismicity~\cite{peresanIdentificationCharacterizationEarthquake2019,
  tengSeismicityDeclusteringHazard2019,
  zaliapinEarthquakeDeclusteringUsing2020}.  However, some deficiencies have
been reported in classifying background and clustered
events~\cite{baylissProbabilisticIdentificationEarthquake2019}---which can be
particularly pronounced for seismicity induced by anthropogenic activity such as
fluid injections~\cite{OK_khajehdehi2023potential}---and there are challenges to
distinguish aftershocks from exogenous episodic
activity~\cite{baroTopologicalPropertiesEpidemic2020,
  karimiAftershockTriggeringSpatial2021, karimiSeparatingPrimarySecondary2023}.

Here, we report a novel potential pitfall in aftershock identification using the
NN declustering method with significant implications. We demonstrate that the
method systematically assigns a large number of ``aftershocks" to major events,
producing power-law behavior in the aftershock decay rates even in the absence
of event-event triggering when no causal connections exist between the
events. Consequently, an Omori-Utsu-like and productivity relations can arise
spuriously such that their observation alone does not necessarily indicate
aftershock triggering. However, we show that one can circumvent this potential
pitfall and in fact use it to establish the presence of aftershocks in debated
cases such as natural
swarms~\cite{hainzl2004seismicity,fischer2023fast}. Specifically, we propose a
generalization of the NN method in which novel tunable parameters are
introduced, that yield, in the absence of actual aftershocks, Omori-Utsu-like and
productivity relations depending on these parameters. When compared with
regional catalogs dominated by tectonic aftershock sequences as well as the
Epidemic-Type Aftershock Sequence (ETAS) model, we observe that the resulting
Omori-Utsu relation is generally independent of these parameters. We propose
that such a parameter independency in the generalized NN method is a true
indicator of the presence of clustering and actual aftershock triggering.

%--
\begin{figure*}[t!p]
    \centering
\includegraphics[width=\textwidth]{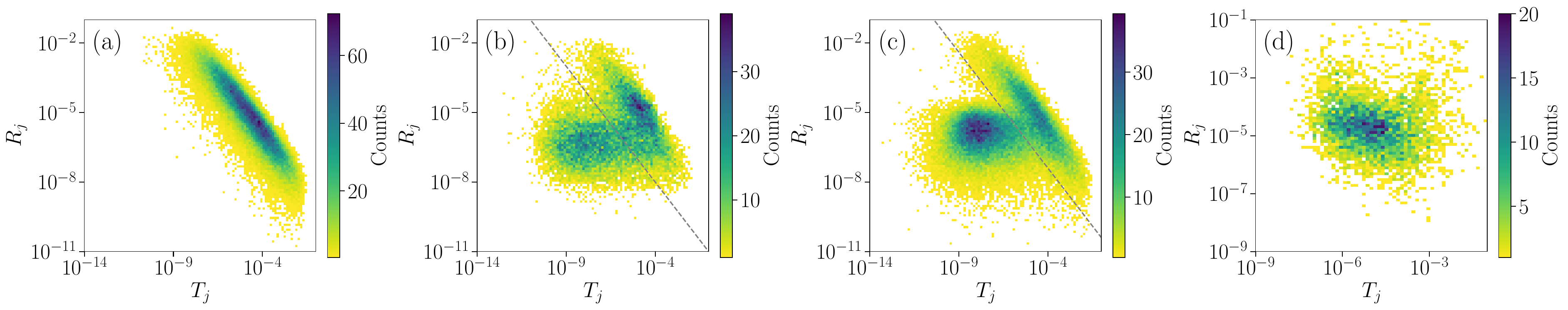}
    \caption{Joint distribution of the rescaled time $T_j$ and space $R_j$ for (a) the null model, (b) the SCSN catalog, (c) the ETAS model and (d) the natural swarm catalog. The gray dashed line indicates the threshold separating background and clustered events, with (b) $\eta_{th}=10^{-12}$ and (c) $\eta_{th}=4\cdot10^{-12}$. We have fixed $b=1$ and (a) and (c) $D=2$, (b) and (d) $D=1.6$.}
    \label{fig:Rij_Tij}
\end{figure*}

%--
\begin{figure*}[t!p]
    \centering
\includegraphics[width=\textwidth]{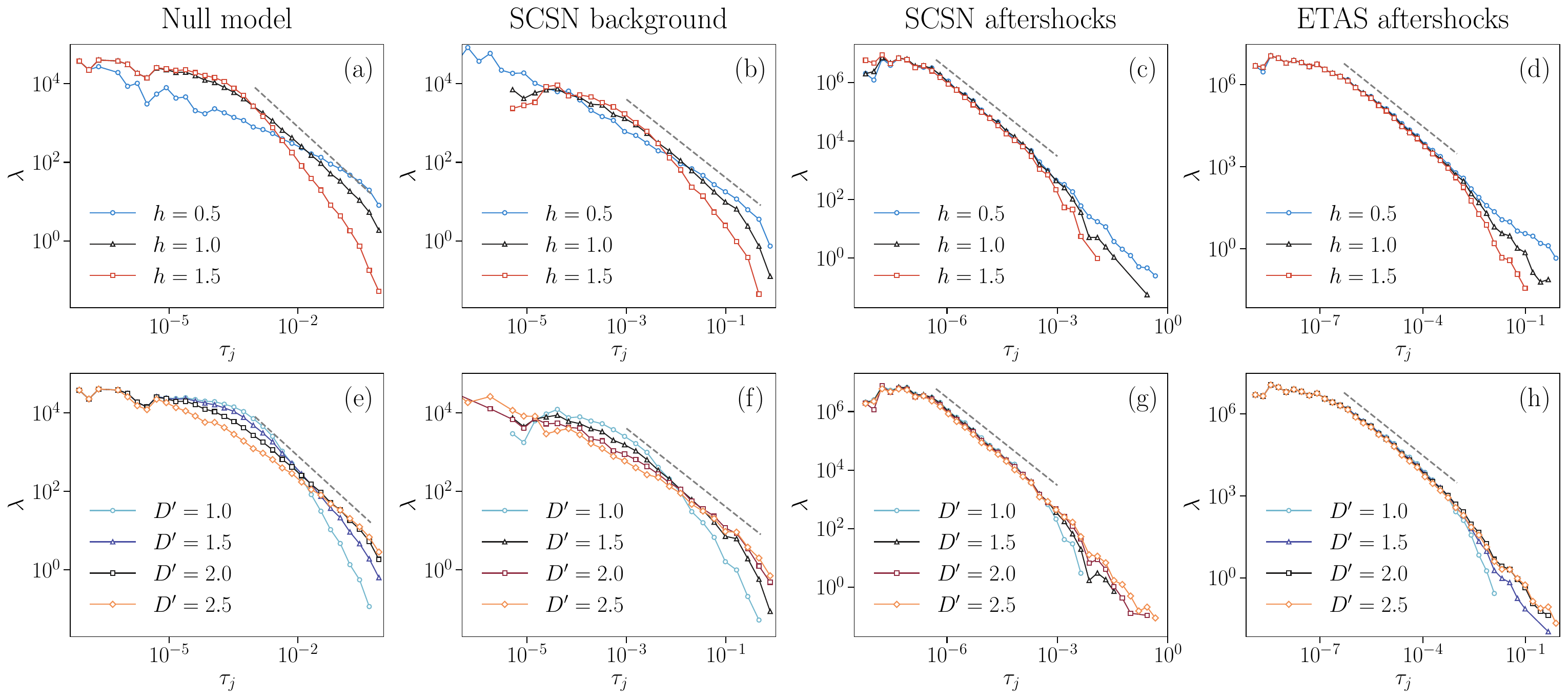}
\caption{Uncorrelated (background) events show spurious Omori-Utsu relations,
  which, unlike true aftershocks, depend on the parameters $h$ and $D'$. Rate of
  the number of ``triggered" events $\lambda(m, \tau)$ after a mainshock of
  magnitude $m\in [5, 6)$ depending on the time interval $\tau$ for (a) and (e)
  the null model, (b) and (f) events labeled as background events in the SCSN
  catalog, (c) and (g) events labeled as aftershocks in the SCSN catalog and (d)
  and (h) events labeled as aftershocks in the ETAS catalog. Other parameters
  are fixed at $h = 1$, (a) and (d) $D' = 2$, (b) and (c) $D'=1.6$, and
  $b' = 1$. The gray dashed line is a guide to the eye $\propto \tau^{-1}$.}
    \label{fig:omori_h_D}
\end{figure*}

%\section{Methodology}

\section{The nearest-neighbor declustering method}

The original NN declustering me\-th\-od is based on the \emph{proximity}
$\eta_{ij}$ in the space-time-magnitude domain from an event $j$ to a previous
(in time) event $i$~\cite{baiesiScalefreeNetworksEarthquakes2004,
  baiesiComplexNetworksEarthquakes2005}, defined as
\begin{equation}
  \eta_{ij} \equiv
  \begin{cases}
    t_{ij} \,r_{ij}^D \,10^{-bm_i}, & \text{if}\ i < j \\
    \infty, & \text{otherwise}
    \end{cases},
%  \label{eq:5}
\end{equation}
where the event times have been ordered such that $t_ i < t_{i+1}$,
$t_{ij} \equiv t_j - t_i$ is the time interval between events $i$ and $j$,
$r_{ij}\equiv\left|\vec{r}_j - \vec{r}_i \right|$ is the spatial distance
between the events locations, $D$ is the fractal dimension associated with the
spatial distribution of the events and $b$ characterizes the magnitude
distribution, as captured by the Gutenberg-Richter (GR) relation,
$P(m) \sim 10^{-b m}$~\cite{gutenbergEarthquakeMagnitudeIntensity1942}.  Using
the proximity measure $\eta_{ij}$, each event $j$ is associated with a
\textit{nearest-neighbor} or \textit{parent} $p_j$, defined as the event in the
past ($p_j < j$) that minimizes the proximity with $j$, namely
$\eta_{p_jj} \leq \eta_{ij}$, $\forall i < j$. Event $j$ is referred to as a
potential aftershock of parent $p_j$, and is characterized by the
\emph{nearest-neighbor proximity} $\eta_j \equiv \eta_{p_jj}$.

To differentiate between true aftershocks and uncorrelated events leading to
small $\eta_j$, Zaliapin \textit{et
  al.}~\cite{zaliapinClusteringAnalysisSeismicity2008,
  zaliapinEarthquakeClustersSouthern2013} proposed to study the \emph{rescaled
  time} $T_j$ and \emph{rescaled space} $R_j$, defined as follows:
\begin{equation}
    T_{j} \equiv \tau_{j}\sqrt{10^{-bm_{p_j}}}, \quad  R_{j} \equiv r_{j}^D\sqrt{10^{-bm_{p_j}}},
\end{equation}
where $\tau_j \equiv t_j - t_{p_j}$ is the time interval between aftershock $j$
and its parent $p_j$, and $r_j$ is the spatial distance between them. The joint
distribution of the rescaled variables, as well as the distribution of
$\eta_{j} =T_{j} R_{j}$, typically exhibit a bimodal form (see, for example,
Figs.~\ref{fig:Rij_Tij}b and c), which facilitates the separation of a
background of Poisson-distributed, uncorrelated events, from a set of clustered,
correlated aftershocks~\cite{zaliapinClusteringAnalysisSeismicity2008,
  zaliapinEarthquakeClustersSouthern2013,guTriggeringCascadesStatistical2013}. A
threshold $\eta_{th}$ can then be introduced, whereby events with smaller and
larger nearest-neighbor proximity $\eta_j$ are classified as \emph{aftershocks}
and \emph{background} events, respectively.  For aftershocks, $\lambda(m, \tau)$ represents the rate of aftershock occurrence at time
$\tau$ after a main event of magnitude $m$, while $\nu(m)$ denotes the average number of aftershocks
generated by a mainshock of magnitude $m$.  In the case of tectonic earthquakes,
$\lambda(m, \tau) \sim \nu(m) \tau^{-p}$ for sufficiently large $\tau$, determining the Omori-Utsu relation, and
$\nu(m) \sim 10^{\alpha m}$, describing the productivity relation, with typical values $p \simeq 1$ and $\alpha \simeq b$.

\section{Generalized NN declustering method}

There are cases where the joint distribution of $T_{j}$ and $R_{j}$ has no clear
bimodal form, making the identification of aftershocks challenging.  This often
occurs in the context of fluid-induced seismicity including natural swarms and
fluid
injections~\cite{maghsoudiIntereventTriggeringMicroseismicity2018,karimiAftershockTriggeringSpatial2021,karimiSeparatingPrimarySecondary2023}.
To assess the robustness of the NN method in estimating aftershocks in such
cases, we consider a variation given by the generalized proximity depending on
three free parameters:
\begin{equation}
  \eta'_{ij} \equiv
  \begin{cases}
    t_{ij}^h \,r_{ij}^{D'} \,10^{-{b'}m_i}, & \text{if}\ i < j \\
    \infty, & \text{otherwise}
    \end{cases}.
  \label{eq:5}
\end{equation}
Here, the exponents $D'$ and $b'$ can be different from the actual fractal
dimension $D$ and the $b$-value of the GR relation, and the novel exponent $h$
controls the importance of the time difference $t_{ij}$. From the generalized
proximities $\eta'_{ij}$, the generalized NN proximities $\eta'_j$, times
$\tau'_j$ and distances $r'_j$ are computed, from which generalized rescaled
times
\begin{equation}
    T'_{j} \equiv \tau'_{j}\sqrt{10^{-b'm_{p_j}}}, \quad  R'_{j} \equiv {r'}_{j}^{D'}\sqrt{10^{-b'm_{p_j}}},
\end{equation}
can be produced.

We employ the proximity threshold $\eta_{th}$ to differentiate between background seismic events and aftershocks~\cite{zaliapinClusteringAnalysisSeismicity2008, zaliapinEarthquakeClustersSouthern2013}. While this method is recognized for its stability and accuracy~\cite{zaliapinEarthquakeClustersSouthern2013}, it is important to note that the two types of events often overlap, making complete separation unattainable. Future research could investigate more sophisticated techniques, such as random thinning~\cite{zaliapinEarthquakeDeclusteringUsing2020}, to enable a more detailed analysis.

\section{Results}

\subsection{Null model of uncorrelated seismic activity}

\begin{figure*}[t!p]
    \centering
\includegraphics[width=\textwidth]{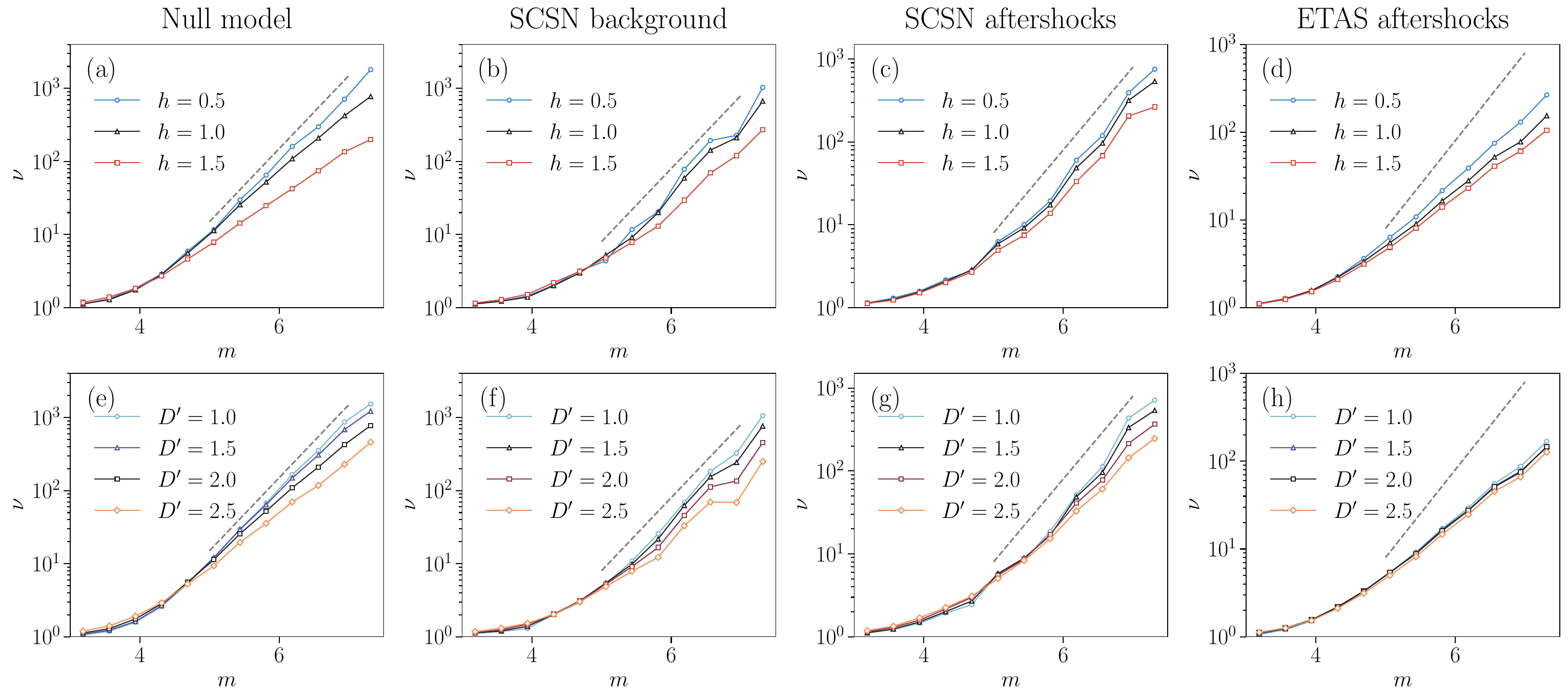}
\caption{Uncorrelated (background) events show spurious productivity
  relations. As in Fig.~\ref{fig:omori_h_D} but showing the average number of
  ``triggered" events $\nu(m)$ as a function of mainshock magnitude $m$.  The
  gray dashed line is a guide to the eye $\propto 10^{m}$.}
    \label{fig:productivity_h_D}
\end{figure*}

To analytically assess the effects of the parameters in the generalized NN
declustering method in the absence of actual aftershocks, we consider a null
model representing uncorrelated seismic activity. This model consists of $N$
events occurring at times $t_i$, randomly drawn from a probability distribution
$P_t(t)$, and positions on the plane $\vec{r}_i = (x, y)$, where the coordinates
are randomly drawn from the joint probability distribution
$P(x, y) = p_x(x) p_y(y)$. The magnitude of each event $m_i$ is also randomly
drawn from the GR distribution with a fixed value $b$. Choosing the
distributions $p_x(x)$ and $p_y(y)$ uniform in some interval $[0, L]$, the
fractal dimension of the null model is $D=2$.  See Appendix~\ref{sec:null_model}
for a detailed definition of the null model.
%~\footnote{See Supporting Information at~\url{http://link.aps.org/supplemental/XXX} for details regarding the data sets, analytical results and supplemental figures.}

To get an intuition of the effect of the parameters $h$ and $b'$ first, we can
consider the case $D' = 0$, indicating no geographical dependence. In this case,
assuming that the distribution of event times $P_t(t)$ is uniform, we can
approximate the time differences between consecutive events as constant, and
recover a rate of ``aftershock" occurrence, averaged over all mainshock
magnitudes, in the limit of large $\tau$ and $N$, given by (see
Appendix-\ref{sec:analytics_Omori_law_null}):
\begin{equation}
    \lambda(\tau) \sim \tau^{-h b'/b}.
\end{equation}
For the case of the productivity ratio as a function of mainshock magnitude, analogous calculations lead to the asymptotic value (see Appendix-\ref{sec:analytics_Omori_law_null})
\begin{equation}
    \nu(m) \sim 10^{b'm/h}.
    \label{eq:nu}
\end{equation}
If the parameters $h$ and $b'$ take their natural values in the original NN method, namely $h=1$ and $b'=b$, we recover the Omori-Utsu relation with exponent $p=1$ and a productivity relation with exponent $\alpha=b$. However, variations of $h$ and $b'$ lead in the null model to continuously varying exponents $p$ and $\alpha$.

While in the more realistic case $D' >0$ we do not expect the exponents $p$ and
$\alpha$ to take the very same values as for $D'=0$, we can conjecture that they
will show a continuous dependence on the method's parameters. In
Figs.~\ref{fig:omori_h_D}a,e we check this fact by numerically computing aftershock rates $\lambda (m,t)$ in the uncorrelated null model for different values of $h$
and $D'>0$, respectively (for details of the simulations refer to
Appendix~\ref{sec:null_model}, Fig.~\ref{fig:Rij_Tij}a and Supplemental
Figs.~S1 and~S4). Aftershock rates
throughout the manuscript are normalized to account for different maximum time
intervals of parents (see Appendix~\ref{sec:rate_aftershocks_calculation} for
computational details). While we show $\lambda(m, \tau)$ for mainshocks with
magnitudes in the range $m\in [5, 6)$, all our results are robust across
different ranges (Supplemental
Fig.~S5). Figs.~\ref{fig:omori_h_D}a,e indeed
confirms that the NN declustering method leads to spurious Omori-Utsu relations
for the uncorrelated null model. Additionally, we find the exponent $p$ of the
Omori-Utsu relation exhibits a clear dependence on the parameters $h$ and
$D'$. In contrast, there is no significant dependency on $b'$ when $D'=2$ (see
Supplemental Fig.~S2a).

\subsection{SCSN catalog}

We hypothesize that the dependence of $p$ on the parameters $h$ and $D'$ are a
signature of uncorrelated event sequences, whereas true aftershock sequences
exhibit robust values of $p$. To test this, we analyzed an empirical dataset of
earthquakes from the Southern California Earthquake Data Center, the SCSN
Catalog
(1932-2024)~\cite{californiainstituteoftechnologycaltechSouthernCaliforniaSeismic1926}
(see Appendix-\ref{sec:data_set_description} and Supplemental
Fig.~S1 for details). From the joint distribution of
rescaled variables, we selected appropriate threshold values $\eta'_{th}$ to
distinguish between background events and aftershocks for different parameters
(Fig.~\ref{fig:Rij_Tij}b and Supplemental
Fig.~S6). We then analyzed the dependency of
the Omori-Utsu relation on the parameters $h$ and $D'$ for events labeled
background and aftershocks separately, as shown in Figs.~\ref{fig:omori_h_D}b, f
and c, g, respectively (see Supplemental
Figs.~S7 and
S8 for $\lambda(m, \tau)$ with different
$m$). While background events exhibit a behaviour analogous to our simulations
of the null model, we find that the exponent $p$ for aftershocks is robust
across different values of $h$ and $D'$ parameters. Furthermore, as for the null
model, the exponent $p$ does not change substantially under variations of the
$b'$ parameter for both background events and aftershocks (Supplemental
Fig.~S2).

\subsection{ETAS model}

To provide further support for our hypothesis, we analyzed a synthetic
earthquake catalog with aftershocks generated by the ETAS
model~\cite{ogataSpaceTimePointProcessModels1998} (see
Appendix-\ref{sec:data_set_description}, Fig.~\ref{fig:Rij_Tij}c and
Supplemental Figs.~S1 and S9 for details). Indeed, aftershocks identified by the
NN method exhibit also robust behavior of the Omori-Utsu exponent $p$ under
variations of the parameters $h$ and $D'$ (Figs.~\ref{fig:omori_h_D}d and h, and
Supplemental Fig.~S10).

\subsection{Productivity relations}

Apart from the Omori-Utsu relation, our analytical results for $D' = 0$ also
predicted a spurious productivity relation in the uncorrelated null model
(Eq.~\eqref{eq:nu}). Indeed, we find that both the null model for $D' > 0$ and
background events in the SCSN catalog show spurious productivity relations for
different values of $h$ and $D'$ (Fig.~\ref{fig:productivity_h_D}). The
estimated exponent $\alpha$ of the spurious productivity relation varies with
the parameters $h$ and $D'$ but only slightly. Hence, the distinction to the
case with true aftershocks --- exemplified here by the events labeled as
aftershocks in the SCSN and ETAS catalogs --- is less pronounced compared to the
Omori-Utsu relation. A similar behavior is observed for variations with the
parameter $b'$ (Supplemental Fig.~S3).

\subsection{Natural swarm catalog}

\begin{figure}[t!p]
\centering
\subfloat[]{%
\includegraphics[width=0.5\columnwidth]{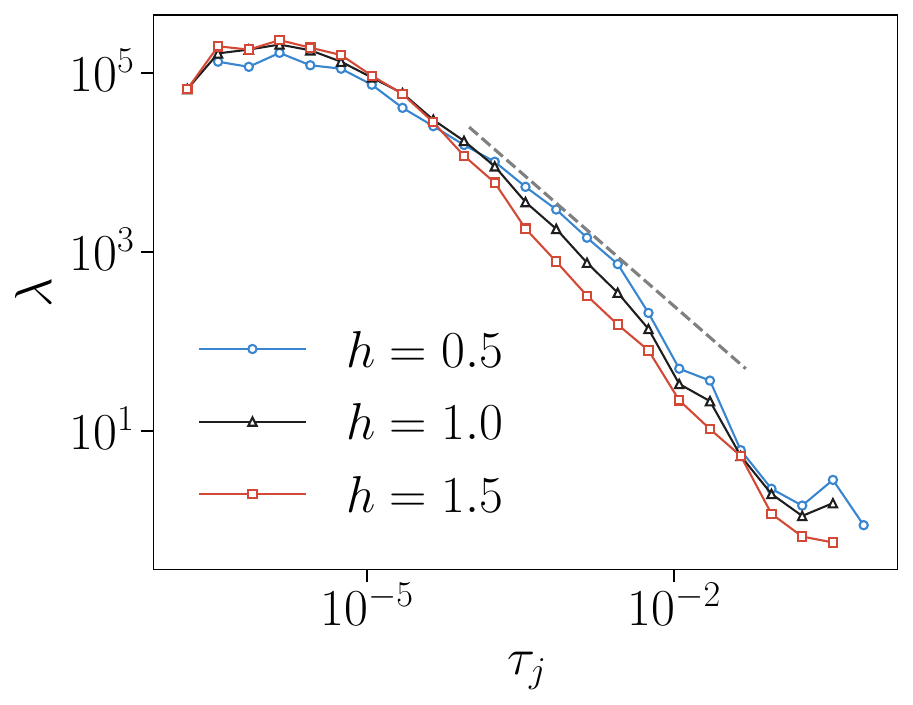}%
}
\subfloat[]{%
\includegraphics[width=0.5\columnwidth]{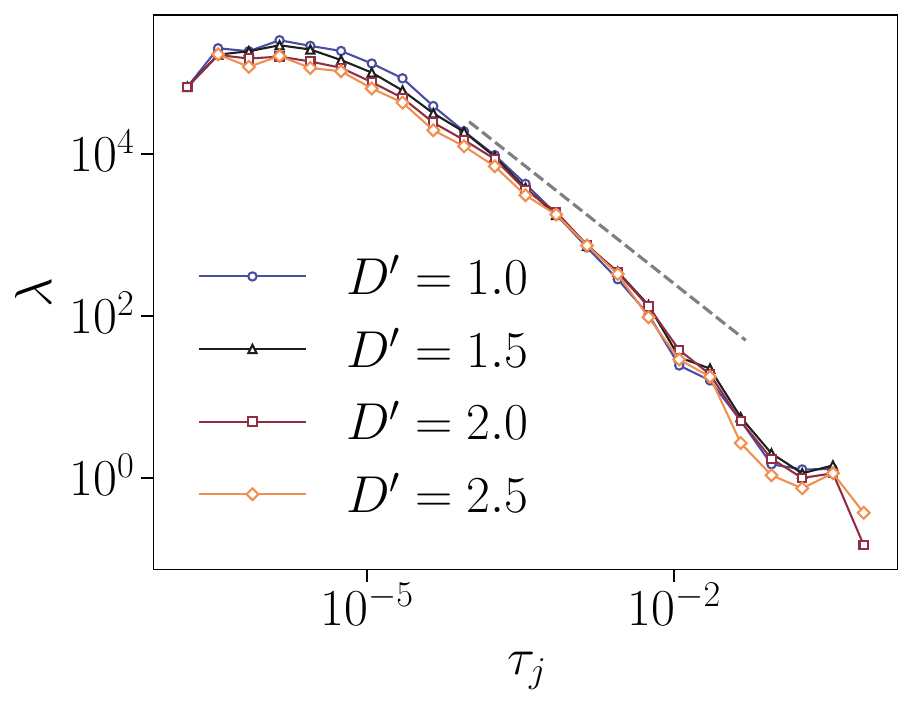}%
}
\caption{Robust Omori-Utsu relation in the natural swarm catalog.  As in Fig.~\ref{fig:omori_h_D} but for mainshocks with magnitude $m\in [2, 3)$. Other parameters in each panel are fixed at $h = 1$, $D' = 1.6$ and $b' = 1$. The gray dashed line is a guide to the eye $\propto \tau^{-1}$.}\label{fig:omori_shelly}
\end{figure}

As mentioned above, some empirical seismic catalogs---often in the context of
fluid-induced seismicity including natural swarms and fluid injections---do not
exhibit a bimodal form in the joint distribution of $T_{j}$ and $R_{j}$. This is
often attributed to an overlap of uncorrelated and correlated regions in the
joint distribution of a given
dataset~\cite{maghsoudiIntereventTriggeringMicroseismicity2018,OK_khajehdehi2023potential}. In
extreme cases, it might be even possible that only one mode is present,
comprising either aftershocks or background events.  Having established the
dependence of $p$ on the parameters $h$ and $D'$ as a signature of background
events, we are now in a position to test whether background events dominate or
not. As a case study, we use a high-quality seismic catalog of natural swarms
from the Long Valley Caldera,
California~\cite{shellyFluidfaultingEvolutionHigh2016} (see
Appendix~\ref{sec:data_set_description}, Fig.~\ref{fig:Rij_Tij}d and Supplemental
Figs.~S1 and~S11 for
details). As Fig.~\ref{fig:omori_shelly} shows, the $p$-value in the Omori-Utsu
relation is robust upon variations of the parameters $h$ and $D'$ (see also
Supplemental Fig.~S12), analogous to the events
labeled as aftershocks for the different catalogs in
Fig.~\ref{fig:omori_h_D}. This indicates that the unique cluster of events in
the swarm catalog is dominated by aftershocks.

\section{Conclusions}

The correct identification of aftershocks is a crucial issue for hazard assessment in seismology, with additional relevance in other fields such as condensed matter physics and material sciences. Our findings show that the often used NN declustering method can lead to spurious statistical properties mimicking the Omori-Utsu and productivity relations in catalogs lacking any sort of aftershock activity. By proposing and analyzing a generalized NN method, depending on a set of method parameters, we find that natural and synthetic earthquake catalogs with aftershocks exhibit robust statistical relations largely independent of the method's parameters, while catalogs without true aftershocks lead to relations that depend in a predictable way on those parameters. We conclude that robustness of the exponent $p$ in the Omori-Utsu relation with respect to the parameters of the modified NN method represents a true indicator of the presence of aftershocks, which allows to tackle cases where the original NN method does not provide a clear distinction between aftershocks and background events. For the specific cases study of natural swarms in the Long Valley Caldera in California, this approach allows us to establish that aftershock triggering plays a dominating role. In conclusion, our results represent an advancement in the statistical detection and prediction of aftershock activity, with direct applications in seismology and other research areas where event-event triggering behavior is observed.

\begin{acknowledgments}
  A.P. and R.P.-S. acknowledge financial support from project
  PID2022-137505NB-C21, funded by MICIU/AEI/10.13039/501100011033, and by “ERDF:
  A way of making Europe”.  J.B. acknowledges financial support from project
  PID2022- 136762NA-I00 funded by MICIU/AEI/10.13039/501100011033 and FEDER, UE.
  J.D. was supported by the Natural Sciences and Engineering Research Council of
  Canada (RGPIN/05221-2020).  We thank \'A. Corral and \'A. Gonz\'alez for
  helpful discussions.

\end{acknowledgments}

\appendix

\section{Null model of uncorrelated seismic activity}

\label{sec:null_model}

We consider a null model of earthquake event generation defined as follows:
\begin{itemize}
    \item The times of the different events $i$, $t_i$,  are drawn independently from a probability distribution $P_t(t)$.
    \item The position of event $i$, $\vec{r}_i = (x, y)$, is extracted from a distribution $P(x, y) = p_x(x)p_y(y)$, independently for each event.
\item Each event has associated a uniform random number $\mu_i = U(0, 1)$ that
  represents the value of the cumulated distribution of magnitudes evaluated at
  the value of the magnitude associate to event $i$. For a normalized
  distribution of magnitudes
  \begin{equation}
    P(m) = b \ln(10) \; 10^{-b(m - m_0)},
  \end{equation}
  we have a cumulated distribution
  \begin{equation}
    \label{eq:2}
    P_c(m) = \int_m^\infty P(m')dm' = 10^{-b(m - m_0)} \equiv \mu.
  \end{equation}
  Thus, for an event with $\mu_i$, the associated magnitude is~\cite{rossSimulation2022}
  \begin{equation}
    \label{eq:3}
    m_i = m_0 - \frac{\log_{10} \mu_i}{b},
  \end{equation}
  where $m_0$ is the minimal magnitude in the model.
\end{itemize}

In the simulations presented in the main text, we use $N=22814$ events (as in
the SCSN catalog, see Sec.~SM-\ref{sec:data_set_description}), and we choose
distributions $P_tt)$, $p_x(x)$ and $p_y(y)$ uniform in the interval $[0,1)$.
For the magnitude, we fix $m_0 = 3$ and $b=1$. To improve statistics, numerical
results are averaged over $3$ independent model samples.

\section{Analytics of the null model}
\label{sec:analytics_Omori_law_null}

Here we focus in the calculation of the probability distribution of times between parents and aftershocks, which has the same functional form  as the Omori-Utsu relation. We consider the generalized proximity, Eq.~\eqref{eq:5}, namely
\begin{equation}
  \eta_{ij}' \equiv
  \begin{cases}
    t_{ij}^h \,r_{ij}^{D'} \,10^{-b' m_i}, & \text{if}\ i < j \\
    \infty, & \text{otherwise}
    \end{cases}.
  \label{eq:BPmodified}
\end{equation}
Given the value of the magnitude $m_i$ in Eq.~\eqref{eq:3}, we have
\begin{equation}
    10^{-b' m_i} = 10^{-b' m_0} 10^{b' \log_{10}(\mu_i)/b} \equiv \mu_i^{1/\beta} \; 10^{-b' m_0},
\end{equation}
where we have defined $\beta = b/b'$. The generalized proximity takes now the form
\begin{equation}
  \eta_{ij}' \equiv
  \begin{cases}
    t_{ij}^h \,r_{ij}^{D'} \, \mu_i^{1/\beta} 10^{-b' m_0}, & \text{if}\ i < j \\
    \infty, & \text{otherwise}
    \end{cases}.
  \label{eq:5bis}
\end{equation}
To simplify calculations, we disregard the constant factor $10^{-b' m_0}$ in
Eq.~\eqref{eq:5bis}, which does not affect the calculation of predecessors.

If the event times are distributed uniformly, they are equivalent to a Poisson
process~\cite{kingmanPoissonProcesses1992}. This is a homogeneous process that
we can approximate as an equi-spaced distribution of points, that we can
normalize to take the integer values
\begin{equation}
  \label{eq:5bis2}
  t_0 = 0, \quad t_1 = 1, \quad t_2 = 2, \quad \ldots,\quad t_{N-1} = N-1.
\end{equation}
With this approximation
\begin{equation}
  \label{eq:6}
  t_{ij} = t_j - t_i = j - i,
\end{equation}
that is, for a given event $n$, the distances of the candidate parents
$i = 0, \ldots, n-1$ are $t_{in} = 1, 2, \ldots, n$.

\subsection{A particular case: No geographical dependence}

We  consider the analytically solvable case of no geographical dependence, which translates into  $D' = 0$. In this case, for an event $n$,
the list of proximities to the candidate parents is
\begin{equation}
  \label{eq:28}
  \eta_m' = m^h \times \mu^{1/\beta}, \quad m=1,2,\ldots, n,
\end{equation}
where $\mu$ are random independently drawn numbers extracted from the uniform
distribution $U(0, 1)$. The random variable $z = \mu^{1/\beta}$ can be obtained from the distribution $P(\mu) = 1$ applying the rules of change of variables, namely
\begin{equation}
    P_z(z) = P(\mu(z)) \frac{d \mu}{d z},
\end{equation}
which leads to
\begin{equation}
    P_z(z) = \beta z^{\beta - 1}.
    \label{eq:prob_z}
\end{equation}
Notice that, since $\mu \in [0,1]$, then $z \in [0,1]$ also. Now, since $\eta_m' = m^h z$, we have that the distribution of $\eta'$, conditioned to a fixed value $m$, is given by
\begin{equation}
    \label{eq:eta_zero}
    P(\eta' | m)  = \frac{\beta}{m^h} \left( \frac{\eta'}{m^h}\right)^{\beta-1}.
\end{equation}
The maximum value of $\eta'$ is of course $m^h$.

\subsection{Rate of aftershocks after a mainshock}

To compute $\lambda (m,t)$, we need to compute the probability that an event has a parent at some given temporal distance $t$. With our selection of times, this distance will be an integer number $r$. Considering the event $n$ at time, it has $n$ candidates as parents, located at times $n-1$, $n-2$, \ldots, $0$, that are at temporal distances $1$, $2$, \ldots, $n$. In terms of the proximity, a candidate at distance $r$ will be a parent if the proximity $\eta_{rn}'$ is smaller than the proximity $\eta_{in}'$ for all $i \neq r$, $i<n$. Since all quantities $\mu$ (or $z$) are randomly and independently distributed, we have that the proximity of the candidates to $n$ are $\eta_{m}'$, $m = 1, 2, \ldots, n$, where the random numbers $\eta_m'$ are extracted from the distribution Eq.~\eqref{eq:eta_zero}. The situation to choose the parent of event $n$ is thus equivalent to randomly drawing the numbers
\begin{equation}
  \label{eq:17}
  \eta_1', \eta_2',\ldots, \eta_n',
\end{equation}
drawn from the distribution $P(\eta' | m)$ in Eq.~\eqref{eq:eta_zero}, and choosing the smallest one. The probability that the event $r$, at distance $r$ from event $n$, has the smallest value $\eta_r' \equiv \eta'$ is equal to the probability that  $\eta_j' > \eta'$, $j \neq r$.  Defining the cumulative distribution
\begin{equation}
    P_c(\eta'|m) = \int_{\eta'}^{m^h} P(\eta''|m) d\eta'' = 1 - \left( \frac{\eta'}{m^h}\right)^\beta,
\end{equation}
we have that the probability that the candidate $r$ has the smallest proximitiy to an event $n$, with a given minimal $\eta'$, is given by
\begin{equation}
    P(r | n, \eta') = \frac{\prod_{m=1}^n  P_c(\eta' |
    m)}{ P_c(\eta' | r)} =
   \frac{\prod_{m=1}^n  \left[ 1 - \left(\dfrac{\eta'}{m^h}\right)^\beta
 \right]}{ 1 - \left(\dfrac{\eta'}{r^h} \right)^\beta}.
\end{equation}
The probability that any aftershock is at a distance $r$ of its predecessor is
obtained by averaging over all events and all possible values of the minimal
proximity $\eta'$. This probability takes  the form
\begin{widetext}
\begin{equation}
  \label{eq:32}
  P(r) = \frac{\beta}{N r^h} \sum_{n=r}^{N-1} \int_0^1 d\eta' P(\eta'|n) P(r|n, \eta') =
   \frac{\beta}{N r^{h\beta}} \sum_{n=r}^{N-1} \int_0^1 d\eta' \eta'^{\beta-1}
\frac{\prod_{m=1}^n  \left[ 1 - \left(\dfrac{\eta'}{m^h}\right)^\beta
 \right]}{ 1 - \left(\dfrac{\eta'}{r^h} \right)^\beta},
\end{equation}
where the integral over $\eta'$ extends up to $1$ because the maximum value of $\eta'_{m=1}$ is $1$.

To extract the asymptotic behavior for large $r$ and $N$, we consider
\begin{equation}
  \label{eq:33}
  \prod_{m=1}^n  \left[ 1 - \left( \frac{\eta'}{m^h}\right)^\beta \right] = \exp\left\{ \sum_{m=1}^n  \ln \left[ 1 - \left( \frac{\eta'}{m^h}\right)^\beta \right]
    \right\} \simeq \exp\left\{   -\eta'^\beta \sum_{m=1}^n  \frac{1}{m^{h \beta}}\right\},
\end{equation}
where we have expanded the logarithm since $\eta' \leq 1$ and $m > 1$. Defining  $H_{n,h\beta} = \sum_{m=1}^n m^{-h\beta}$ as the $n$-th generalized harmonic number or order $h\beta$, we can approximate, for large $r$,
\begin{equation}
  \label{eq:36}
  P(r) \simeq \frac{\beta}{r^{h\beta} N} \sum_{n=r}^{N-1}  \int_0^1 d\eta' \eta'^{\beta-1} e^{-\eta' H_{n,h\beta}} =
  \frac{1}{r^{h\beta} N} \sum_{n=r}^{N-1}  \frac{1 - e^{-H_{n,h\beta}}}{H_{n,h\beta}}.
\end{equation}
\end{widetext}

For $r$ large, $n \geq r$ is also large, so we can approximate $H_{n,a}$ by its asymptotic expansion, that takes the form
\begin{equation}
    H_{n,a} \simeq \left \{ \begin{array}{lr}
    \dfrac{n^{1-a}}{1-a} &   \mathrm{for} \; a < 1 \\
    \log n & \mathrm{for} \; a = 1 \\
    \zeta(a) & \mathrm{for} \; a >1
  \end{array}
    \right. ,
    \label{eq:harmonic_expansions}
\end{equation}
where $\zeta(a)$ is the Riemann Zeta function. For $a \leq 1$ the summation in the generalized harmonic number is divergent, and the asymptotic behavior for large $n$ in
Eq.~\eqref{eq:harmonic_expansions} is obtained by approximating the sum by an integral. For $a >1$ the sum is finite and converges for large $n$ to the Riemann Zeta function.

Let us consider the different possibilities, depending on $h \beta$.

(1) $h \beta = 1$.

Approximating the sum over $n$ by an integral, we have
\begin{eqnarray*}
  \label{eq:37}
  P(r) &\simeq& \frac{1}{r N} \int_r^N \frac{1 - e^{-\log(n)}}{\log n} dn \\
  &=&
   \frac{1}{r N} \left[   \int_r^N \frac{1}{\log n} dn  -  \int_r^N
                \frac{1}{n \log (n)} dn  \right] \\
       &=&  \frac{1}{r N} \left[  \mathrm{li}(N) -  \mathrm{li}(r) - \log(\log(N)) +
           \log(\log(r))  \right],
\end{eqnarray*}
where $\mathrm{li}(z)$ is the logarithmic integral function~\cite{abramowitzHandbookMathematicalFunctions1968}. For large $z$,
\begin{equation}
  \label{eq:38}
  \mathrm{li}(z) \simeq \frac{z}{\log(z)}.
\end{equation}
For $N \to \infty$, the $\log(\log(N))$ term is irrelevant, and therefore we can approximate
\begin{equation}
  \label{eq:39}
  P(r) \simeq  \frac{1}{r N} \left[  \frac{N}{\log(N)} - \frac{r}{\log(r)}
  \right] \simeq   \frac{1}{r N}  \frac{N}{\log(N)}.
\end{equation}

(2) $h \beta > 1$

We have now
\begin{eqnarray*}
  P(r) &\simeq& \frac{1}{r^{h \beta} N} \int_r^N \frac{1 - e^{-\zeta(h
                \beta)}}{\zeta(h \beta)}  dn \nonumber  \\
  &=&\frac{1 - e^{-\zeta(h \beta)}}{\zeta(h \beta)} \frac{N - r}{N r^{h \beta}} \simeq r^{-h \beta}.
\end{eqnarray*}

(3) $h \beta < 1$

In this case,
\begin{eqnarray*}
  P(r) &\simeq& \frac{1-h \beta}{r^{h \beta} N} \int_r^N \frac{1 - e^{-n^{1-h
  \beta}/(1-h \beta)}}{n^{1-h \beta}} dn \nonumber\\
  &=& \frac{1- h \beta}{r^{h \beta} N^{1-h \beta}} \int_{r/N}^1
  \frac{1 - e^{- N^{1-h \beta} x^{1-h \beta}/(1-h \beta)}}{x^{1-h \beta}} dx,
\end{eqnarray*}
where we have applied the change of variables $n = x / N$. For $h\beta > 1$, the argument in the exponential becomes very small for large $N$, so we can approximate
\begin{eqnarray*}
    P(r) &\simeq& \frac{1-h \beta}{r^{h \beta} N^{1-h \beta}} \int_{r/N}^1  x^{h
                  \beta-1} dx \nonumber \\
  &=& \frac{1-h \beta}{h \beta r^{h \beta} N^{1-h \beta}} \left[ 1 - \left( \frac{r}{N}\right)^{h \beta} \right] \simeq N^{h \beta-1} r^{-h \beta}.
\end{eqnarray*}

That is, in terms of the real time $\tau = r/N$,  the probability that an aftershock takes place at a time $\tau$ of its predecessor scales, for any $h \beta$, as
\begin{equation}
    P(\tau) \sim \tau^{- h\beta},
\end{equation}
and thus, for large $\tau$, averaged over all mainshocks, we obtain the Omori-Utsu-like relation
\begin{equation}
    \lambda(\tau) \sim \tau^{-h \beta}.
    \label{eq_d_0_omori}
\end{equation}

\subsection{Productivity relation}

We can also apply our formalism to estimate $\nu(m)$, defined as the average number of aftershocks generated by an event of magnitude $m$ or, in other words, the number of events that have associated the event $n$ as predecessor.

Consider the event $n$, that has associated a value $\mu$, characteristic of its
magnitude $m$ by the relation Eq.~\eqref{eq:3}. It can be the parent (associated
mainshock) of any of the events on its right, $m = n+1, n+2, \ldots,
N-1$. Consider one of these events $m$ on the right of $n$, and at a temporal
distance $r$ from $n$, such that $m = n + r$. The event $n$ has associated a
proximity to $m$ $\eta_r' = \mu^{1/\beta} r^h$. The probability $p_{n,\mu, r}$
that $m$ is an aftershock of $n$ is equal to the probability that all the parent
candidates of $m$, at distances from $m$ given by $j=1, 2, \ldots, n+r$, have a
distance $\eta'$ larger than $\eta_r'$. Thus we have
\begin{widetext}
\begin{equation}
  \label{eq:41}
  p_{n,\mu, r} = \frac{\prod_{j=1}^{n+r} P_c(\eta_r'| j) }{  P_c(\eta_r' | r)} =
  \dfrac{\prod_{j=1}^{n+r}  \left[  1 - \left(\dfrac{\eta_r'}{j^h} \right)^\beta\right]}{1 - \left(\dfrac{\eta_r'}{r^h} \right)^\beta} =
  \frac{1}{1-\mu} \prod_{j=1}^{n+r}  \left[  1 - \mu \left(\dfrac{r}{j} \right)^{h  \beta}\right].
\end{equation}

\end{widetext}
Notice that, for $n$ to be the predecessor of $m$, $\eta_r'$ must correspond to the minimum proximity, and thus fulfill $\eta_r' \leq 1$, namely $\mu r^{h \beta} \leq 1$, ensuring that all terms in the square brackets in Eq.~\eqref{eq:41} are non-negative.

Each one of the events on the right of $n$, at distances $r = 1, 2, \ldots, N-1-n$, is connected to $n$ with probability $p_{n,\mu,r}$. Therefore, the probability that $n$ is connected to $k$ events on its right (i.e. it has $k$  aftershocks) is given by a Poisson binomial distribution, representing the probability distribution of a sum of $N-1-n$ independent Bernoulli trials that have different success probabilities $p_{n,\mu,r}$~\cite{hoeffdingDistributionNumberSuccesses1956}.
The Poisson binomial distribution has a rather complex form, but its average value can be computed as $\sum_r p_{n,\mu,r}$. Therefore, we can write the average productivity of a main event $n$, with magnitude given by $\mu$, as
\begin{eqnarray}
  \label{eq:42}
  \nu(n, \mu) &=&  \frac{1}{1-\mu} \sum_{r=1}^{N-1-n}
  \prod_{j=1}^{n+r} \left(  1 - \mu \left(\frac{r}{j} \right)^{\alpha\beta}
                  \right) \nonumber\\
  &\simeq&
  %\frac{1}{1-\mu} \sum_{r=1}^{N-1-n} \exp\left[ -\mu r^{\h \beta} \sum_{j=1}^{n+r}\frac{1}{j^{h \beta}} \right] =
  \frac{1}{1-\mu} \sum_{r=1}^{N-1-n} \exp\left[ -\mu r^{h \beta} H_{n+r, h \beta} \right],
\end{eqnarray}
where we have expanded the argument of the product taking advantage that $\mu (r/j)^{h \beta} < 1$. Averaging finally over all events that  can act as  a predecessor, in number $N-1$, we have the average productivity as function of the magnitude $m$ (in terms of $\mu$)
\begin{equation}
  \label{eq:43}
   \nu(\mu) =  \frac{1}{(N-1)(1-\mu)} \sum_{n=0}^{N-2}  \sum_{r=1}^{N-1-n}\exp\left[ -\mu r^{h \beta} H_{n+r, h \beta} \right].
\end{equation}
Rearranging the summations in this equation
\begin{equation}
  \label{eq:44}
   \nu(\mu) =  \frac{1}{(N-1)(1-\mu)}  \sum_{r=1}^{N-1} \sum_{n=0}^{N-1-r}  \exp\left[ -\mu r^{h \beta} H_{n+r, h \beta} \right].
\end{equation}
Consider the simplest case $h \beta > 1$. Approximating $H_{n+r, h \beta}$ by its asymptotic value $\zeta(h \beta)$,  considering $N$ large, and approximating sums by integrals,
\begin{eqnarray}
   \nu(\mu) &\simeq&  \frac{1}{N(1-\mu)}  \int_{0}^{N} dr \int_{0}^{N-r} dn
                     e^{ -\mu r^{h \beta} \zeta(\alpha\beta)} \nonumber \\
  &=&
   \frac{1}{N(1-\mu)}  \int_{0}^{N} dr (N-r)  e^{ -\mu r^{h \beta} \zeta(h \beta)} .
\end{eqnarray}
The argument of the integral is dominated by the values of $r$ close to zero, so we can approximate
\begin{eqnarray}
  \label{eq:45}
   \nu(\mu) &\simeq&
   \frac{1}{1-\mu}  \int_{0}^{\infty} dr  e^{ -\mu r^{h \beta} \zeta(h \beta)}
                     \nonumber \\
     &=& \mu^{-1/(h  \beta)}\frac{\zeta(h \beta)^{-1/(h \beta)}}{h \beta(1-\mu)}
                     \int_{0}^{\infty} dz z^{-1 + 1/(h \beta)} e^{-z} \nonumber \\
  &=&
   \mu^{-1/(h \beta)}\frac{\zeta(h \beta)^{-1/(h \beta)}}{h \beta(1-\mu)} \Gamma\left( \frac{1}{h\beta}\right),
\end{eqnarray}
where $\Gamma(z)$ is the Gamma function. Thus, we have  $\nu(\mu) \sim \mu^{-1/(h \beta)}$, or, in terms of the magnitude $m$, expressed in Eq.~\eqref{eq:3}, we obtain the productivity relation
\begin{equation}
    \label{eq:46}
    \nu(m) \sim 10^{\frac{b}{h \beta} (m - m_0)} =
    10^{\frac{b'}{h} (m - m_0)}.
\end{equation}

Calculations for $\alpha \beta \leq 1$ are more complex, but direct numerical simulations confirm that in general, for the null model with $D=0$, $\nu(m) \sim 10^{\frac{b'}{h} m}$.

\section{Rate of aftershocks calculation}\label{sec:rate_aftershocks_calculation}

We calculate the rate of aftershocks depending on the time interval $t_j$ and
spatial distance $r_j^D$. Each variable is binned and normalized, considering
that each parent may have different maximum time intervals and spatial
distances. Specifically, for each variable $z$, we calculate the rate $\nu_b$ at
bin $b$ using the formula:
\begin{equation}
    \nu_b = \frac{n_b}{\Delta z_b N_b},
\end{equation}
where $n_b$ is the number of aftershocks occurring in the bin, $\Delta z_b$ is the width of the bin, and $N_b$ is the total number of parents that could occur in the bin.

For the time interval, bin $b$ falls within the interval $\left[ t_{b}^-, t_{b}^+\right)$. Thus, $N_b$ includes all parents $p$ with time $t_p$ having a possible larger maximum time interval, i.e., those satisfying $ 1 - t_p \geq t_{b}^- $.

For the spatial distance, bin $b$ falls within the interval $\left[ (r^D)_{b}^-, (r^D)_{b}^+\right)$. Consequently, $N_b$ encompasses all parents $p$ with position $\vec{x}_p$ having possible larger maximum distances, i.e., those satisfying $ \max \left( \left| \vec{x}_p \right|^D, \left| \vec{x}_p - (1,0) \right|^D, \left| \vec{x}_p - (0,1) \right|^D, \left| \vec{x}_p - (1,1) \right|^D \right) \geq (r^D)_{b}^- $, considering all four corners of the box.

It is worth noting that normalizing based on the maximum time intervals and spatial distances of the parent, rather than all parents, only impacts the tails of the distribution.

\section{Data sets  description}\label{sec:data_set_description}

\subsection{SCSN catalog}
The SCSN catalog comprises earthquakes from Southern California spanning the years 1932 to 2024~\cite{californiainstituteoftechnologycaltechSouthernCaliforniaSeismic1926}. It provides the position  with latitude and longitude coordinates. We convert it into Universal Transverse Mercator (UTM) coordinates employing the UTM python library~\footnote{B. van Andel, T. Bieniek, T. I. Bø., utm, \url{https://github.com/Turbo87/utm}}.  The UTM system is a map projection that treats the Earth's surface as a perfect ellipsoid and assigns coordinates to locations, disregarding altitude. We have found that calculating distances for Southern California earthquakes using UTM coordinates yields very similar results to those calculated using the Haversine formula directly with latitude and longitude coordinates. We retain all earthquakes with magnitude $m\geq3$. This results in a catalog composed by $N=22814$ events. Then, we normalize the time and position coordinates in the interval $\left[0,1\right)$ using the formula:
\begin{equation}
    x_\text{norm} = \frac{x - x_\text{min}}{x_\text{max} - x_\text{min}},\label{supp:eq:x_norm}
\end{equation}
where we use the same $x_\text{min}$ and $x_\text{max}$ for the two spatial coordinates to maintain the scale proportion.

\subsection{ETAS model}

The statistics of aftershock production are well captured by a Hawkes self-excitation model, commonly known as epidemic type aftershock sequence (ETAS) model \citep{ogataSpaceTimePointProcessModels1998}, the intensity of which is a linear combination of background activity rate $\mu_0$ and history dependent aftershock rate $\Psi_i$ caused by each previous event:
\begin{equation}
\mu (t,\mathbf{r},m) = \mu_0 (t,\mathbf{r},m) + \sum_{t_i<t} \Psi_i (t,\mathbf{r},m| t_i,\mathbf{r}_i,m_i).
\label{eq:muETAS}
\end{equation}
The ETAS model is interpreted as the outcome of a branching process, i.e. an event-event triggering phenomena, where each event is either an independent background event, that we will assign the label of a mainshock, given by rate $\mu_0$ or an aftershock uniquely linked to a single parent event, rendering a forest of independent aftershock sequences. In our synthetic catalog, the aftershock rate in the ETAS model is factorized as:
\begin{equation}
\Psi_i (t,\mathbf{r},m| t_i,\mathbf{r}_i,m_i) = \rho(m) \nu(m_i)\Psi_t (t - t_i)\Psi_r (\mathbf{r} - \mathbf{r}_i).
\label{eq:psiETAS}
\end{equation}

Starting from a background population sampled from a uniform rate $\mu_0 = 10^{-7}~\textrm{s}^{-1}\textrm{km}^{-2}$, the ETAS catalog is generated by building a branching process where each event $i$ marked by $\lbrace m_i,t_i, \mathbf{r}_i \rbrace$ generates a number of aftershocks drawn from a Poisson distribution with parameter $\nu(m_i)$. Each aftershock is then assigned a location and time given $\Psi_r(\mathbf{r}-\mathbf{r}_i)$ and $\Psi_t(t-t_i)$.
Aftershock sizes are independently distributed with the GR relation $\rho(m)=b \ln(10) 10^{-b(m-m_0)}$ with $b=1$. The productivity parameter $\nu(m_i)$, the expected number of direct aftershock events after a parent event of magnitude $m_i$, is given by the so-called productivity relation.
\begin{equation}
\nu (m) = \nu_0 10^{\alpha (m-m_0)},
\label{eq:prod ETAS}
\end{equation}
where $\alpha = 0.5$ and $\nu_0 = 0.118585 ~\textrm{s}^{-1}$ is the expected number of direct aftershocks for an event with $m_0=1$, corresponding to an average branching ratio integrating for all event sizes $n_b=
\frac{\nu_0 10^{-\alpha m_0}}{1-\alpha/b} =
0.75$.

We select a normalized spatial kernel $\Psi_r ({r'})$ isotropic and radially distributed as a Gaussian with $\sigma=1~\textrm{km}$, and a temporal kernel inspired by the Omori-Utsu relation~\cite{omoriAftershocksEarthquakes1894}:
\begin{equation}
\Psi_t (t') = \frac{\theta C^{\theta}}{(t' + C)^{1+\theta}},
\label{eq:OmoriETAS}
\end{equation}
where the exponent $p  = 1+\theta = 1.2$ and $C = 1~\textrm{s}$ is an empirical constant.

We retain earthquakes with magnitude $m \geq 3$. We consider $N=22814$ events. We normalize the time and position coordinates in the interval $\left[0, 1\right)$ using formula~\ref{supp:eq:x_norm}. To improve statistics, numerical results are averaged over 3 independent model samples.

\subsection{Natural swarm catalog}

We use a natural swarm catalog of Long Valley Caldera in California from 2014~\cite{shellyFluidfaultingEvolutionHigh2016}. We retain earthquakes with magnitude $m \geq 0$. This results in a catalog with $N=4703$ events. We normalize time and position coordinates in the interval $[0,1)$ with Eq.~\ref{supp:eq:x_norm}.

\bibliographystyle{apsrev4-1}
\bibliography{all, bibJB}%

%%%%%%%%%%%%%%%%%%%%%%%%%%%%%%%%%%%%%%%%%%%%%%%%%

\clearpage

\newpage

\renewcommand{\thepage}{\arabic{page}} 
\renewcommand{\theequation}{SE\arabic{equation}} 
\renewcommand{\thesection}{\arabic{section}}  
\renewcommand{\thetable}{ST\arabic{table}}  
\renewcommand{\thefigure}{SF\arabic{figure}}
\renewcommand{\thevideo}{SV\arabic{video}}

%\preprint{APS/123-QED}

\setcounter{page}{1}
\setcounter{section}{0}
\setcounter{figure}{0}
\setcounter{equation}{0}

\onecolumngrid
%\appendix
\begin{center}
\textbf{\large Supplemental Material\\~\\}
\end{center}

\section*{Supplementary Figures}

\begin{figure*}[h]
\subfloat[]{%
  \includegraphics[width=0.25\textwidth]{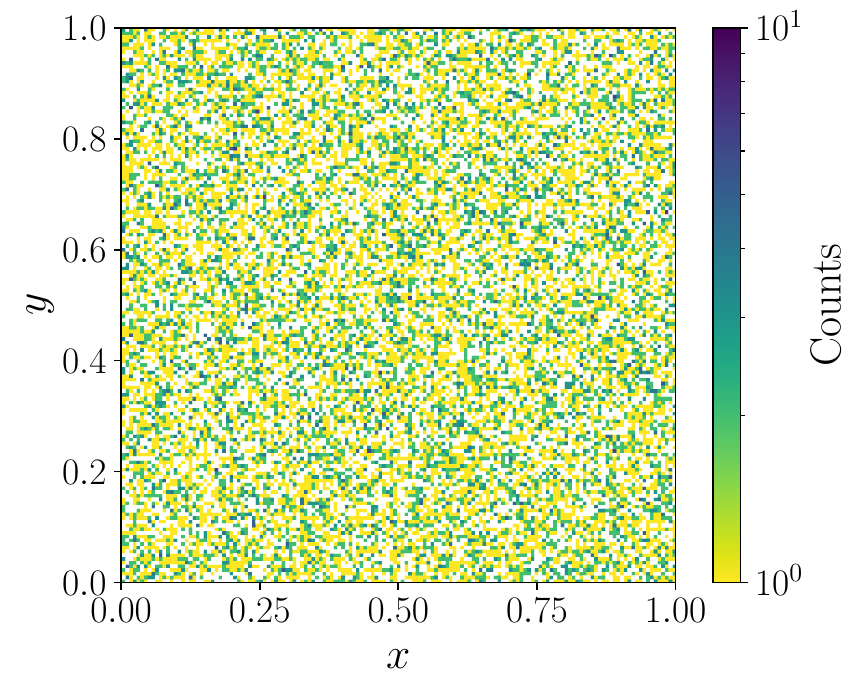}%
}
\subfloat[]{%
  \includegraphics[width=0.25\textwidth]{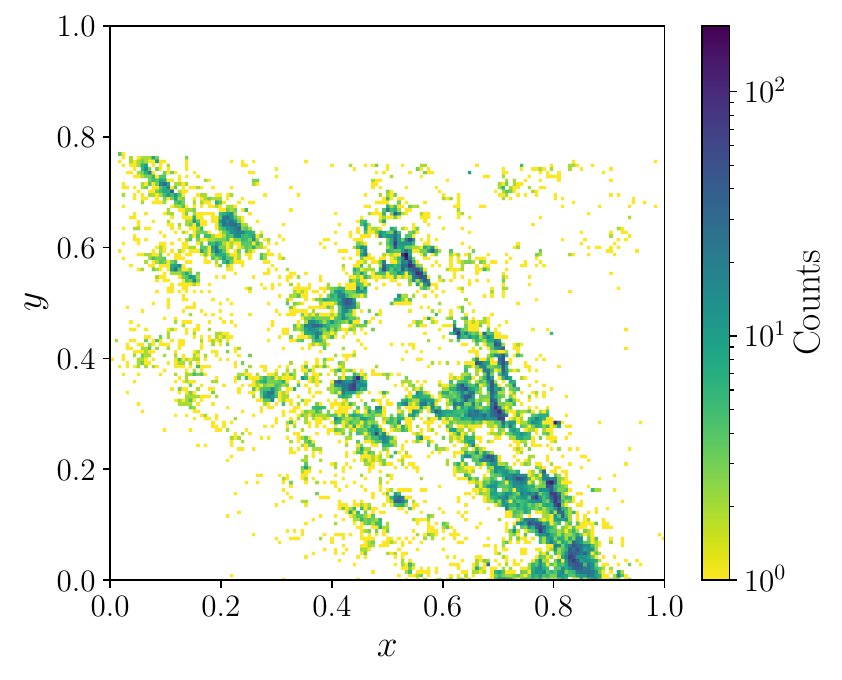}%
}
\subfloat[]{%
\includegraphics[width=0.25\textwidth]{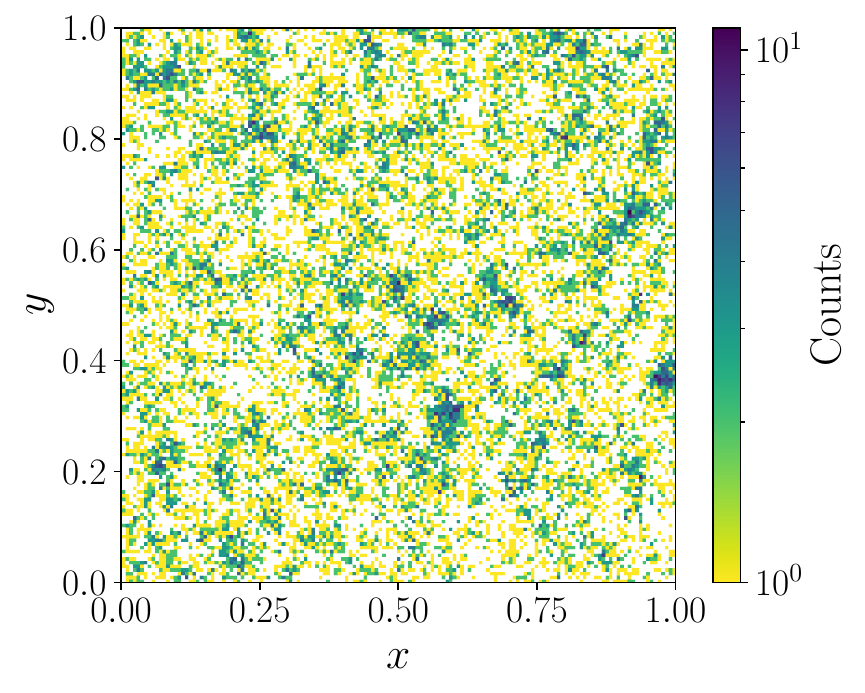}%
}
\subfloat[]{%
\includegraphics[width=0.25\textwidth]{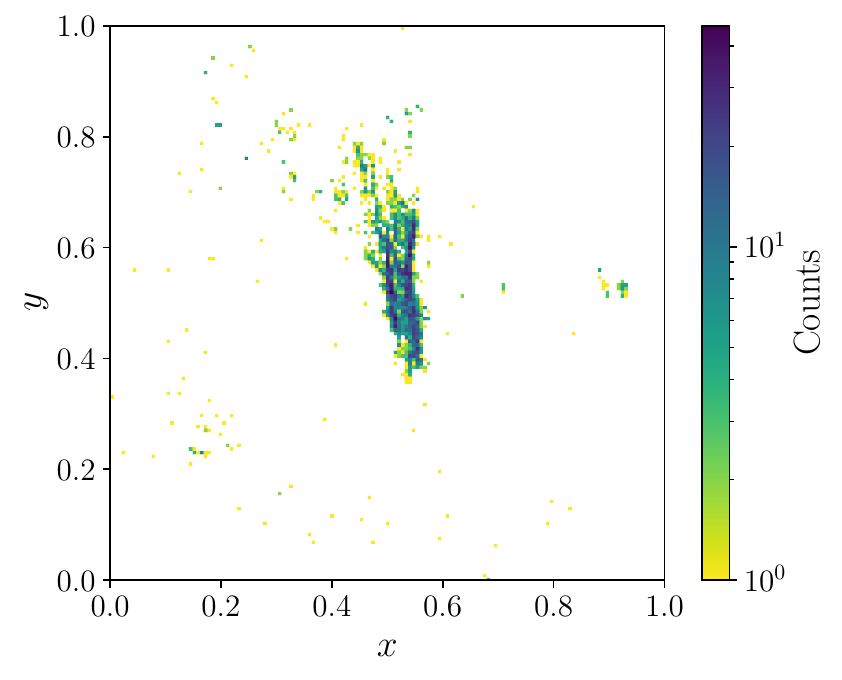}%
}

\subfloat[]{%
  \includegraphics[width=0.26\textwidth]{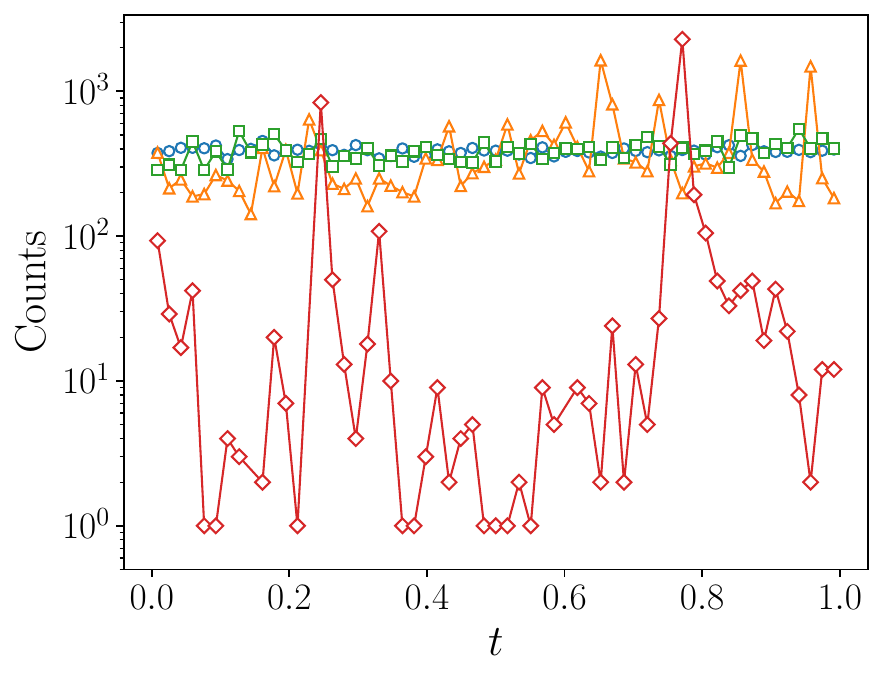}%
}
\subfloat[]{%
  \includegraphics[width=0.26\textwidth]{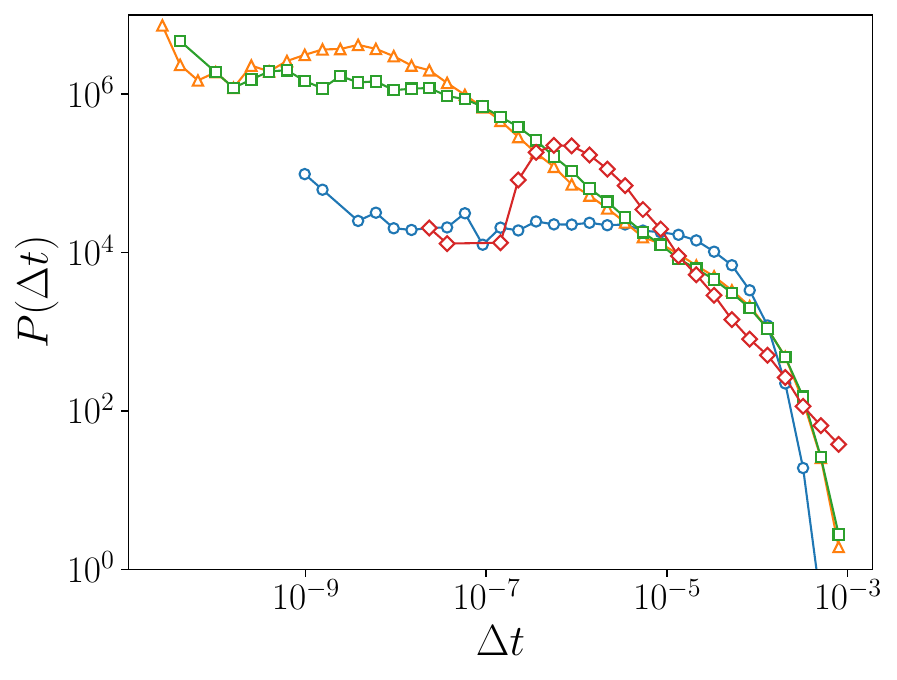}%
}
\subfloat[]{%
  \includegraphics[width=0.4\textwidth]{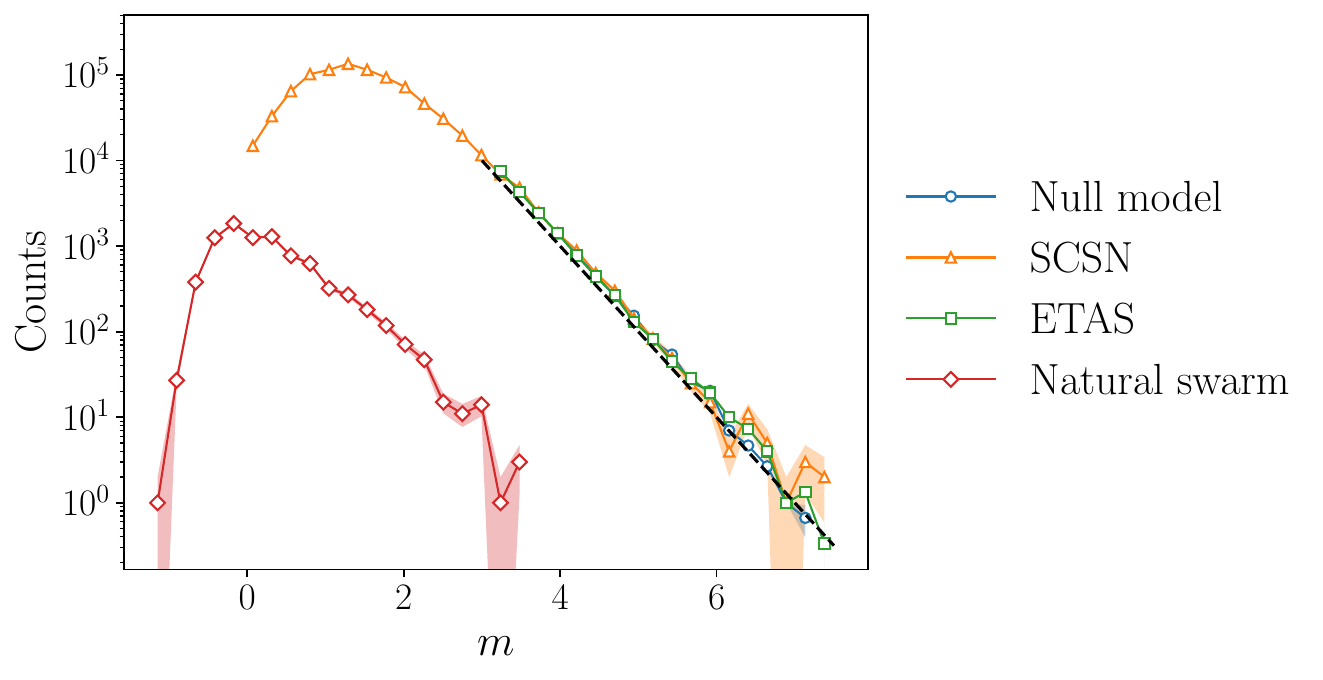}%
}
\caption{Description statistics for the different data sets. Counts for the position in (a) the null model, (b) the SCSN catalog, (c) the ETAS model and (d) the natural swarm catalog, (e) the time, (f) PDF of the inter-event time and (g) counts of the magnitude of events in the different data sets. The results of the null model and the ETAS model in (a), (c) and (e) correspond to a single run, while in (f) and (g) correspond to the average of all runs. The black dashed line in (g) is proportional to $10^{-m}$.} \label{supp:fig:dataSets}
\end{figure*}

%--
\begin{figure*}[t!p]
    \centering
\includegraphics[width=0.95\textwidth]{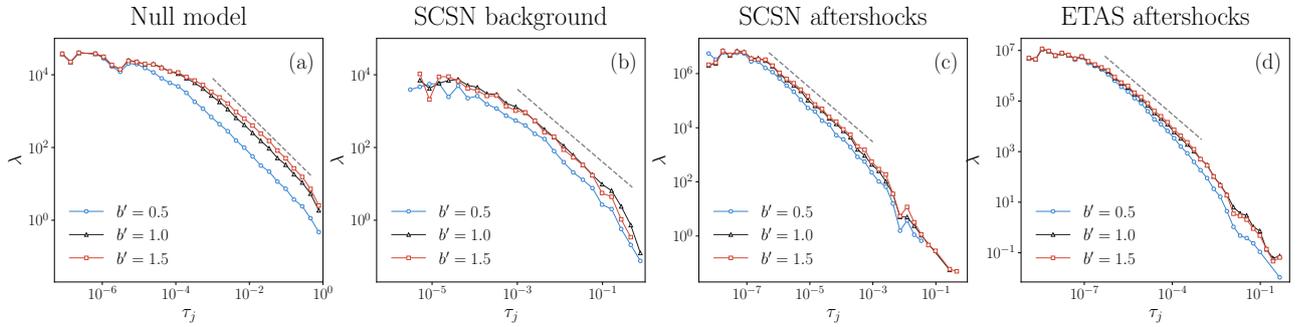}
    \caption{Rate of the number of aftershocks $\lambda(m, \tau)$ after a mainshock of magnitude $m\in [5, 6)$ depending on the time interval $\tau$ for (a) the null model, (b) events labeled as background events in the SCSN catalog, (c) events labeled as aftershocks in the SCSN catalog and (d) events labeled as aftershocks in the ETAS catalog. We keep other parameters fixed at $h = 1$, (a) and (d) $D' = 2$, (b) and (c) $D'=1.6$, and $b' = 1$. The gray dashed line is proportional to $\tau^{-1}$.}
    \label{supp:fig:omori_b}
\end{figure*}

%--
\begin{figure*}[t!p]
    \centering
\includegraphics[width=0.95\textwidth]{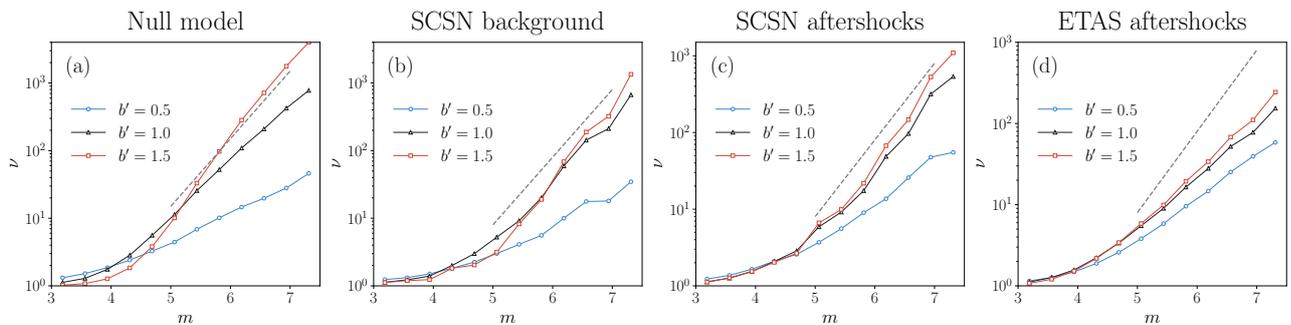}
    \caption{Average number of aftershocks $\nu(m)$ per mainshock magnitude $m$ for (a) the null model, (b) events labeled as background events in the SCSN catalog, (c) events labeled as aftershocks in the SCSN catalog and (d) events labeled as aftershocks in the ETAS catalog. We keep other parameters fixed at $h = 1$, (a) and (d) $D' = 2$, (b) and (c) $D'=1.6$. The gray dashed line is proportional to $10^{m}$.}
    \label{supp:fig:productivity_b}
\end{figure*}

%%%%---
\begin{figure*}[t!p]
\subfloat[]{%
\includegraphics[width=0.2\textwidth]{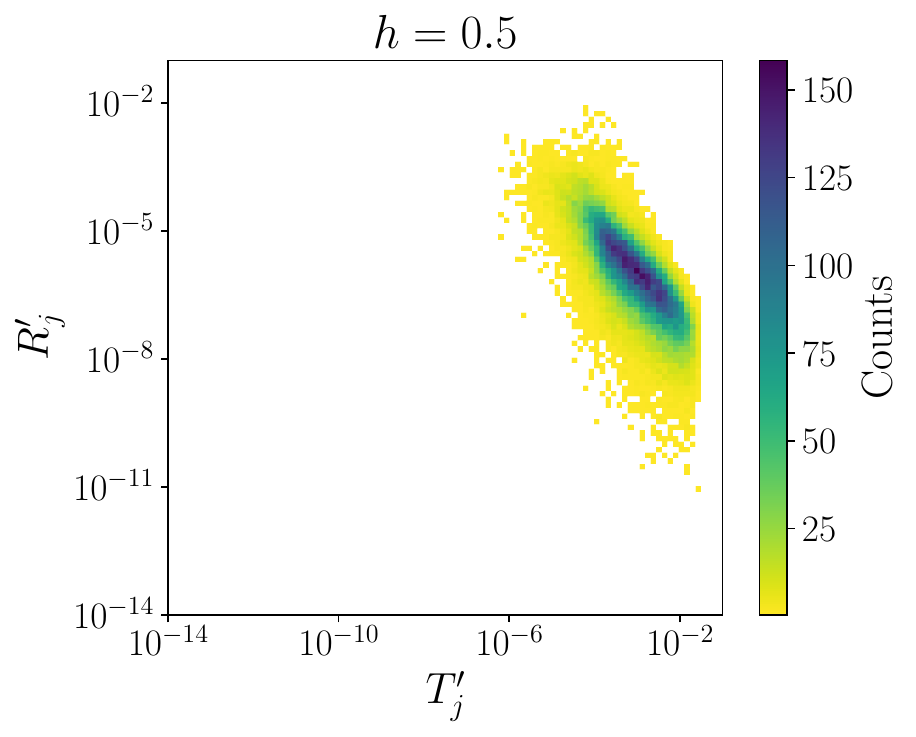}%
}
\subfloat[]{%
\includegraphics[width=0.2\textwidth]{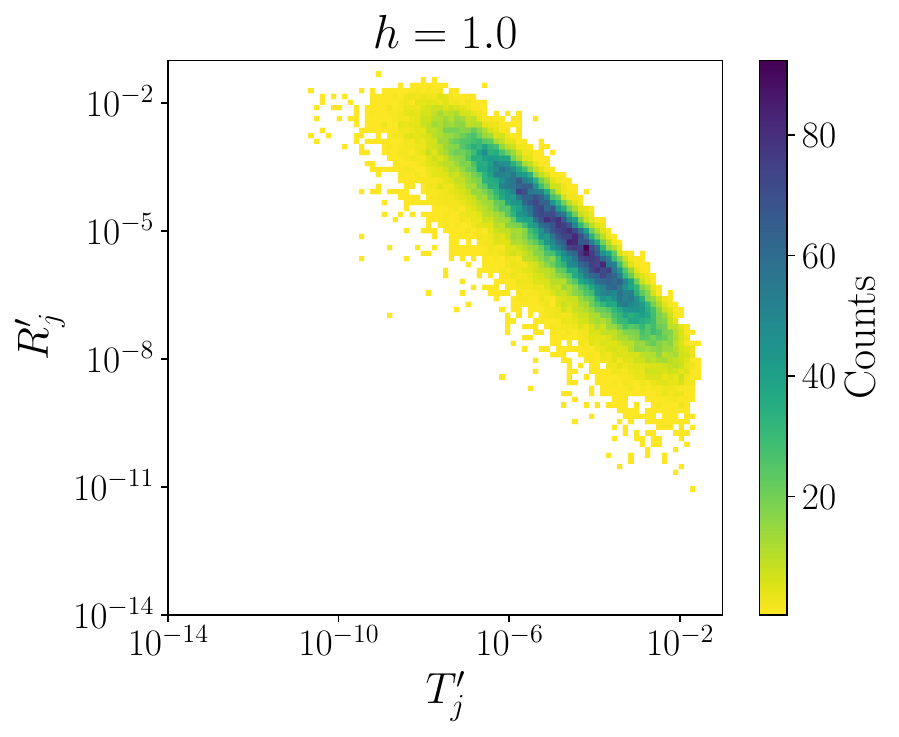}%
}
\subfloat[]{%
\includegraphics[width=0.2\textwidth]{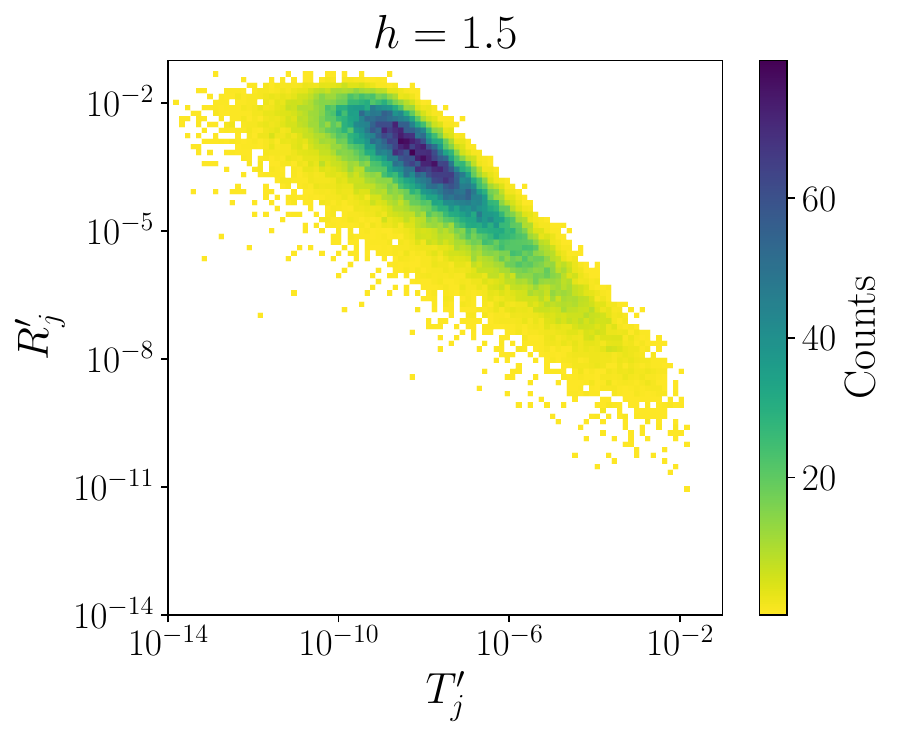}%
}

\subfloat[]{%
\includegraphics[width=0.2\textwidth]{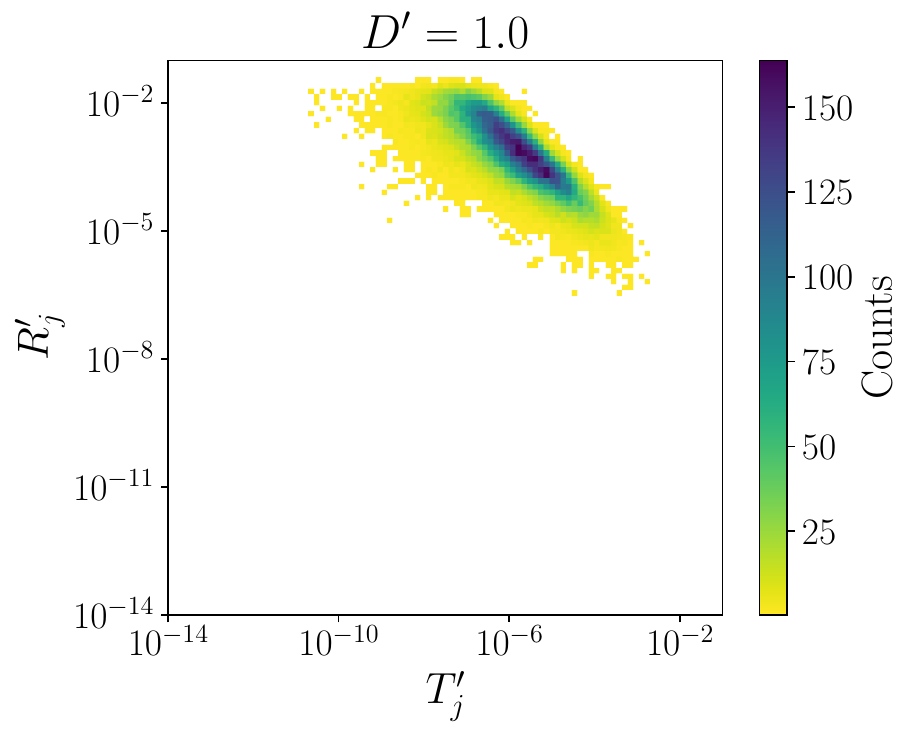}%
}
\subfloat[]{%
\includegraphics[width=0.2\textwidth]{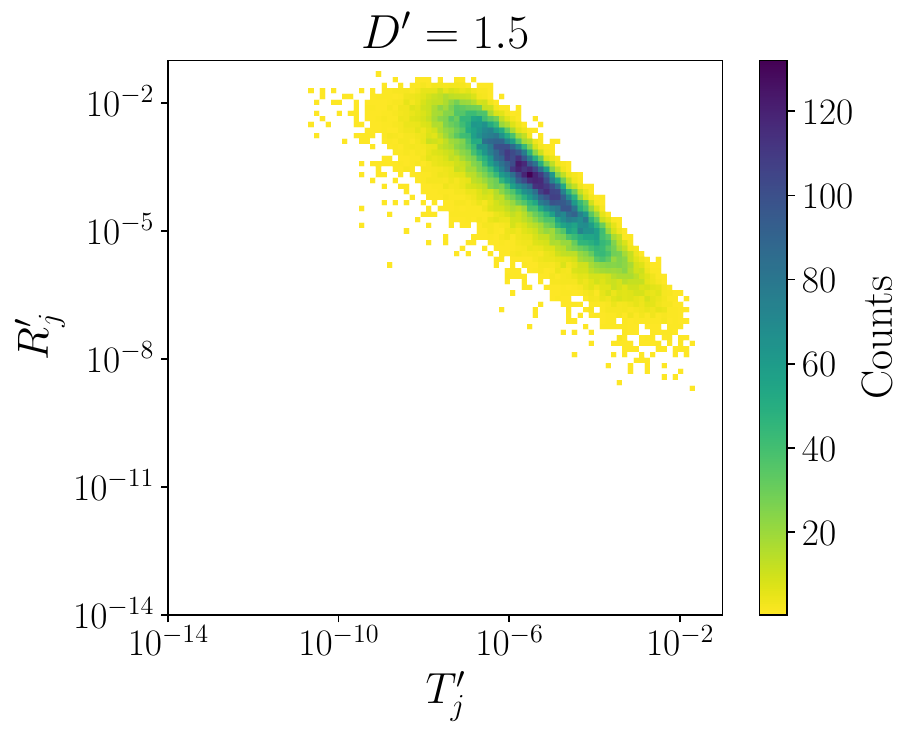}%
}
\subfloat[]{%
\includegraphics[width=0.2\textwidth]{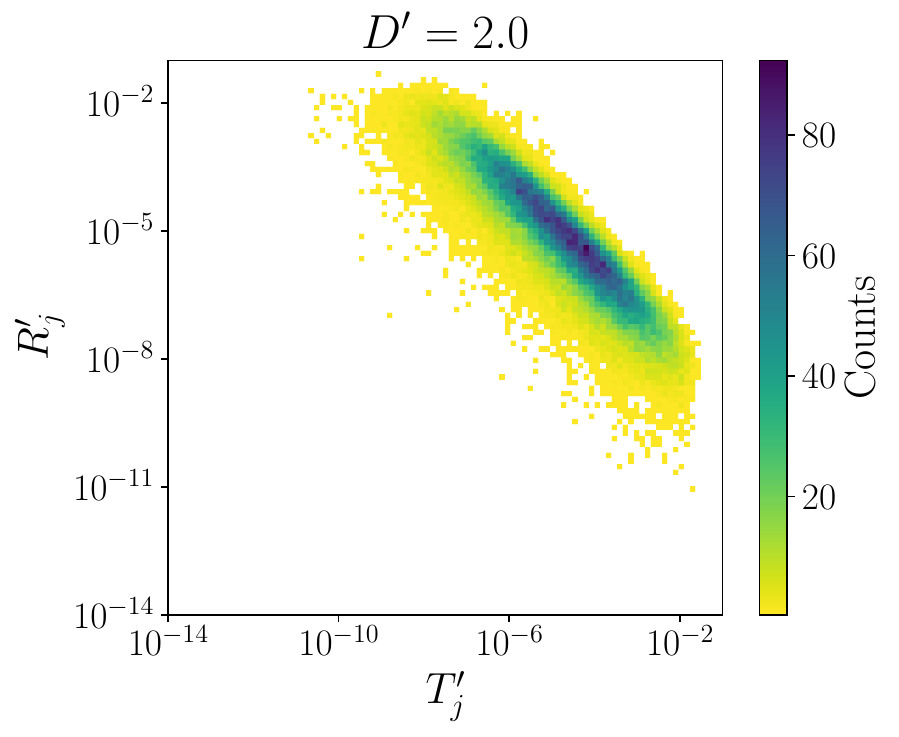}%
}
\subfloat[]{%
\includegraphics[width=0.2\textwidth]{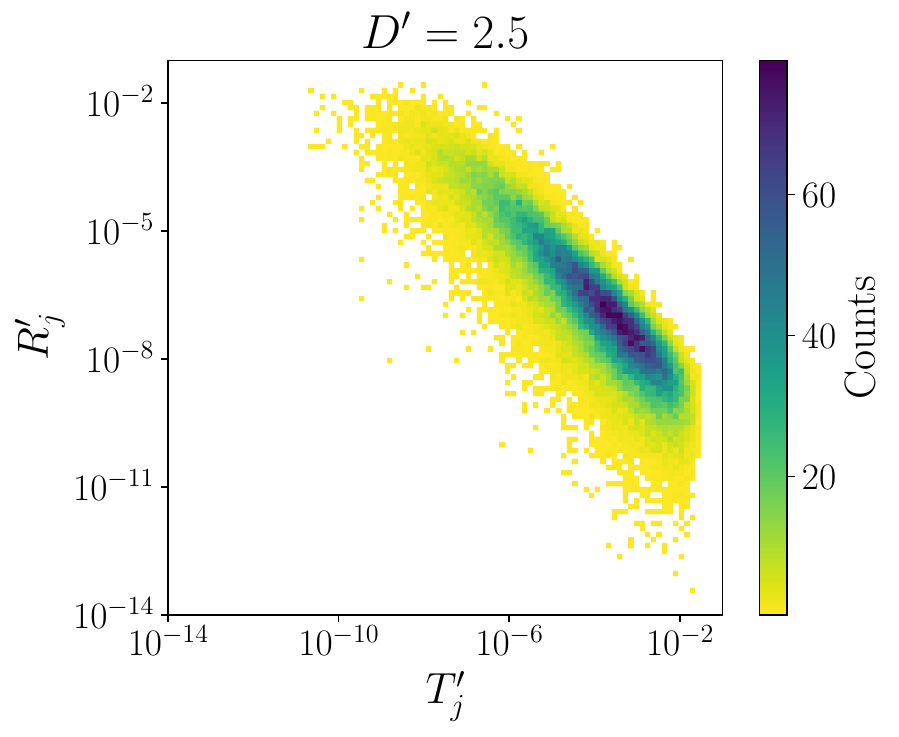}%
}

\subfloat[]{%
\includegraphics[width=0.2\textwidth]{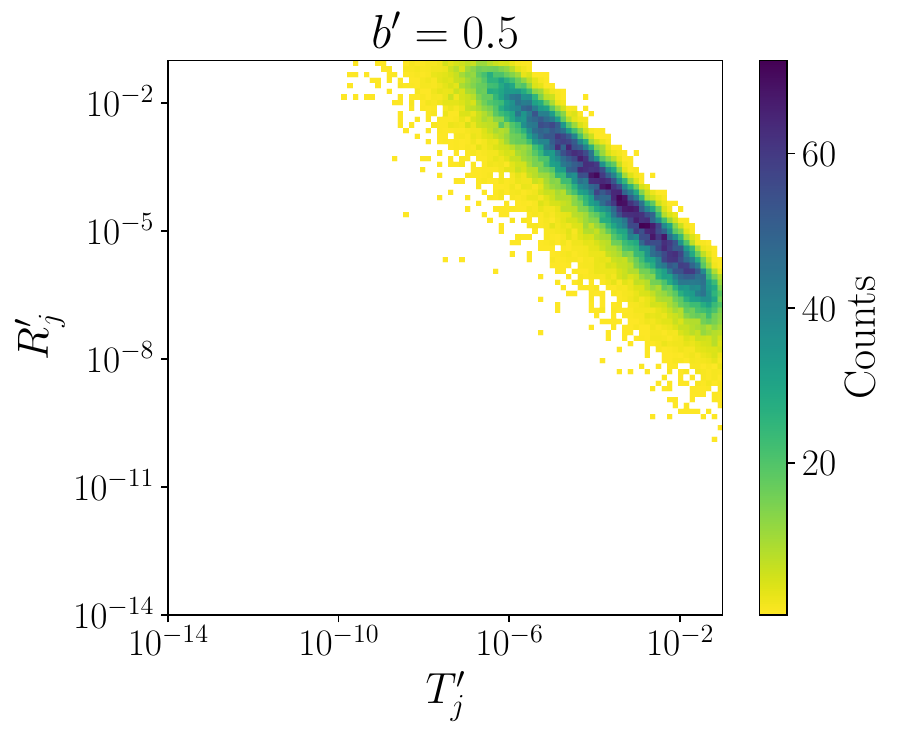}%
}
\subfloat[]{%
\includegraphics[width=0.2\textwidth]{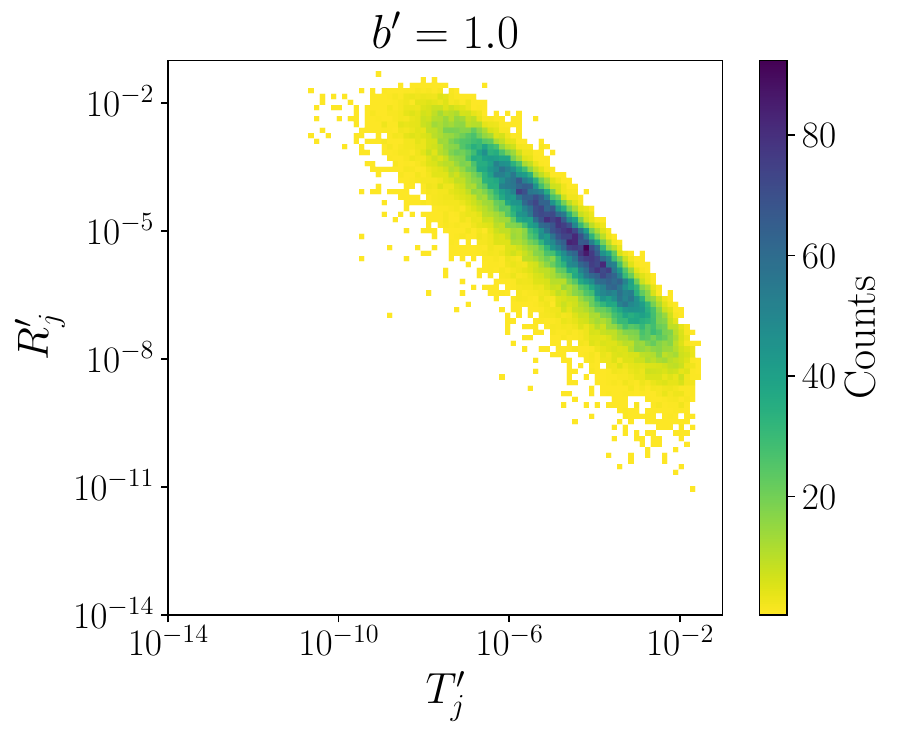}%
}
\subfloat[]{%
\includegraphics[width=0.2\textwidth]{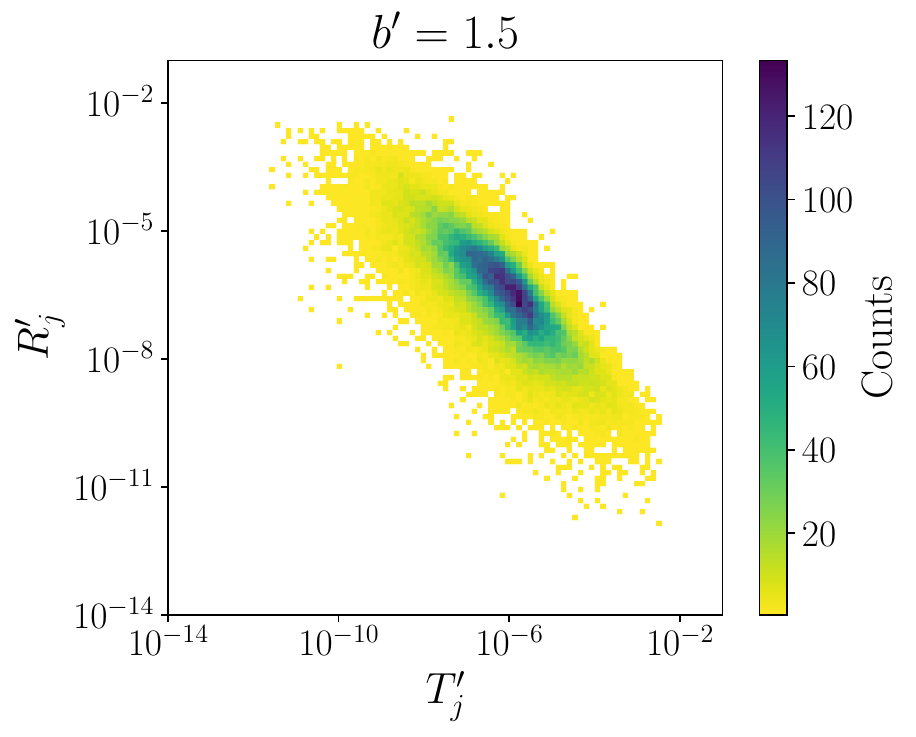}%
}
\caption{Joint distribution of the rescaled space $R'_j$ and time $T'_j$ for different $h$, $D'$ and $b'$ values in the \textbf{null model}. We keep other parameters fixed at $h = 1$, $D' = 2$ and $b' = 1$.} \label{supp:fig:rescaledDiagram_Null}
\end{figure*}

%---
\begin{figure*}[t!p]
\subfloat[]{%
\includegraphics[width=0.25\textwidth]{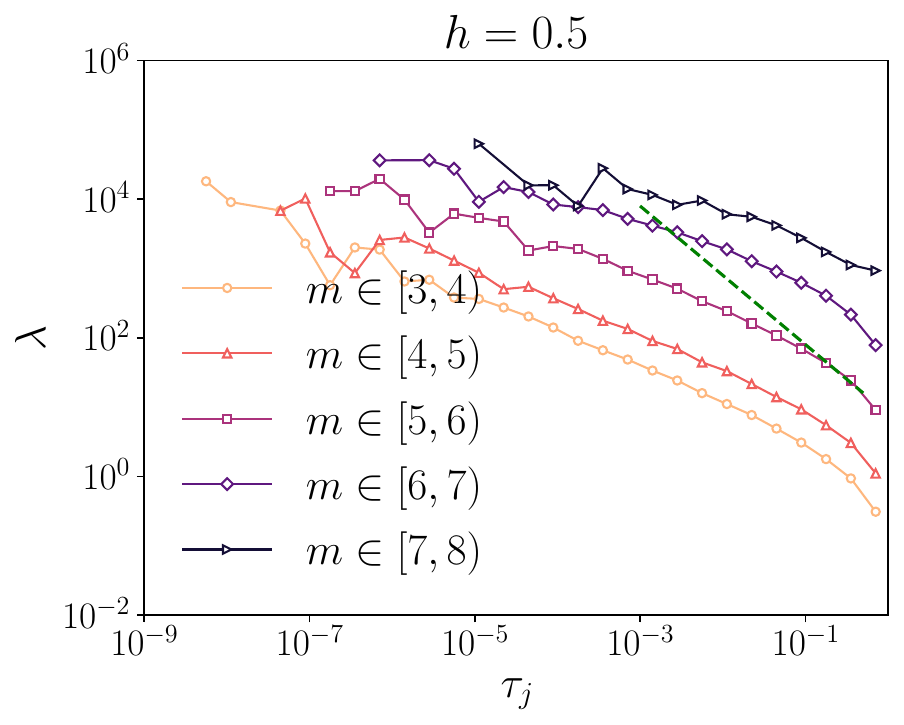}%
}
\subfloat[]{%
\includegraphics[width=0.25\textwidth]{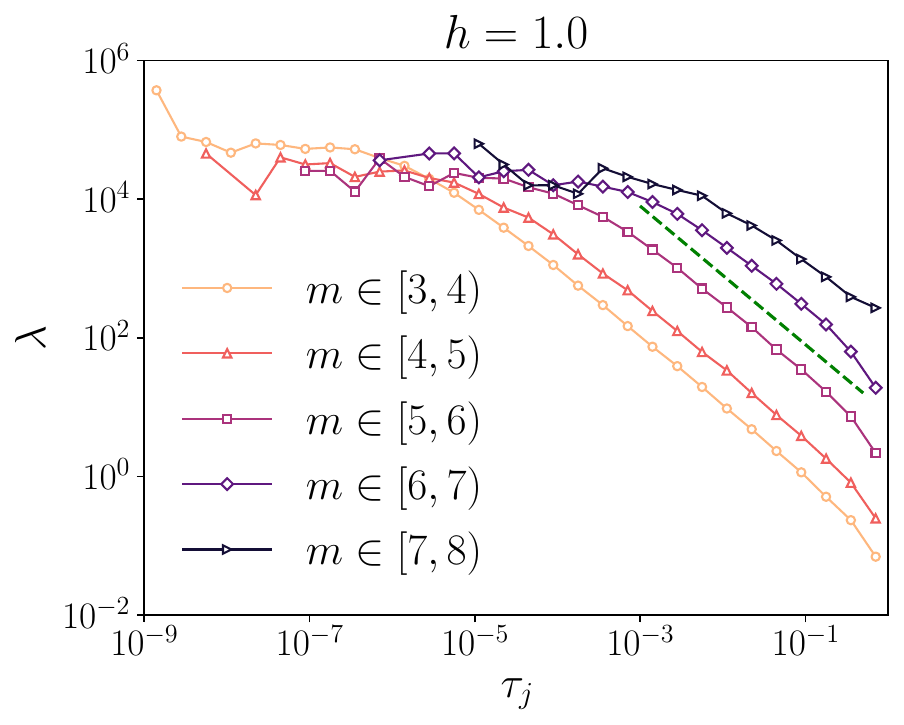}%
}
\subfloat[]{%
\includegraphics[width=0.25\textwidth]{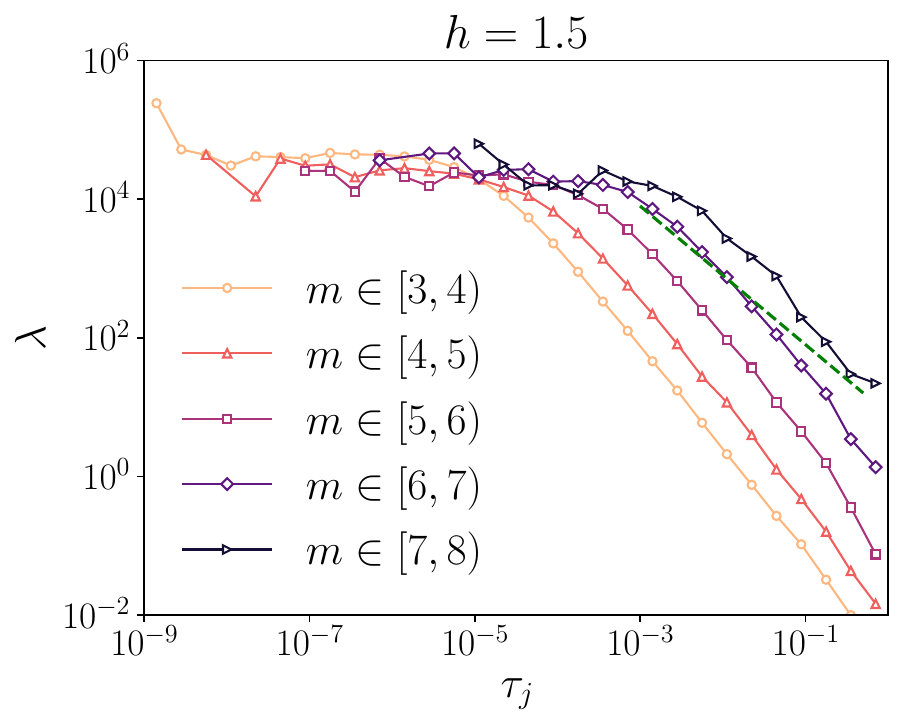}%
}

\subfloat[]{%
\includegraphics[width=0.25\textwidth]{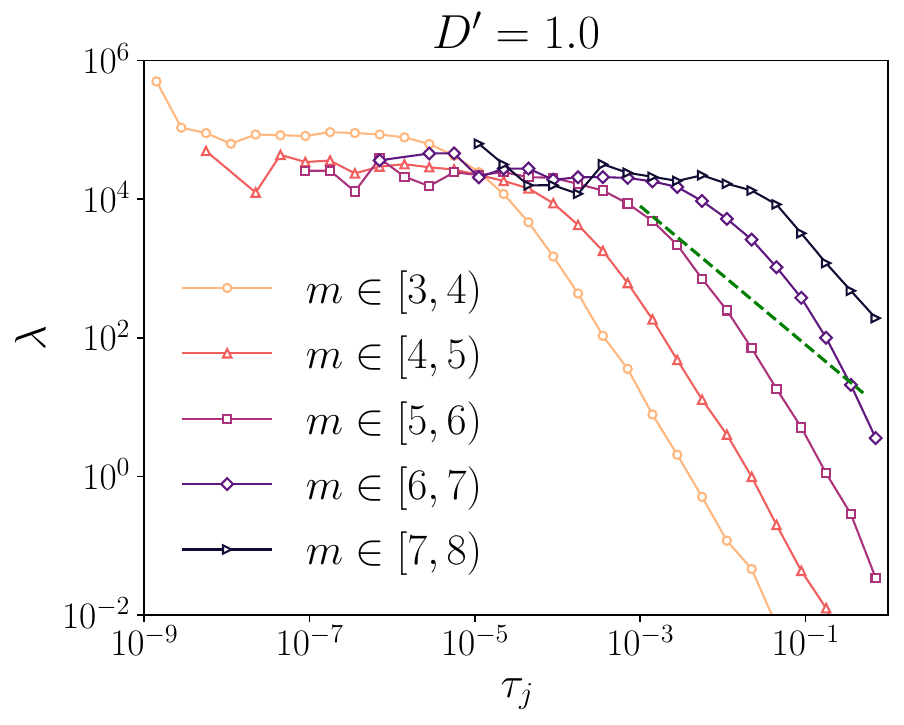}%
}
\subfloat[]{%
\includegraphics[width=0.25\textwidth]{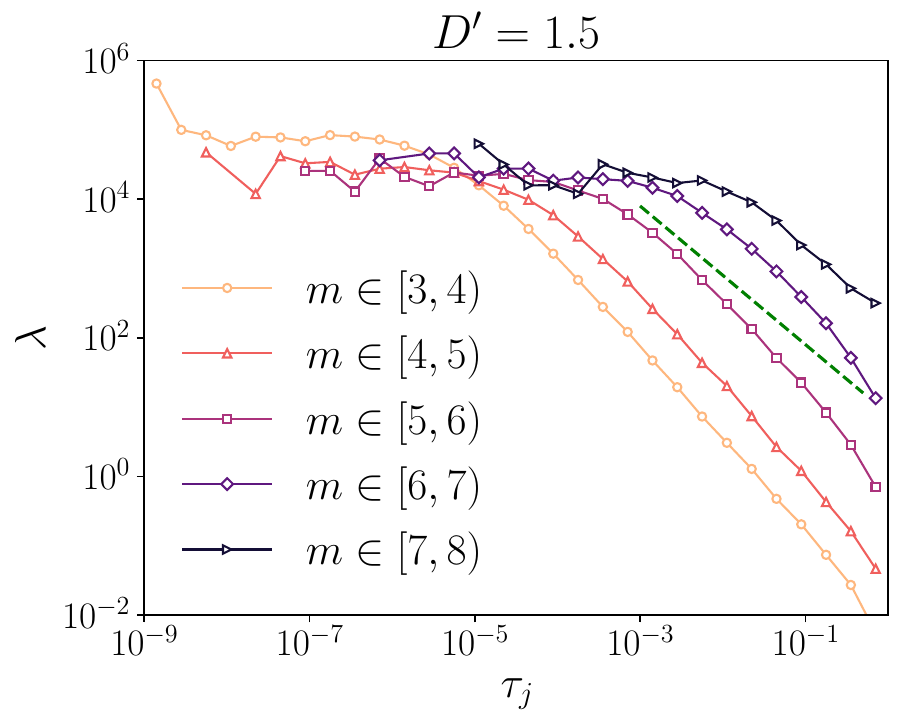}%
}
\subfloat[]{%
\includegraphics[width=0.25\textwidth]{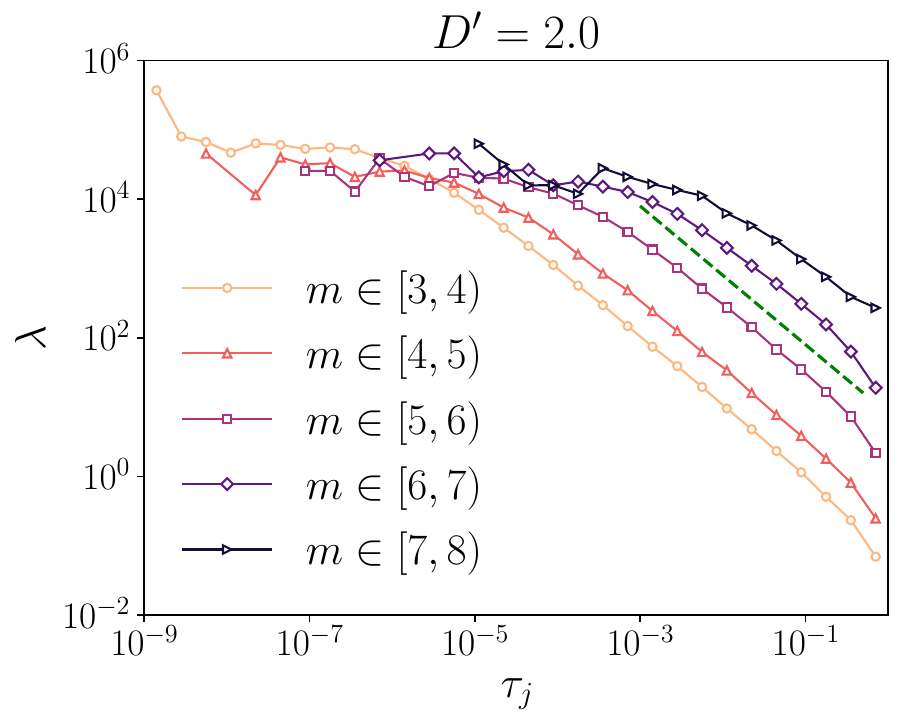}%
}
\subfloat[]{%
\includegraphics[width=0.25\textwidth]{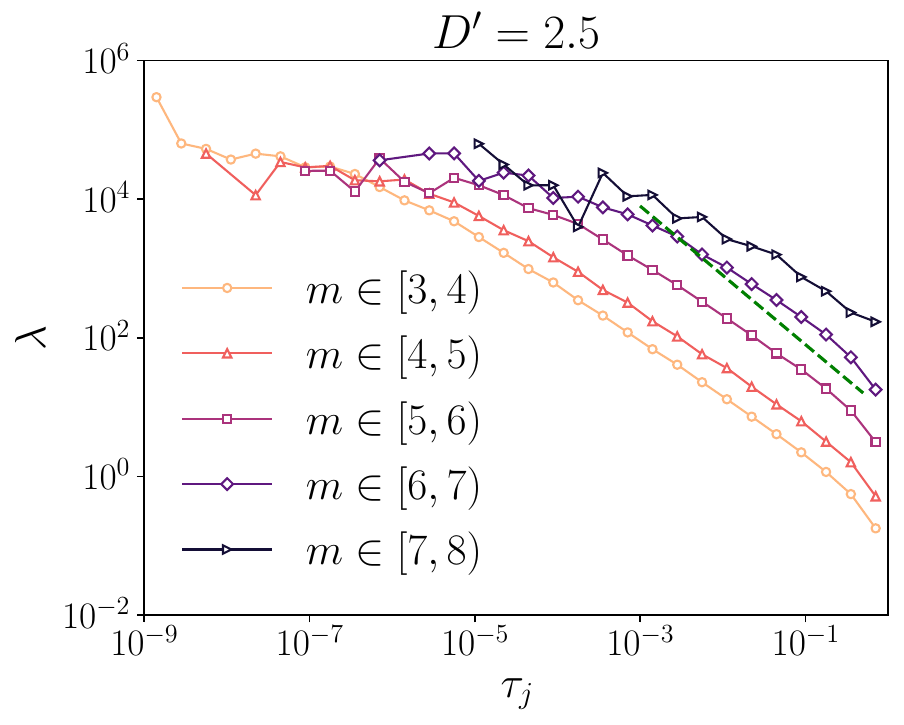}%
}

\subfloat[]{%
\includegraphics[width=0.25\textwidth]{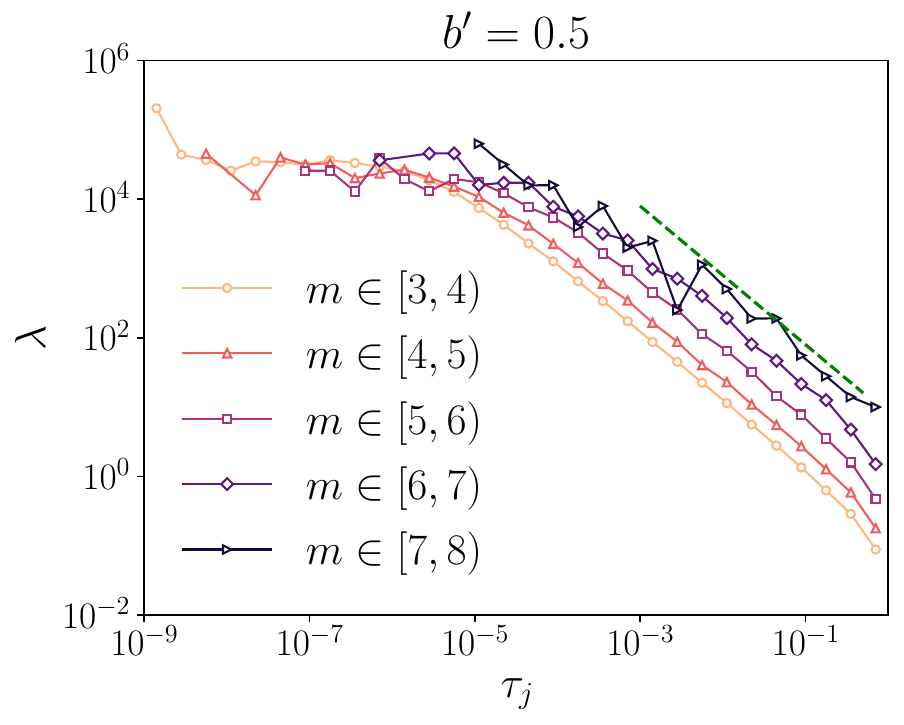}%
}
\subfloat[]{%
\includegraphics[width=0.25\textwidth]{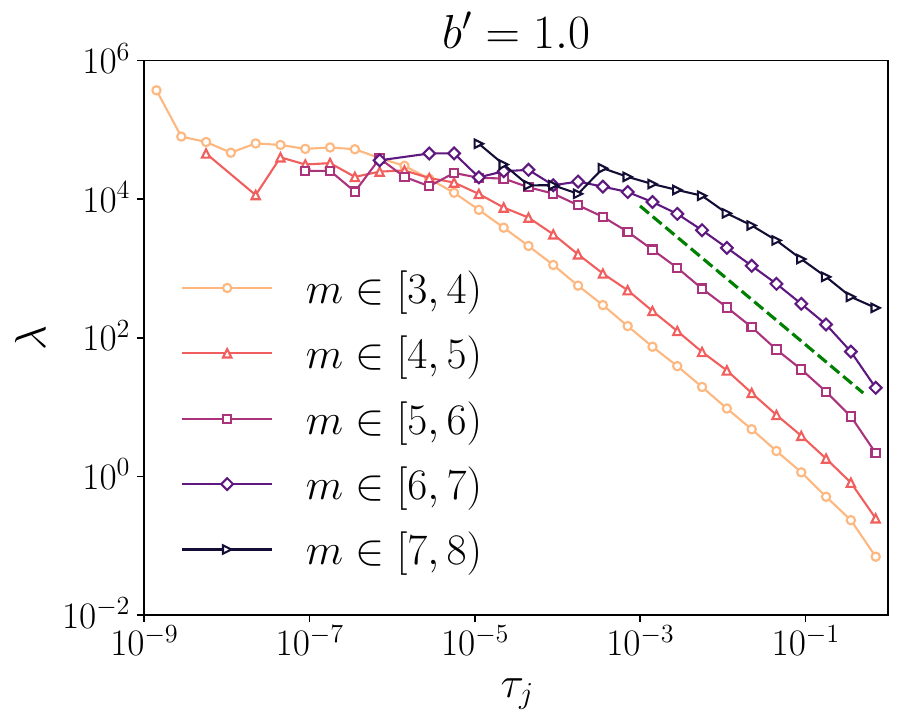}%
}
\subfloat[]{%
\includegraphics[width=0.25\textwidth]{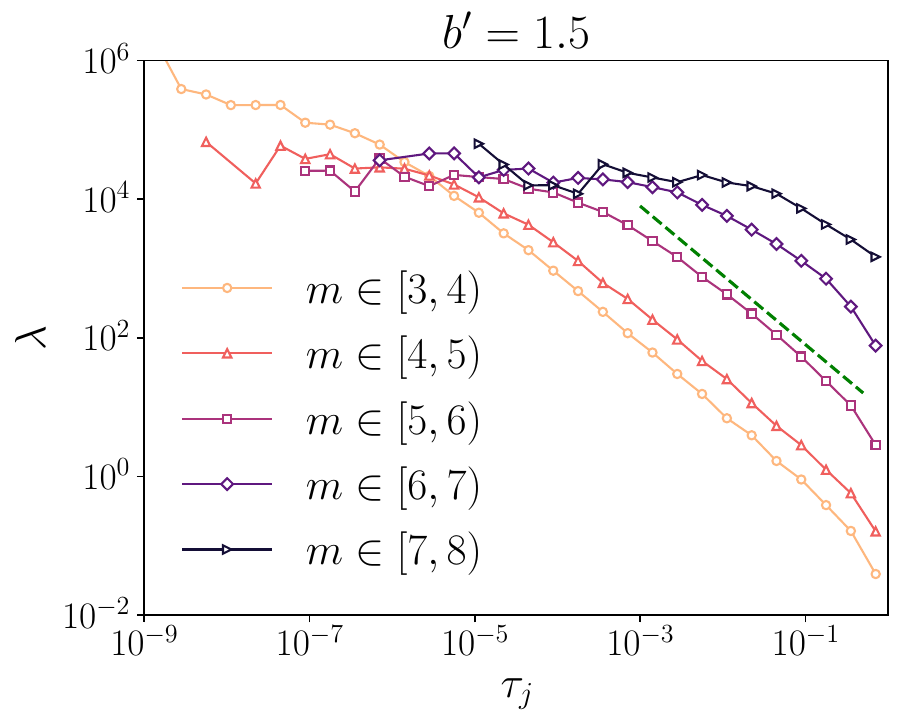}%
}
\caption{Rate of the number of aftershocks $\lambda (\tau_j)$ for the \textbf{null model} for different magnitudes $m$ varying (a)-(c) $h$, (d)-(g) $D'$ and (h)-(j) $b'$ values. We keep other parameters fixed at $h = 1$, $D' = 2$ and $b' = 1$. The green dashed line is a power law with exponent $-1$.}\label{supp:fig:Omori_Null_all_m}
\end{figure*}

%%%%---
\begin{figure*}[t!p]
\subfloat[]{%
\includegraphics[width=0.2\textwidth]{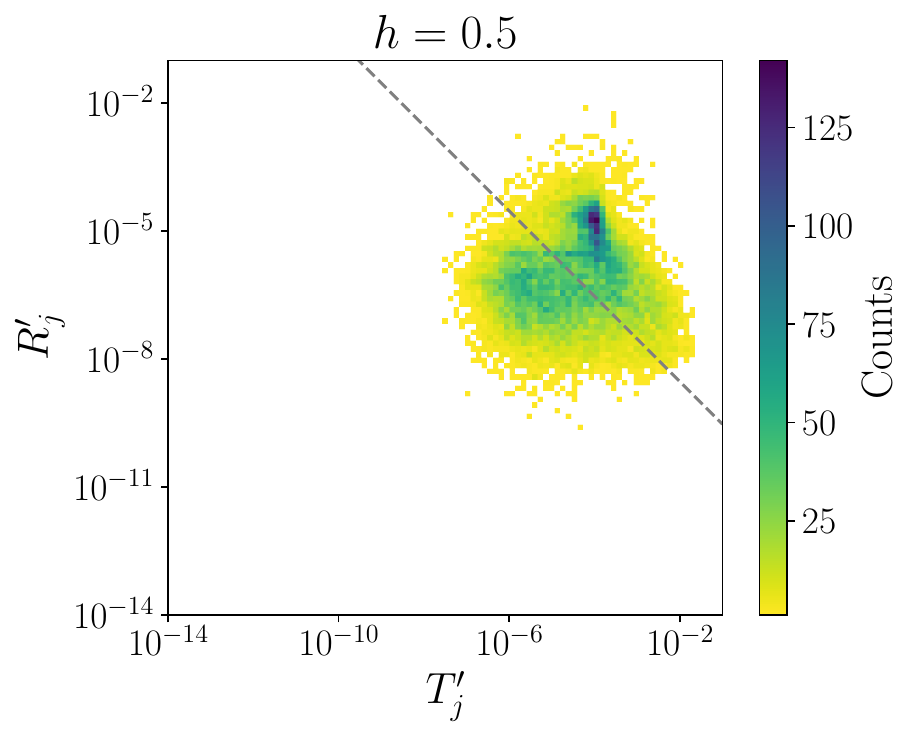}%
}
\subfloat[]{%
\includegraphics[width=0.2\textwidth]{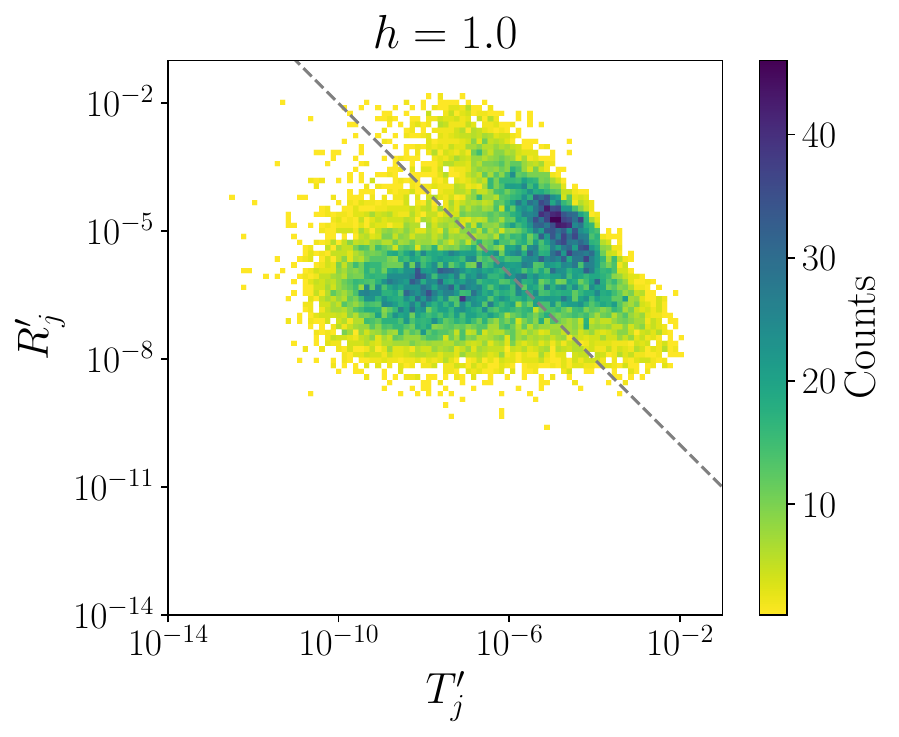}%
}
\subfloat[]{%
\includegraphics[width=0.2\textwidth]{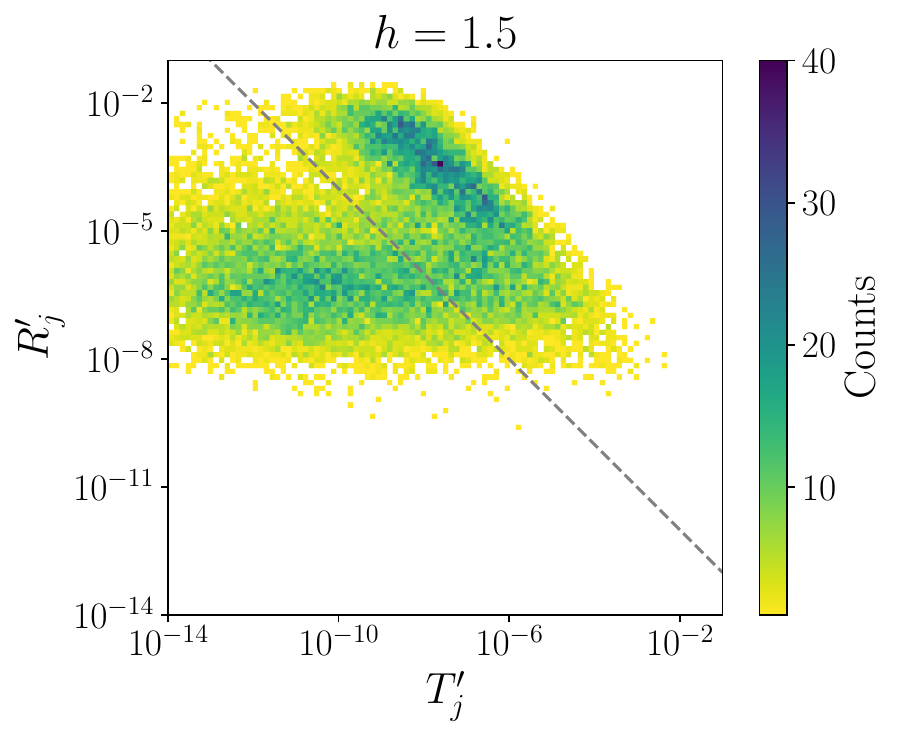}%
}

\subfloat[]{%
\includegraphics[width=0.2\textwidth]{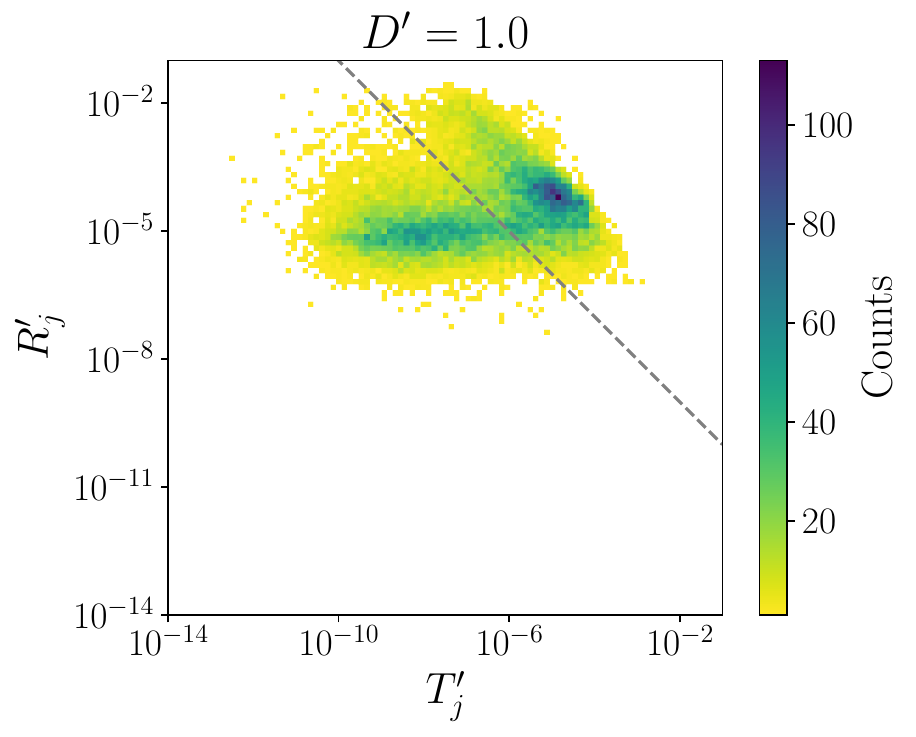}%
}
\subfloat[]{%
\includegraphics[width=0.2\textwidth]{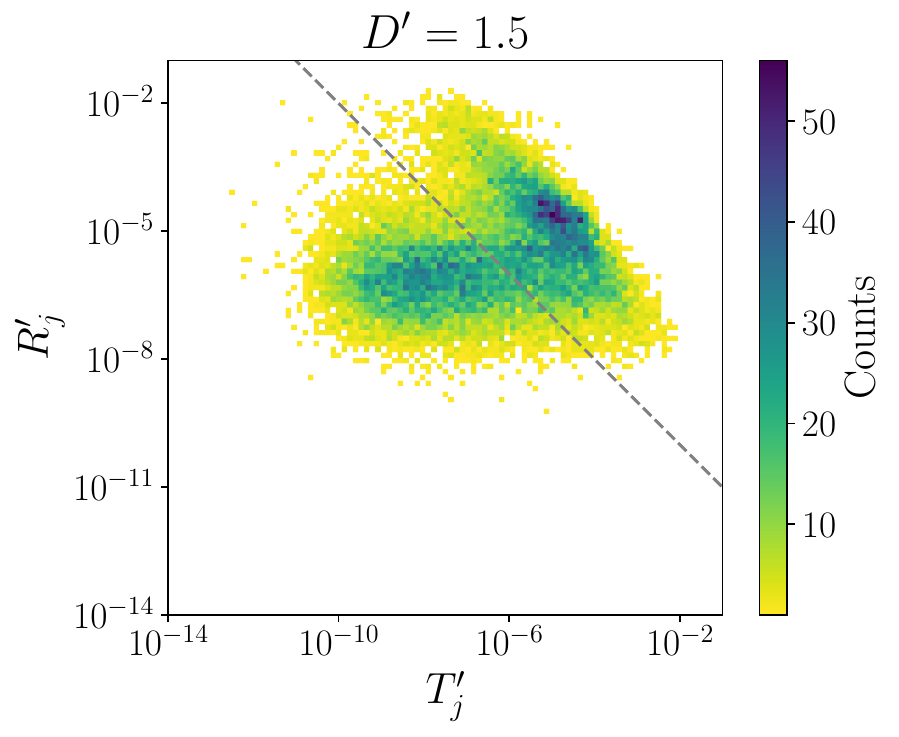}%
}
\subfloat[]{%
\includegraphics[width=0.2\textwidth]{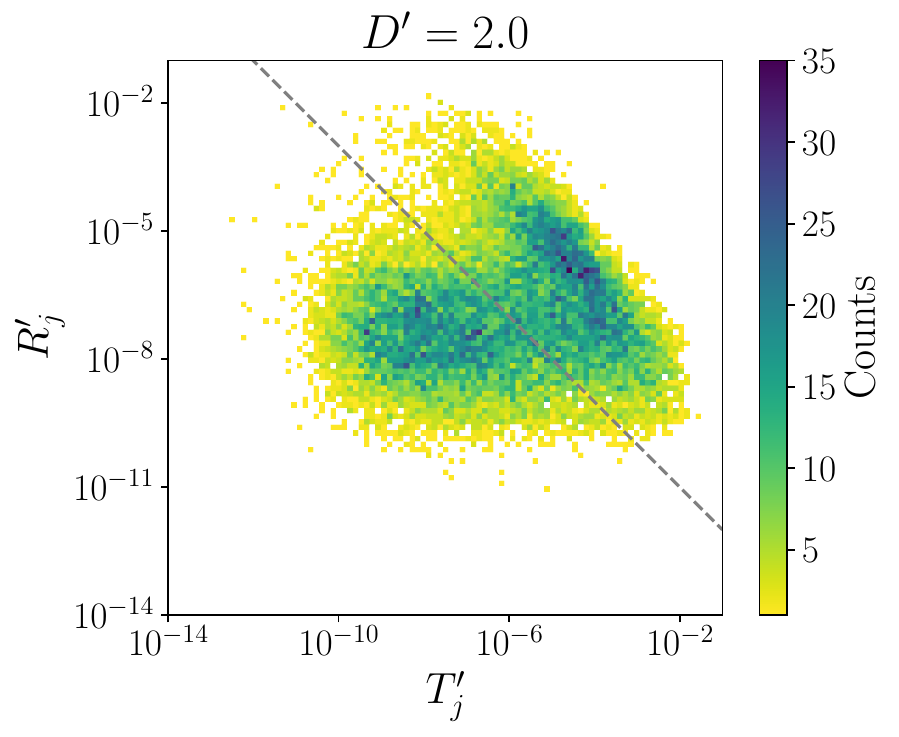}%
}
\subfloat[]{%
\includegraphics[width=0.2\textwidth]{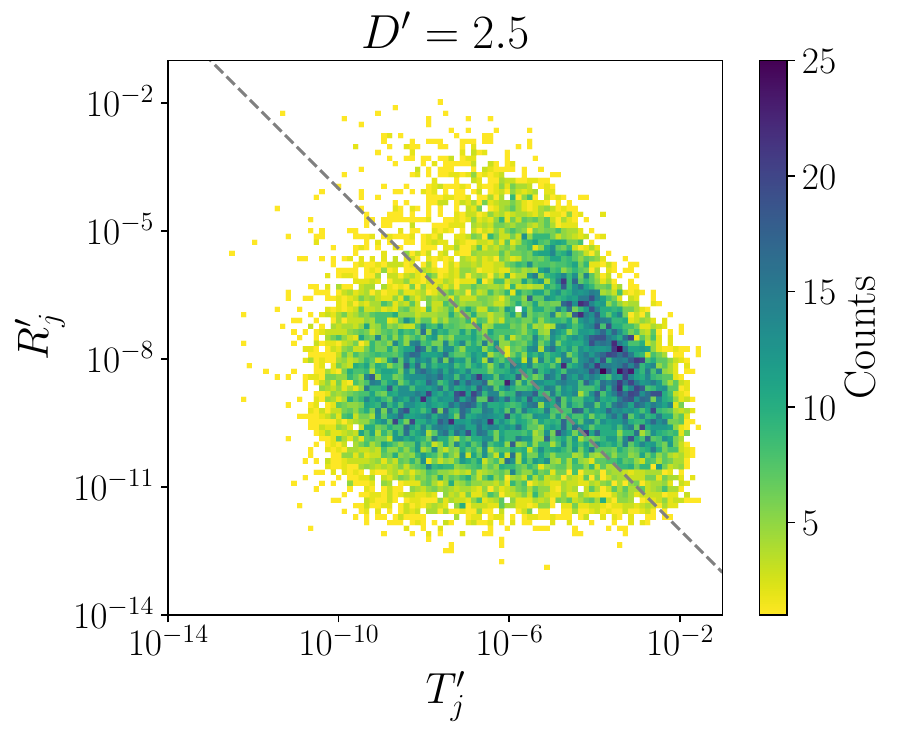}%
}

\subfloat[]{%
\includegraphics[width=0.2\textwidth]{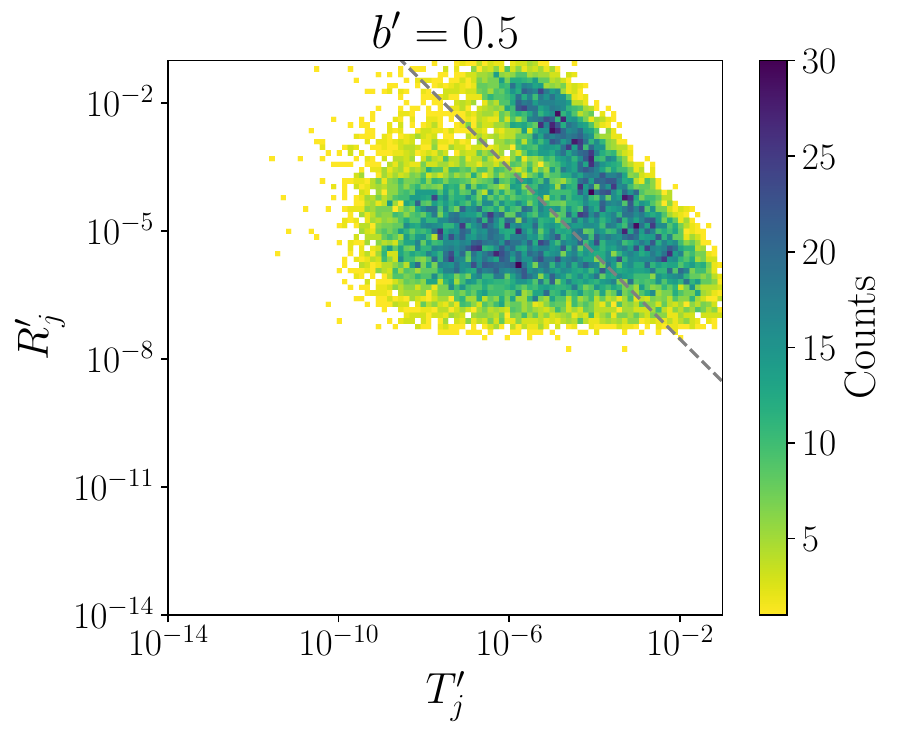}%
}
\subfloat[]{%
\includegraphics[width=0.2\textwidth]{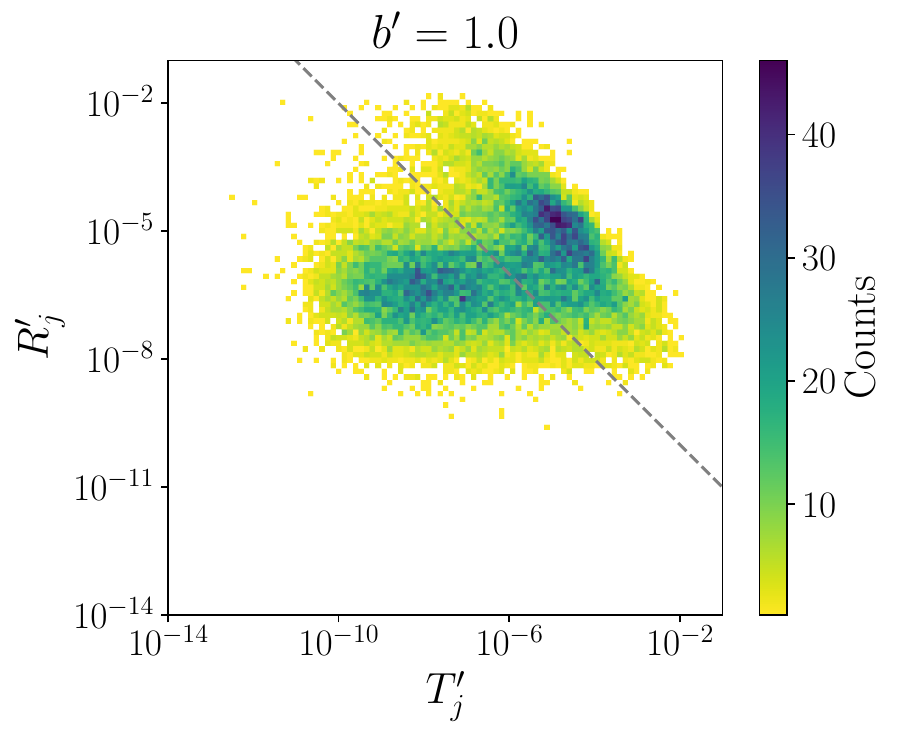}%
}
\subfloat[]{%
\includegraphics[width=0.2\textwidth]{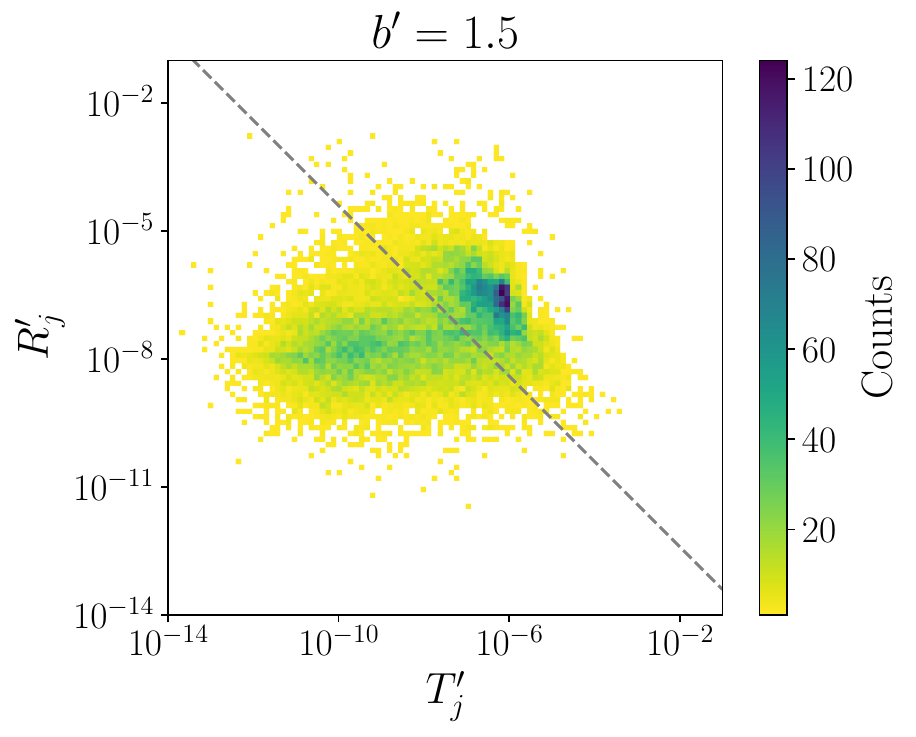}%
}
\caption{Joint distribution of the rescaled space $R'_j$ and time $T'_j$ for different $h$, $D'$ and $b'$ values in the \textbf{SCSN catalog}.  We keep other parameters fixed at $h = 1$, $D' = 1.6$ and $b' = 1$. The gray dashed line represents
(a) $\eta'_{th} = 3\cdot10^{-11}$, (b) $\eta'_{th} = 10^{-12}$,
(c) $\eta'_{th} = 10^{-14}$,
(d) $\eta'_{th} = 10^{-11}$,
(e) $\eta'_{th} = 10^{-12}$,
(f) $\eta'_{th} = 10^{-13}$,
(g) $\eta'_{th} = 10^{-14}$,
(h) $\eta'_{th} = 3\cdot10^{-10}$
(i) $\eta'_{th} = 10^{-12}$,
and (j) $\eta'_{th} = 4\cdot10^{-15}$.} \label{supp:fig:rescaledDiagram_SCSN}
\end{figure*}

%---
\begin{figure*}[t!p]
\subfloat[]{%
\includegraphics[width=0.25\textwidth]{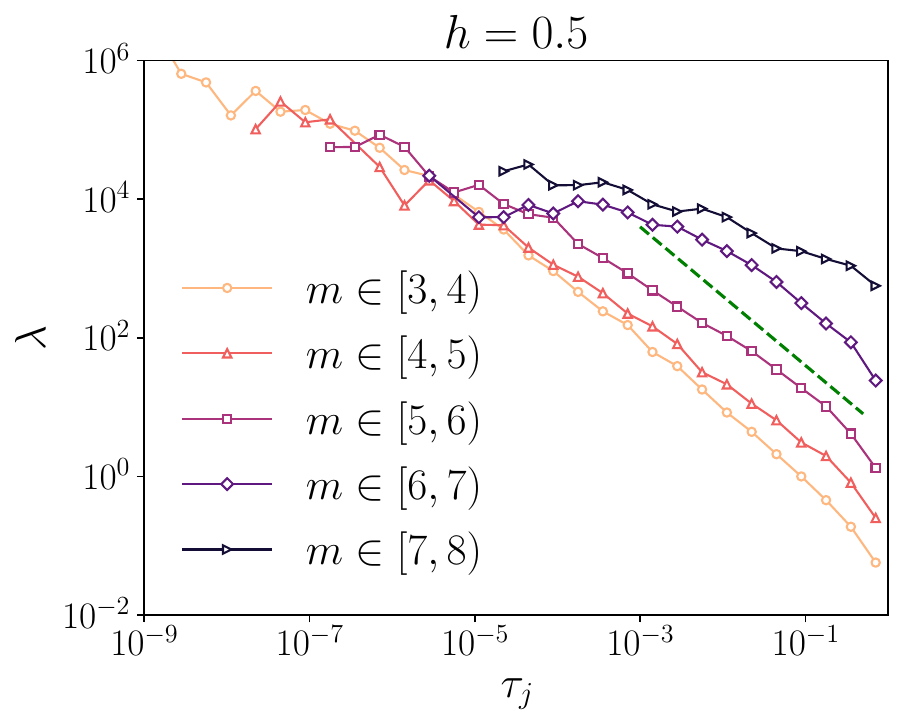}%
}
\subfloat[]{%
\includegraphics[width=0.25\textwidth]{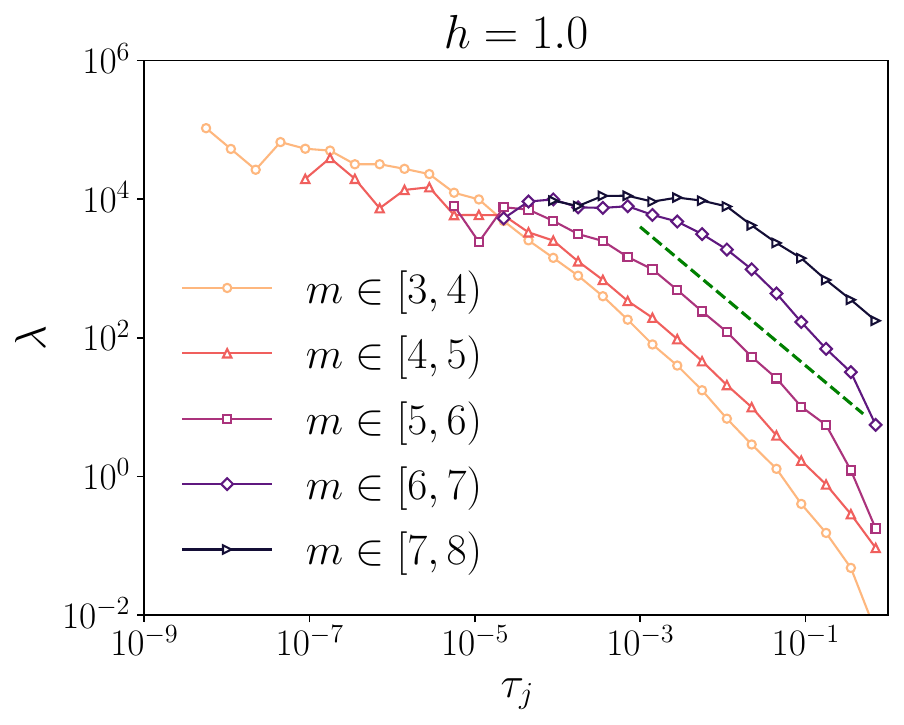}%
}
\subfloat[]{%
\includegraphics[width=0.25\textwidth]{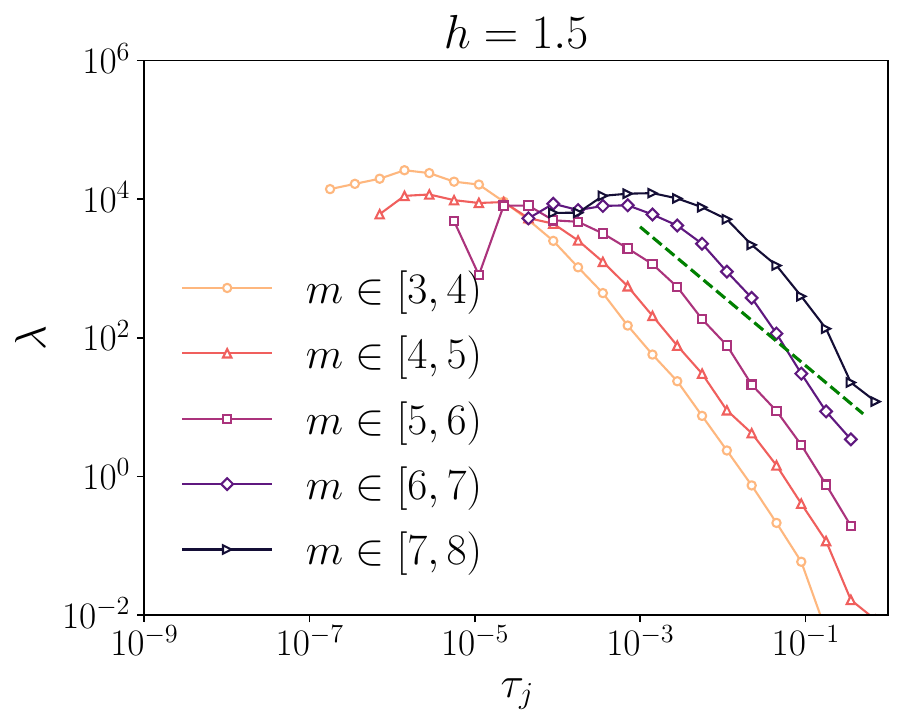}%
}

\subfloat[]{%
\includegraphics[width=0.25\textwidth]{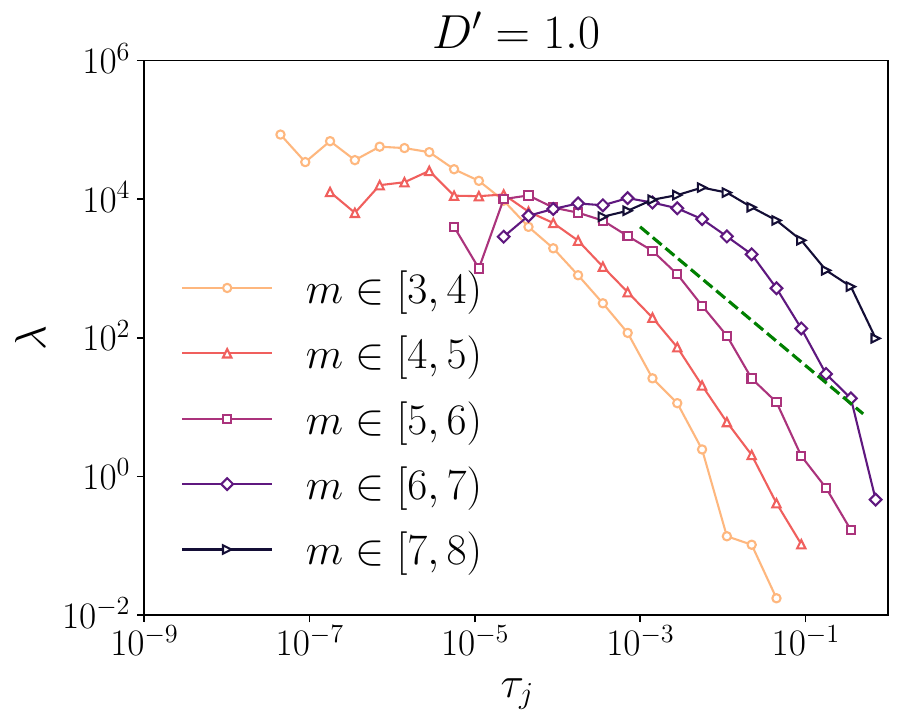}%
}
\subfloat[]{%
\includegraphics[width=0.25\textwidth]{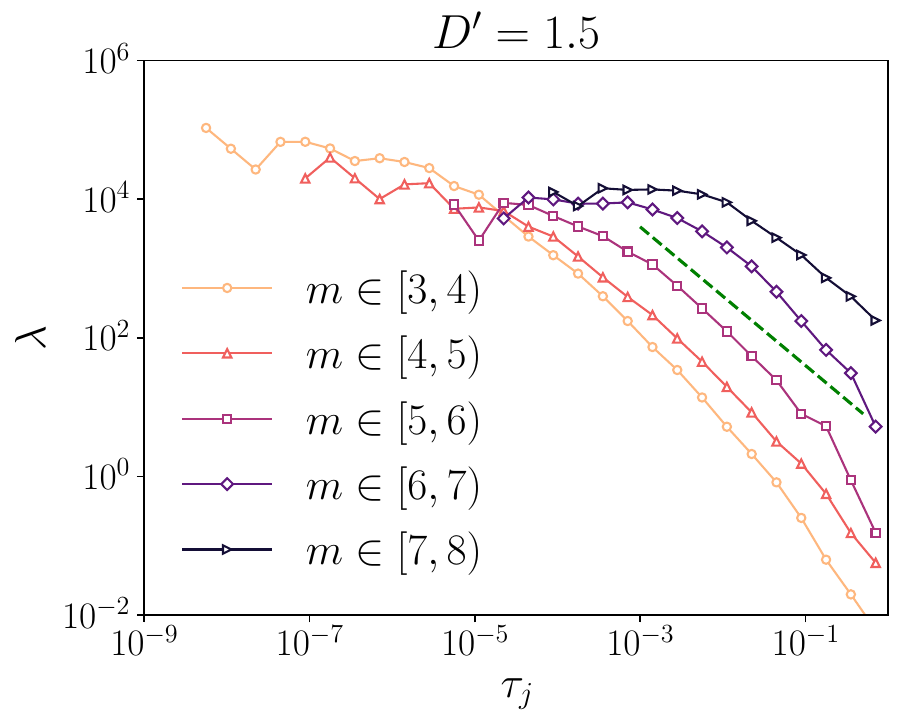}%
}
\subfloat[]{%
\includegraphics[width=0.25\textwidth]{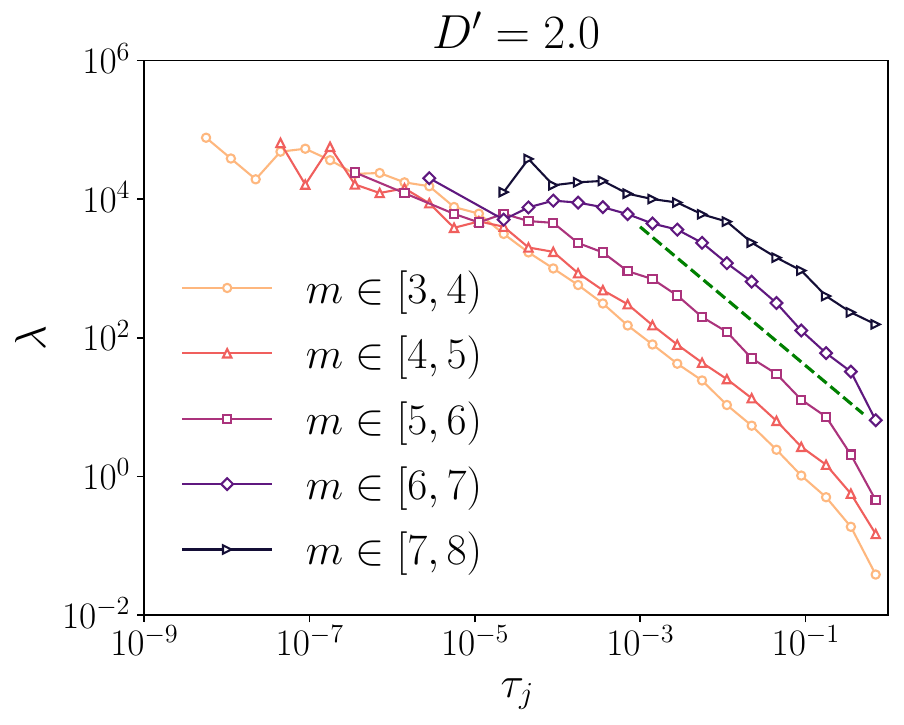}%
}
\subfloat[]{%
\includegraphics[width=0.25\textwidth]{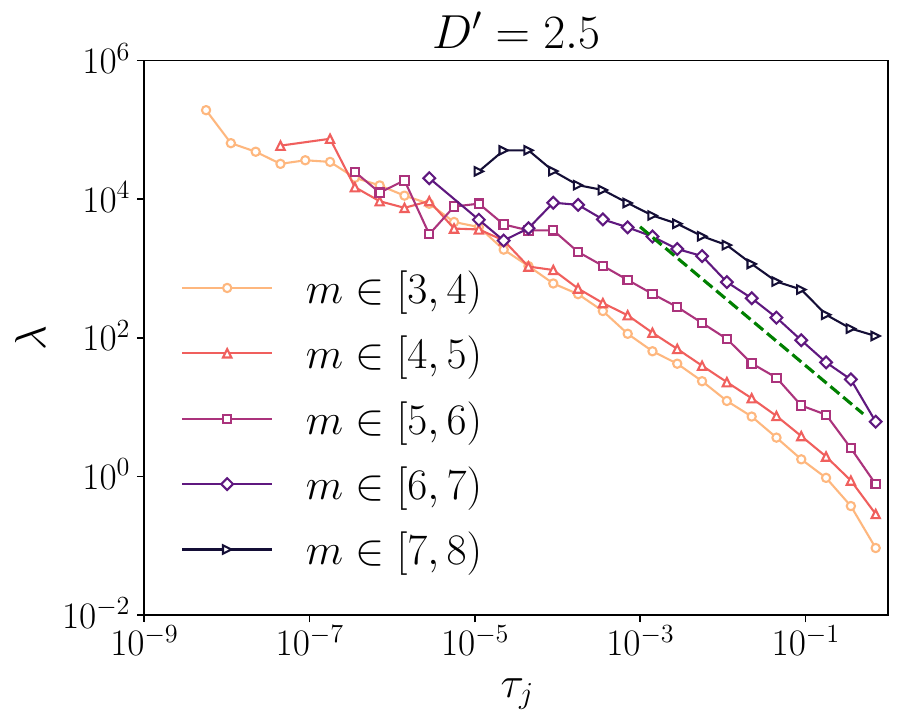}%
}

\subfloat[]{%
\includegraphics[width=0.25\textwidth]{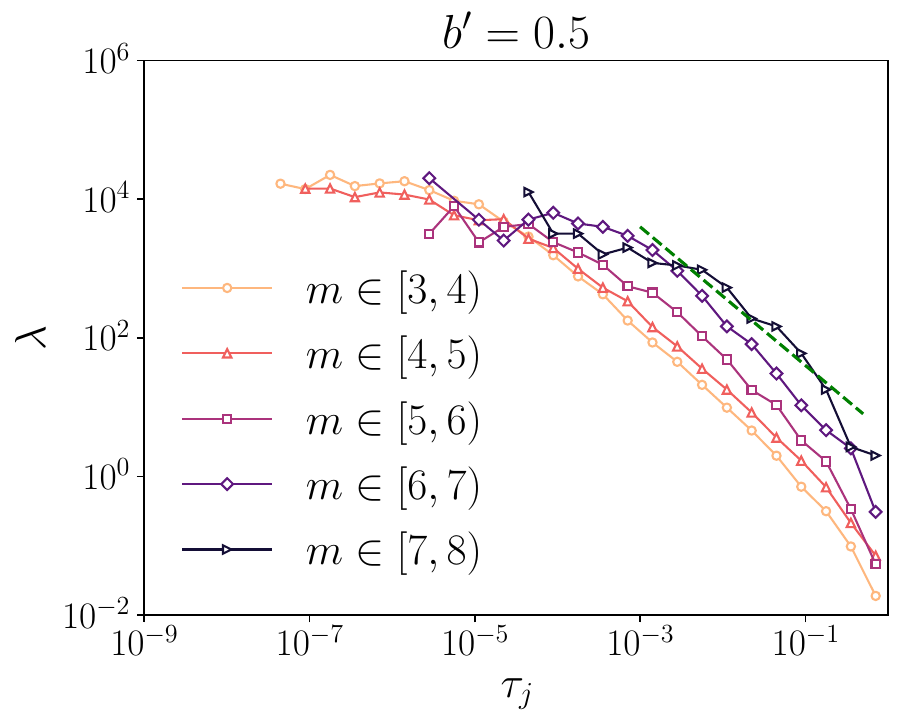}%
}
\subfloat[]{%
\includegraphics[width=0.25\textwidth]{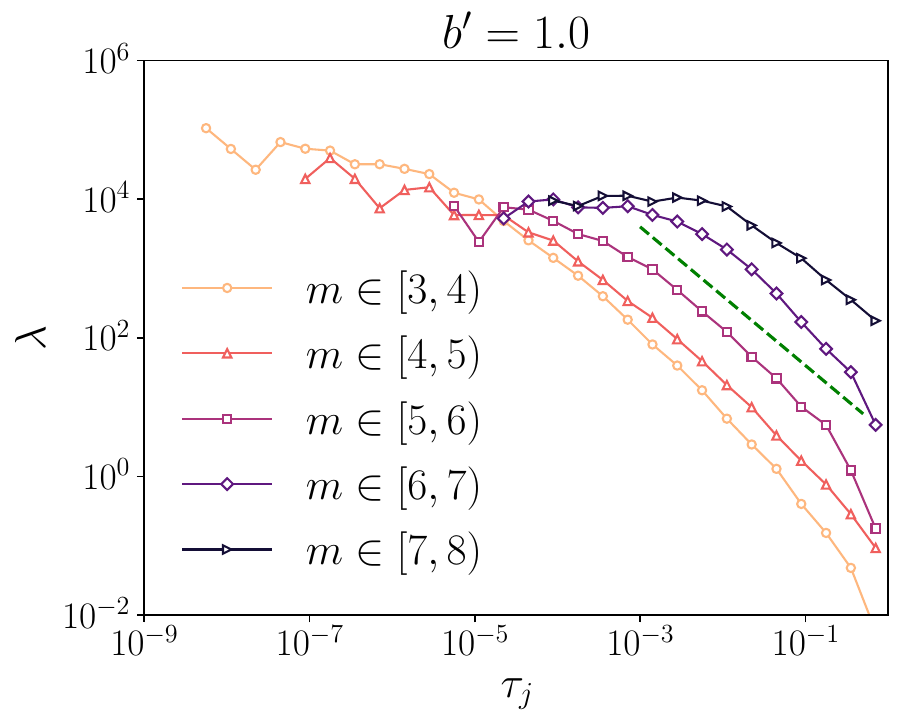}%
}
\subfloat[]{%
\includegraphics[width=0.25\textwidth]{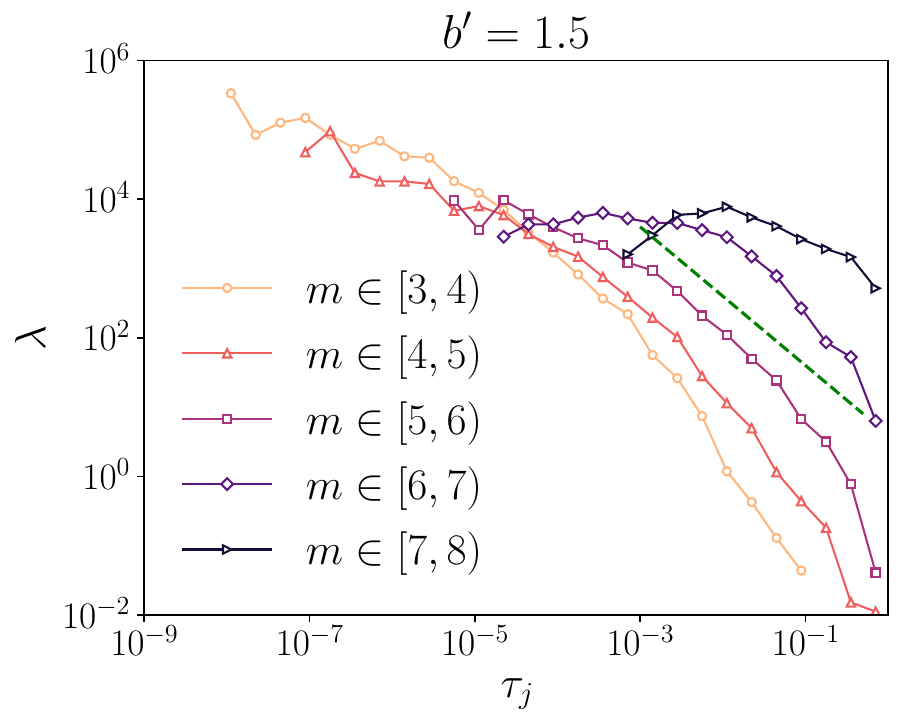}%
}
\caption{Rate of the number of aftershocks $\lambda (\tau_j)$ for events labeled as \textbf{background events in the SCSN catalog} for different magnitudes $m$ varying (a)-(c) $h$, (d)-(g) $D'$ and (h)-(j) $b'$ values. We keep other parameters fixed at $h = 1$, $D' = 1.6$ and $b' = 1$. The green dashed line is a power law with exponent $-1$.}\label{supp:fig:Omori_SCSN_background_all_m}
\end{figure*}

%---
\begin{figure*}[t!p]
\subfloat[]{%
\includegraphics[width=0.25\textwidth]{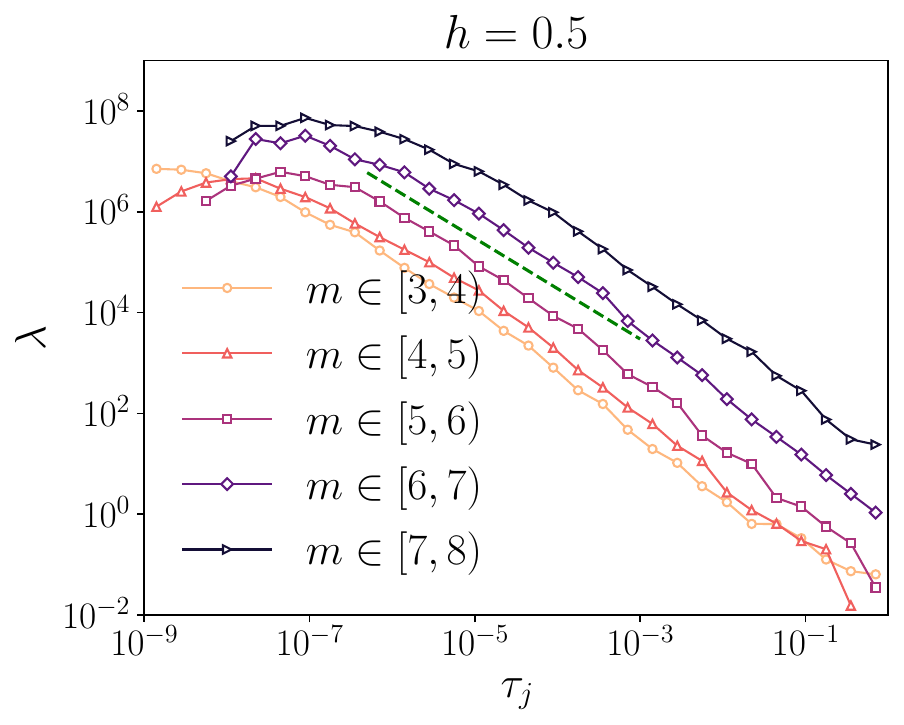}%
}
\subfloat[]{%
\includegraphics[width=0.25\textwidth]{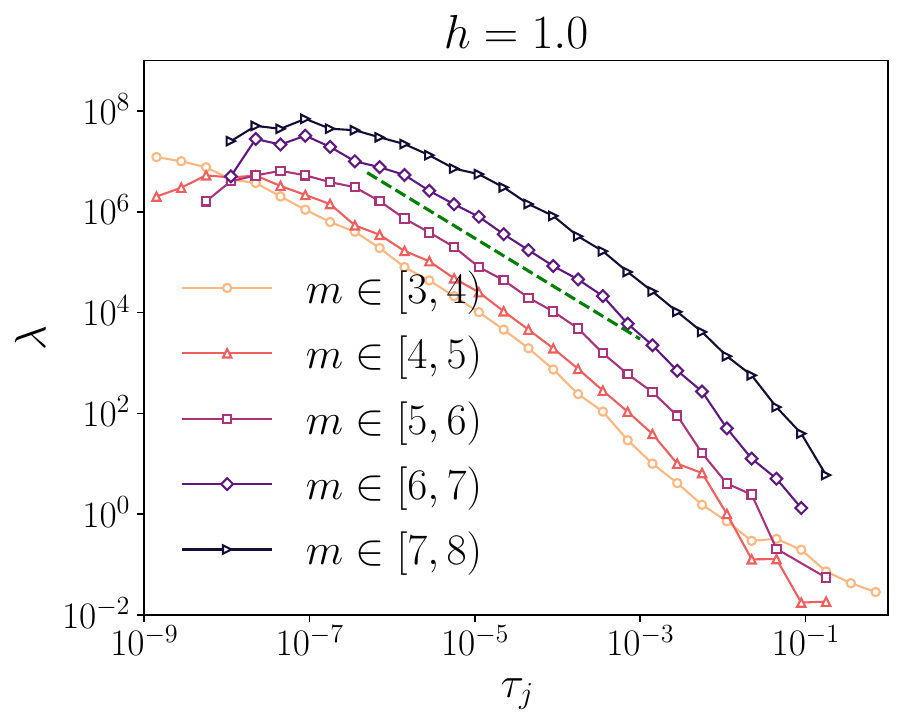}%
}
\subfloat[]{%
\includegraphics[width=0.25\textwidth]{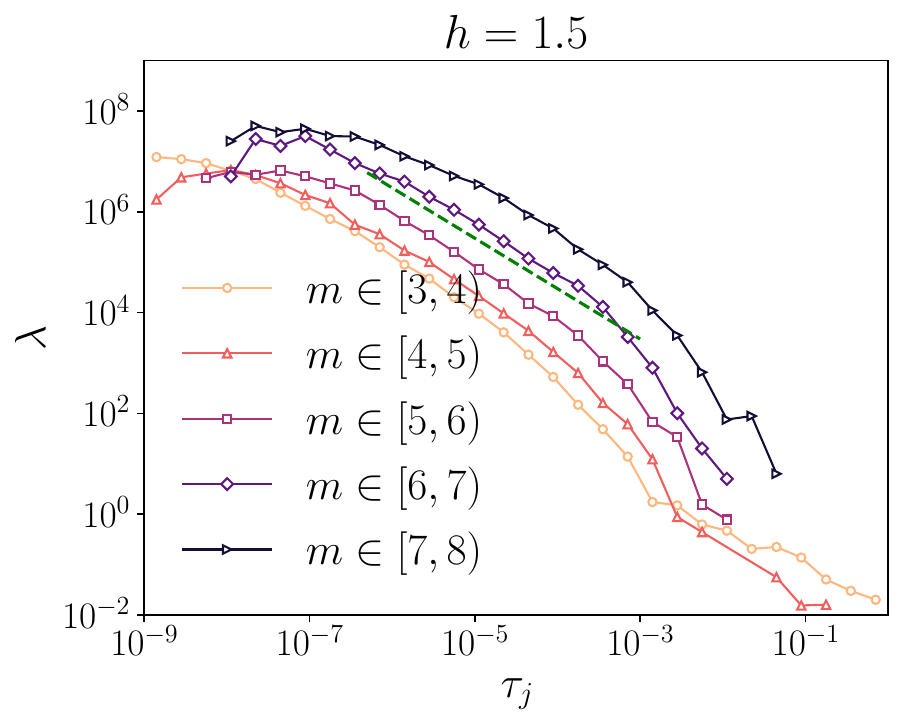}%
}

\subfloat[]{%
\includegraphics[width=0.25\textwidth]{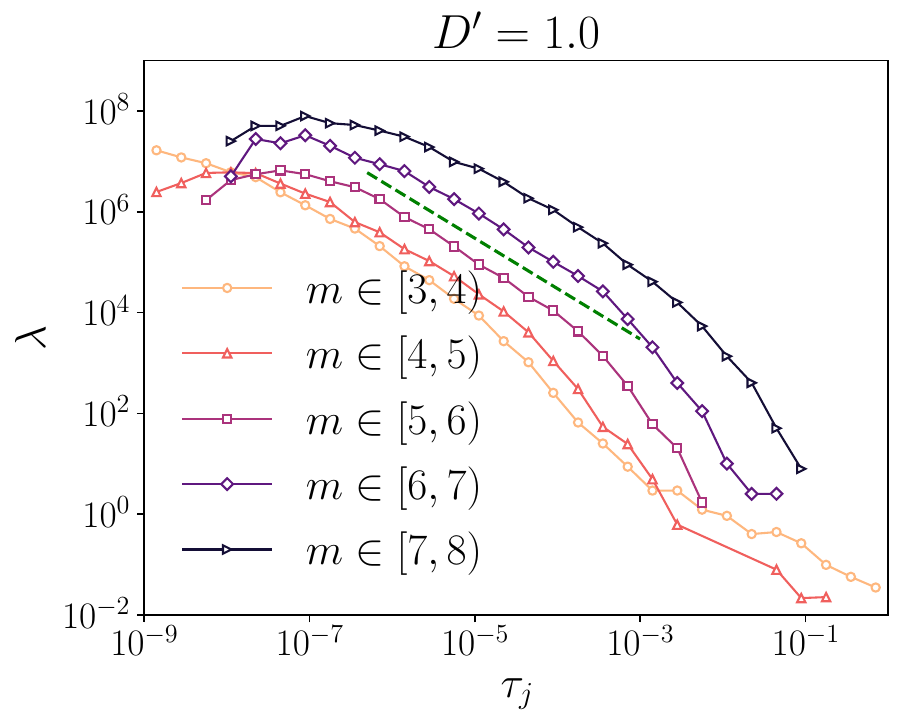}%
}
\subfloat[]{%
\includegraphics[width=0.25\textwidth]{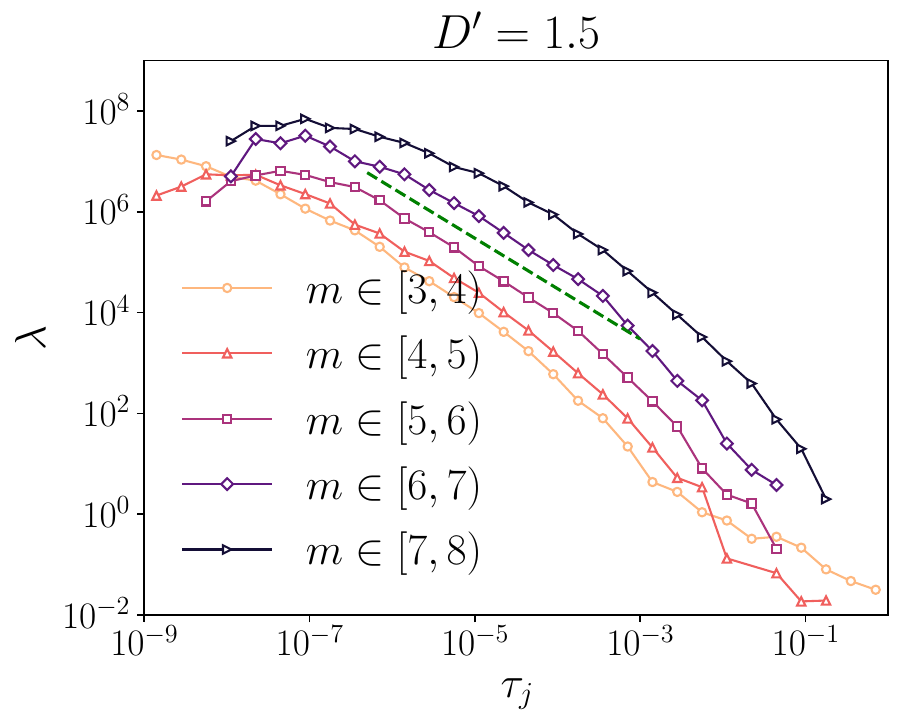}%
}
\subfloat[]{%
\includegraphics[width=0.25\textwidth]{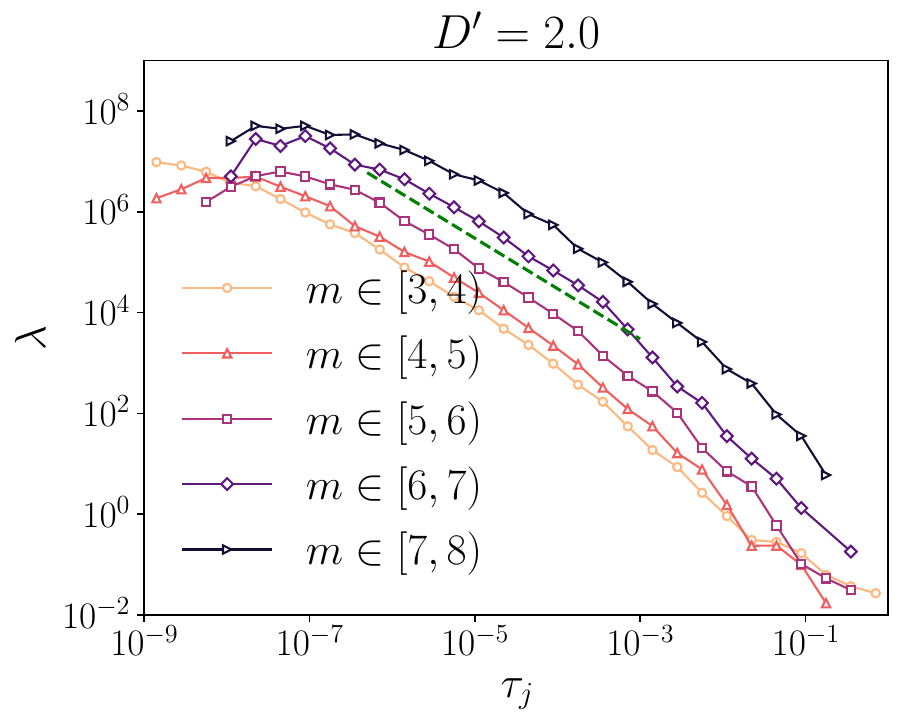}%
}
\subfloat[]{%
\includegraphics[width=0.25\textwidth]{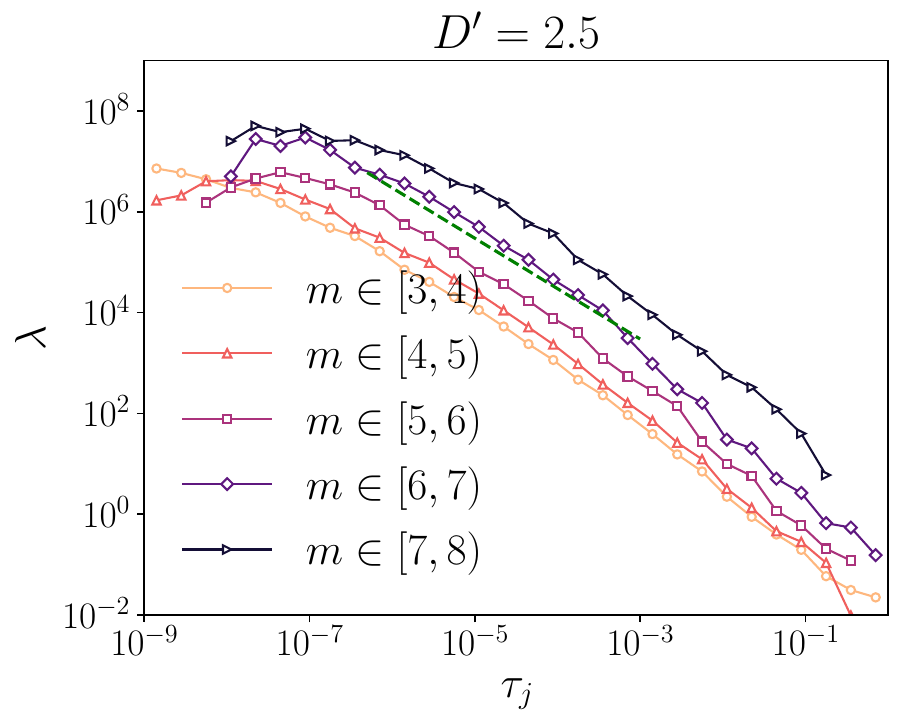}%
}

\subfloat[]{%
\includegraphics[width=0.25\textwidth]{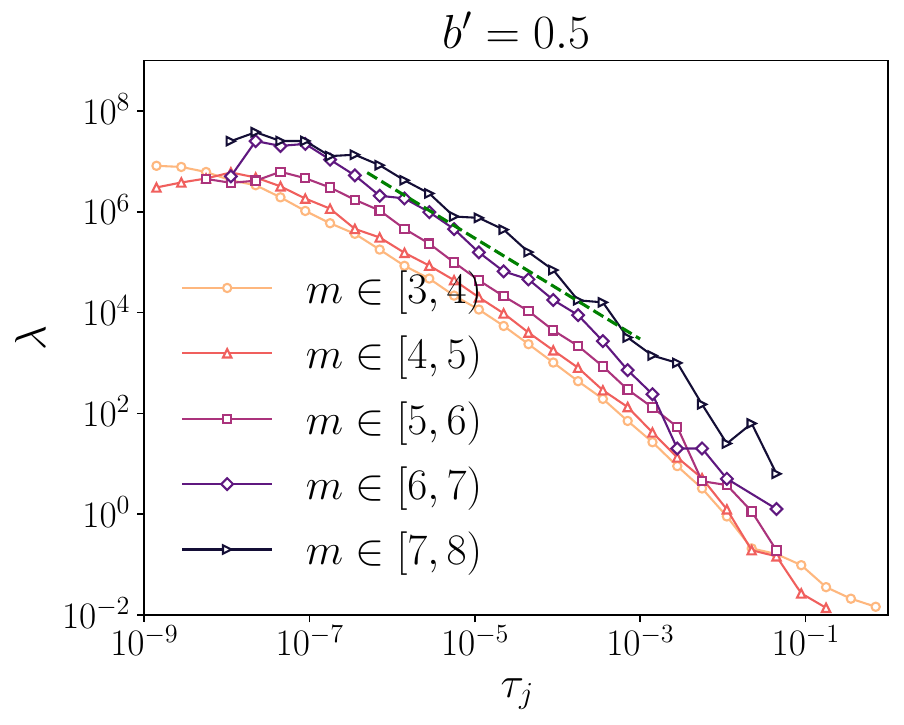}%
}
\subfloat[]{%
\includegraphics[width=0.25\textwidth]{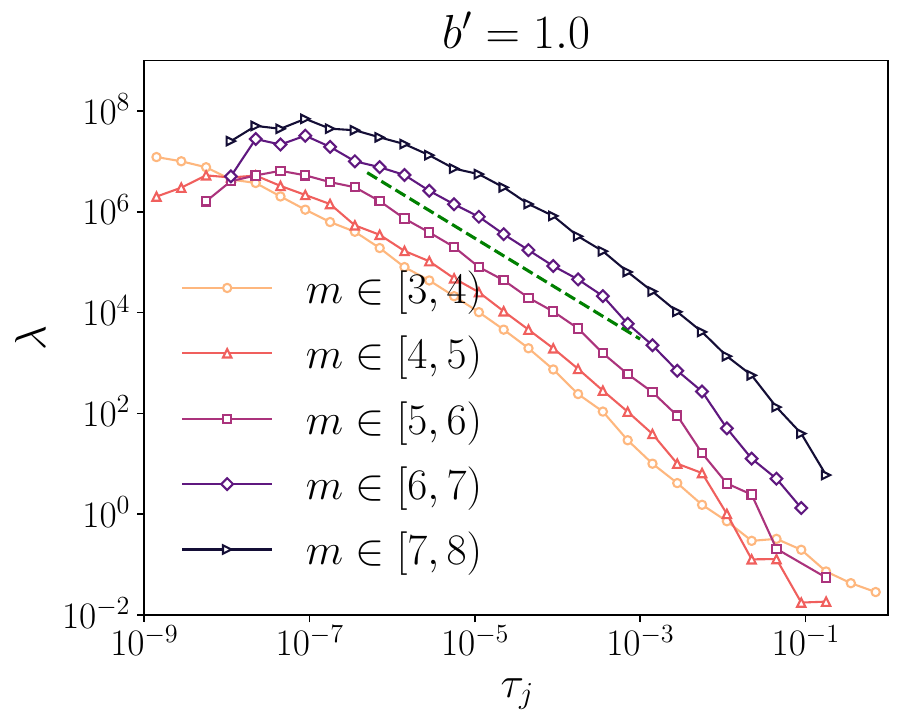}%
}
\subfloat[]{%
\includegraphics[width=0.25\textwidth]{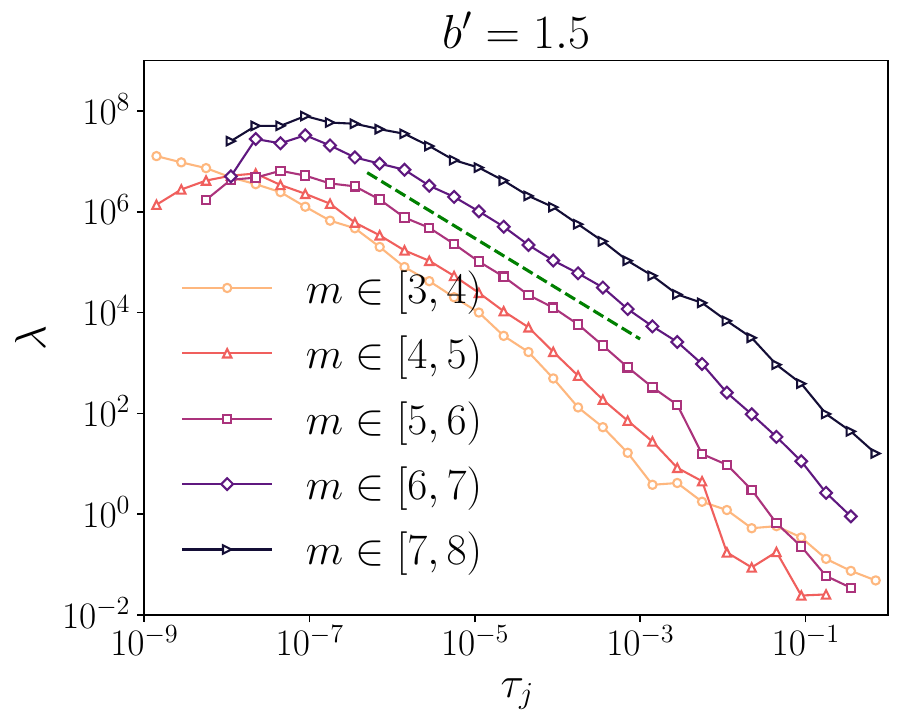}%
}
\caption{Rate of the number of aftershocks $\lambda (\tau_j)$ for events labeled as \textbf{aftershocks in the SCSN catalog} for different magnitudes $m$ varying (a)-(c) $h$, (d)-(g) $D'$ and (h)-(j) $b'$ values. We keep other parameters fixed at $h = 1$, $D' = 1.6$ and $b' = 1$. The green dashed line is a power law with exponent $-1$.}\label{supp:fig:Omori_SCSN_all_m}
\end{figure*}

%%%%---
\begin{figure*}[t!p]
\subfloat[]{%
\includegraphics[width=0.2\textwidth]{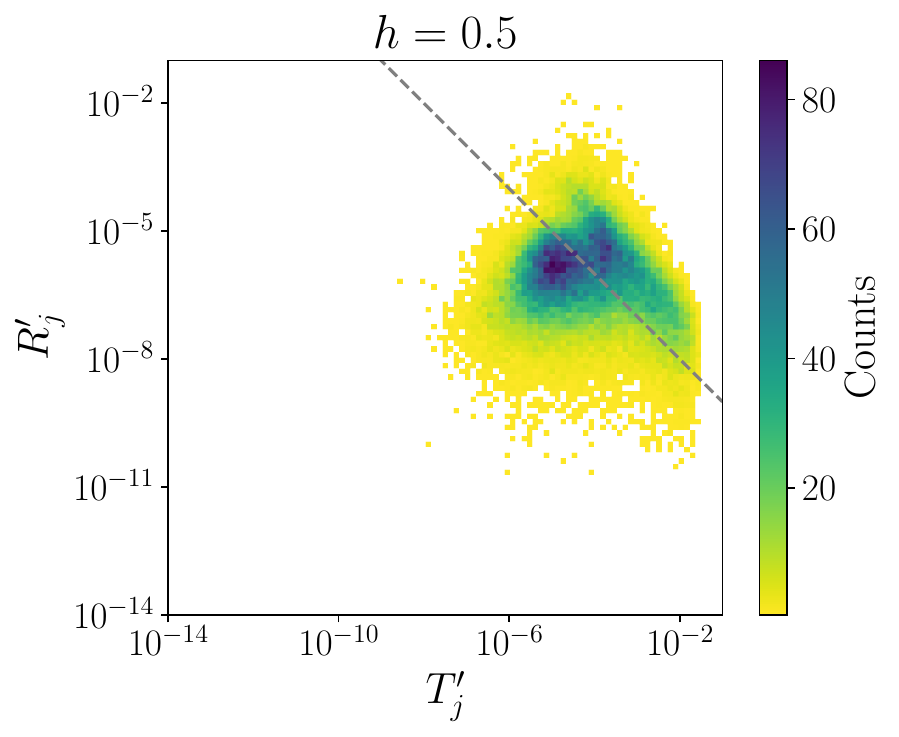}%
}
\subfloat[]{%
\includegraphics[width=0.2\textwidth]{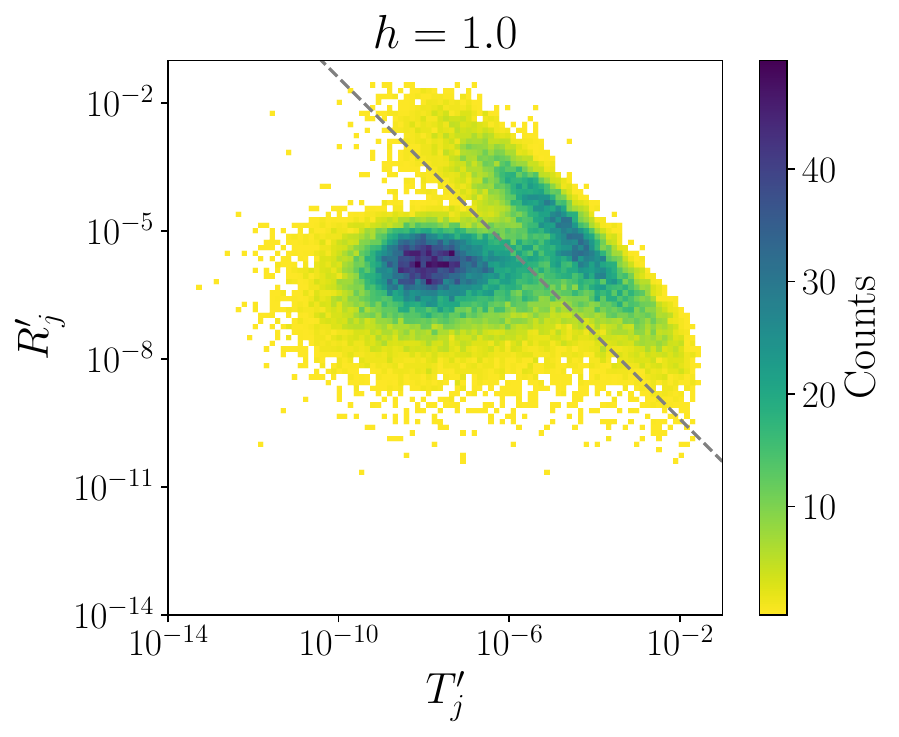}%
}
\subfloat[]{%
\includegraphics[width=0.2\textwidth]{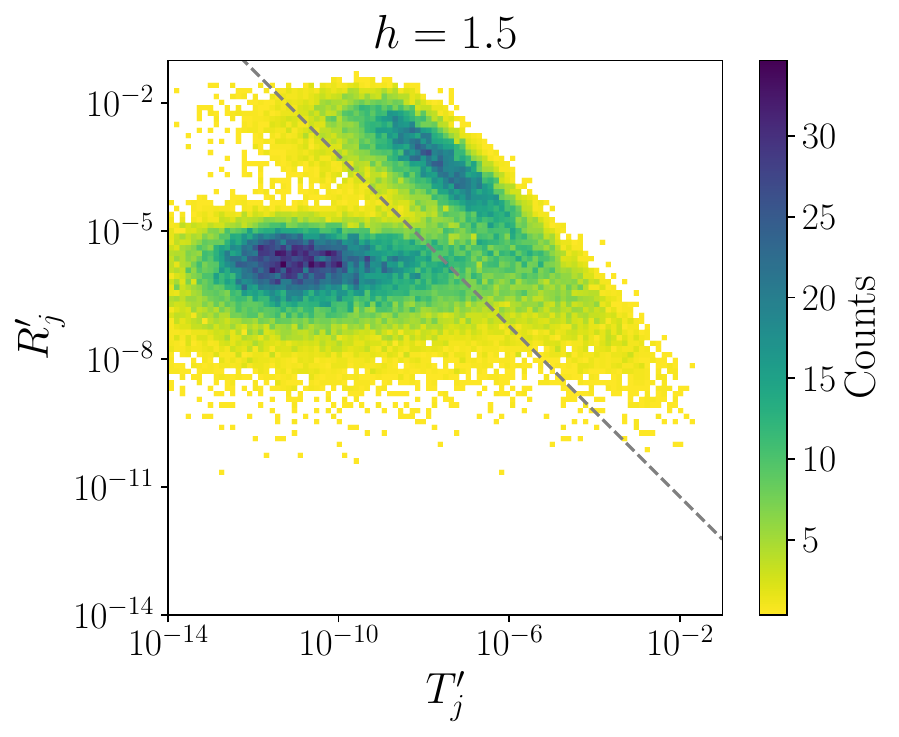}%
}

\subfloat[]{%
\includegraphics[width=0.2\textwidth]{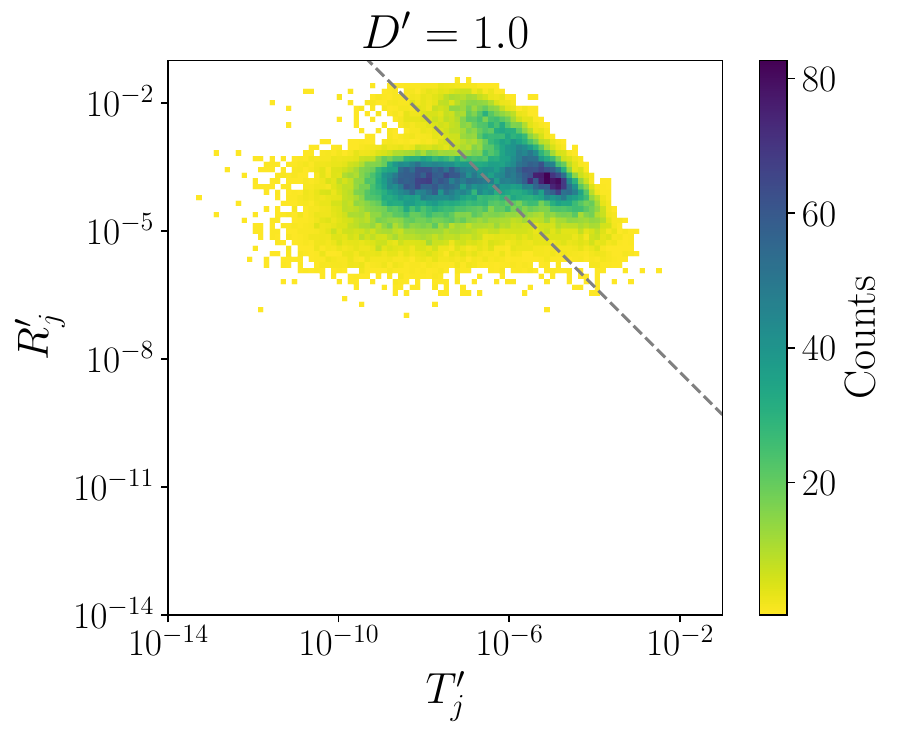}%
}
\subfloat[]{%
\includegraphics[width=0.2\textwidth]{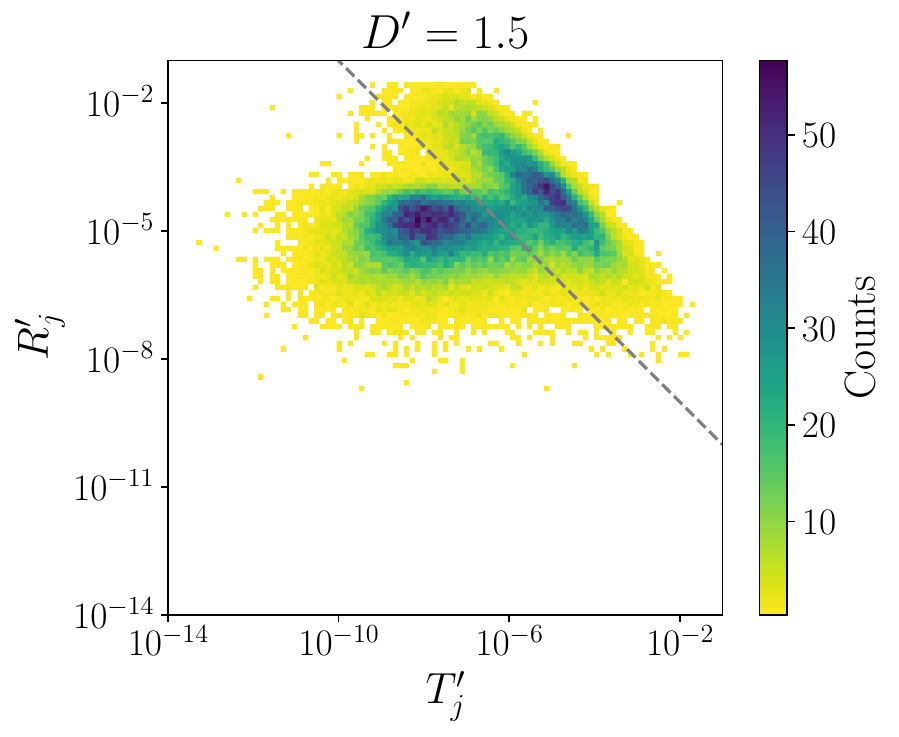}%
}
\subfloat[]{%
\includegraphics[width=0.2\textwidth]{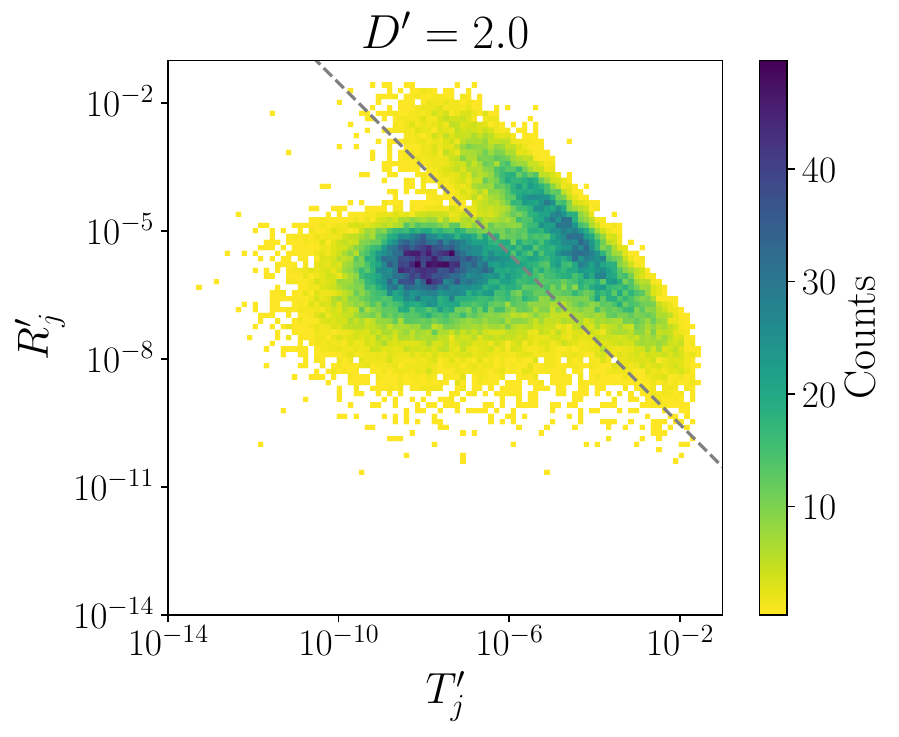}%
}
\subfloat[]{%
\includegraphics[width=0.2\textwidth]{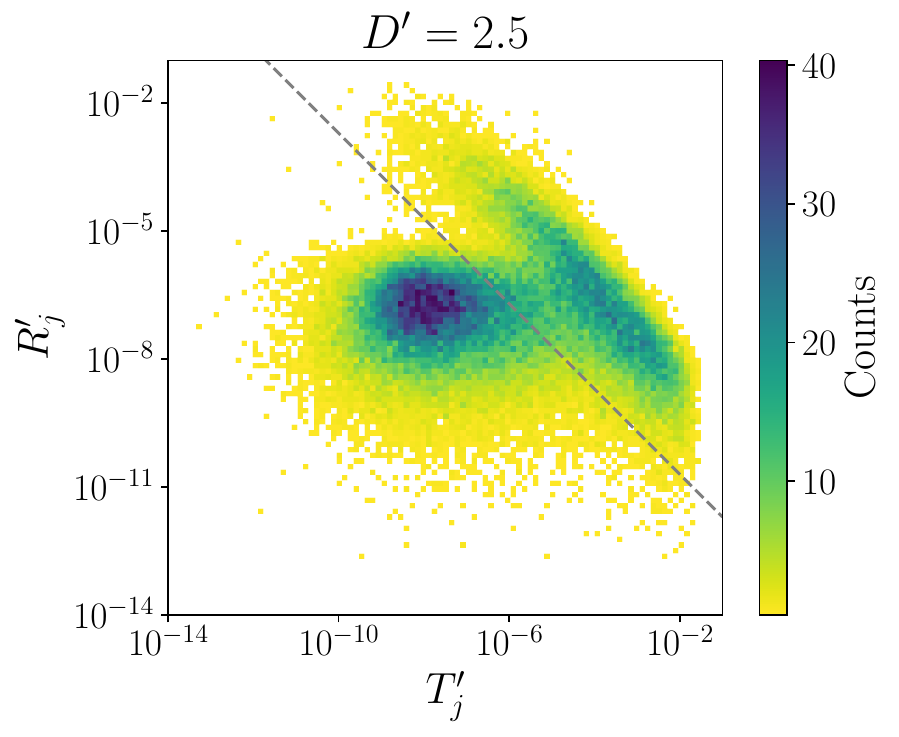}%
}

\subfloat[]{%
\includegraphics[width=0.2\textwidth]{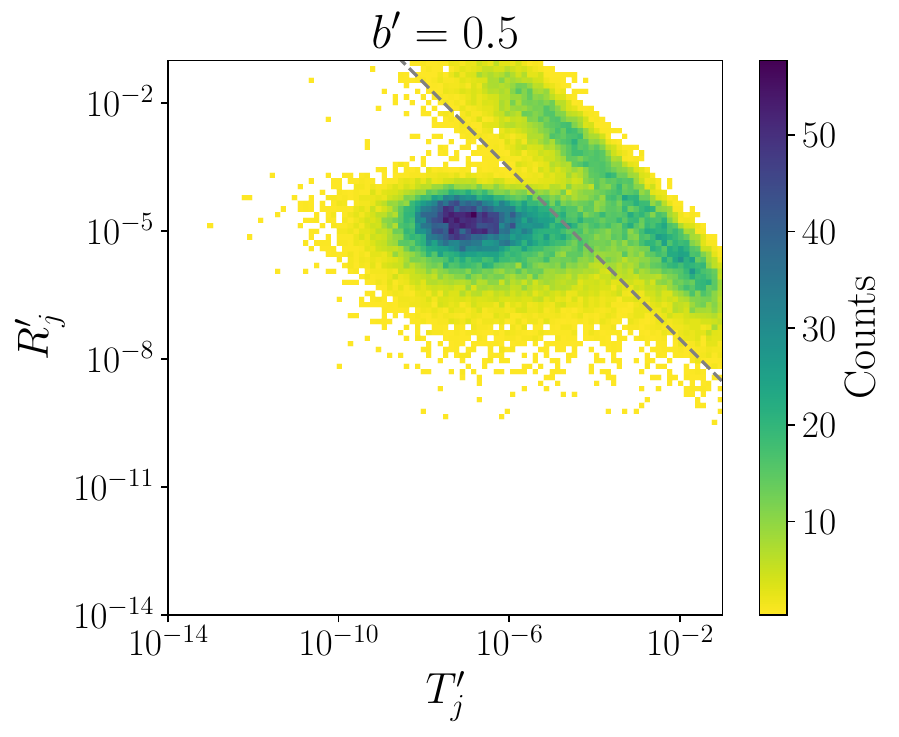}%
}
\subfloat[]{%
\includegraphics[width=0.2\textwidth]{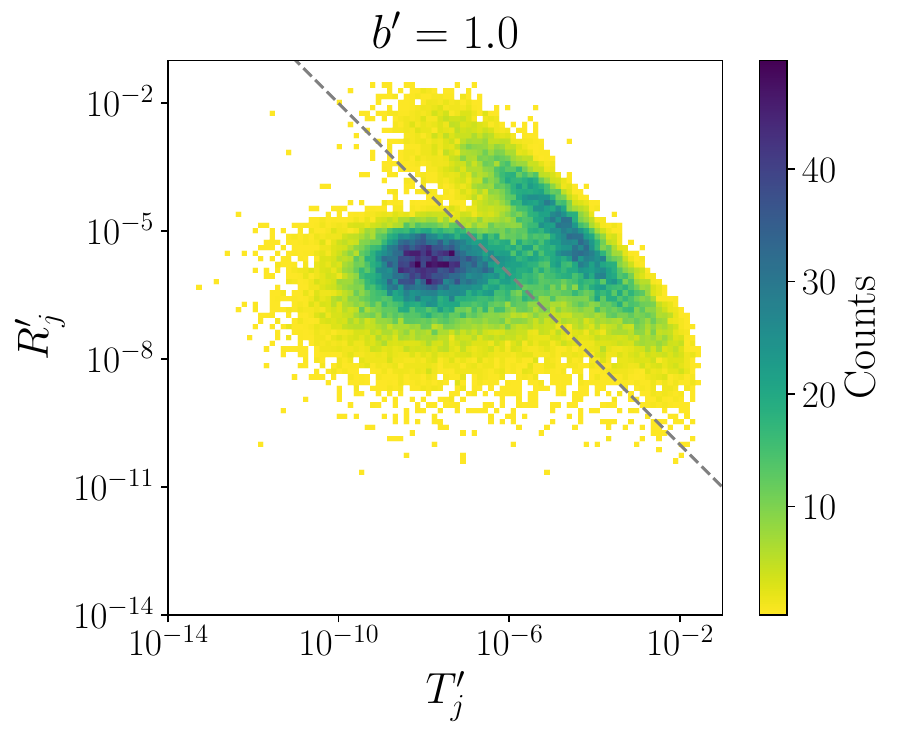}%
}
\subfloat[]{%
\includegraphics[width=0.2\textwidth]{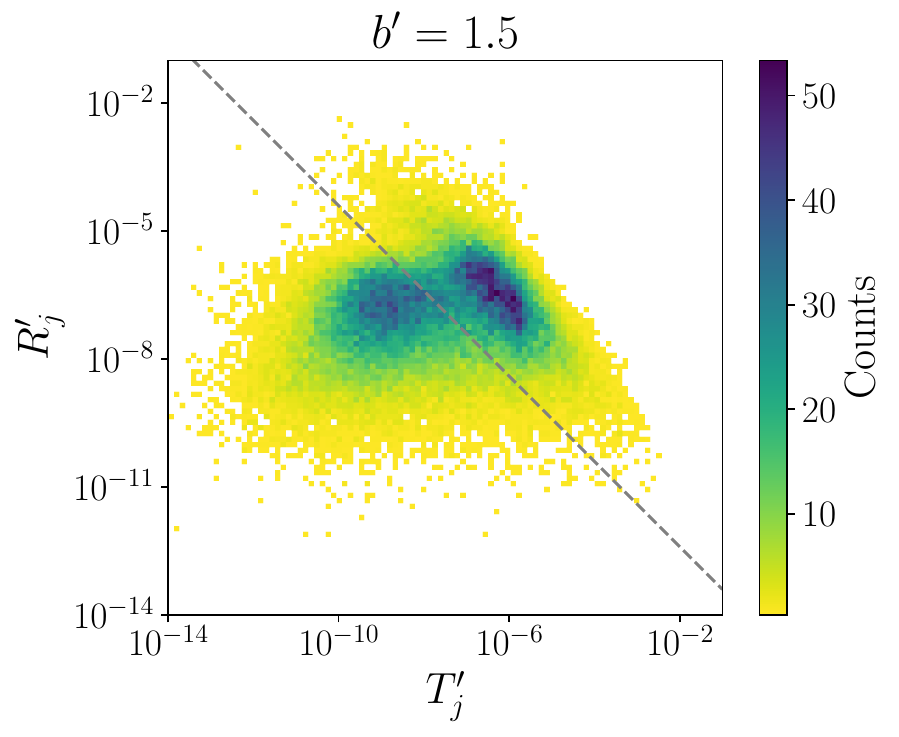}%
}
\caption{Joint distribution of the rescaled space $R'_j$ and time $T'_j$ for different $h$, $D'$ and $b'$ values in the \textbf{ETAS catalog}. We keep other parameters fixed at $h = 1$, $D' = 2$ and $b' = 1$. The gray dashed line represents
(a) $\eta'_{th} = 10^{-10}$,
(b) $\eta'_{th} = 4\cdot10^{-12}$,
(c) $\eta'_{th} = 6\cdot10^{-14}$,
(d) $\eta'_{th} = 5\cdot10^{-11}$,
(e) $\eta'_{th} = 10^{-11}$,
(f) $\eta'_{th} = 3\cdot10^{-12}$,
(g) $\eta'_{th} = 2\cdot10^{-13}$,
(h) $\eta'_{th} = 3\cdot10^{-10}$
(i) $\eta'_{th} = 4\cdot10^{-12}$,
and (j) $\eta'_{th} = 4\cdot10^{-15}$.} \label{supp:fig:rescaledDiagram_ETAS}
\end{figure*}

%---
\begin{figure*}[t!p]
\subfloat[]{%
\includegraphics[width=0.25\textwidth]{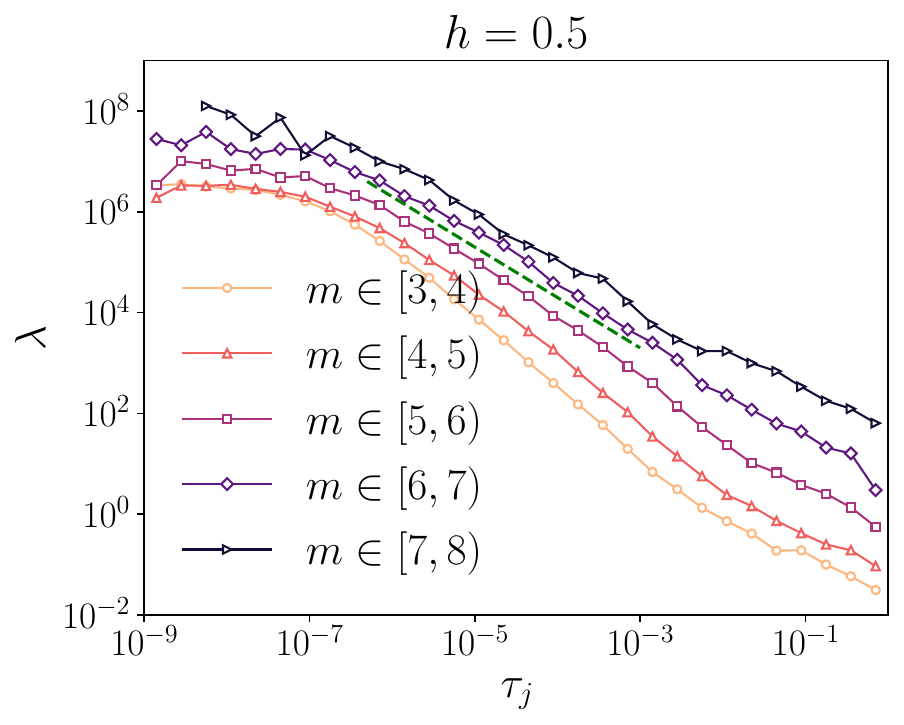}%
}
\subfloat[]{%
\includegraphics[width=0.25\textwidth]{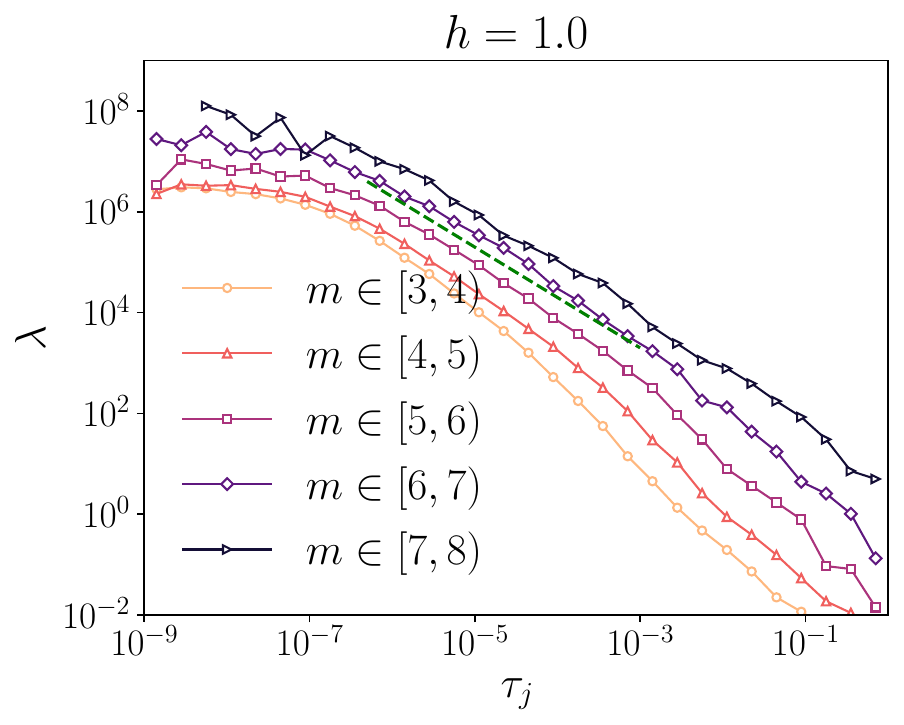}%
}
\subfloat[]{%
\includegraphics[width=0.25\textwidth]{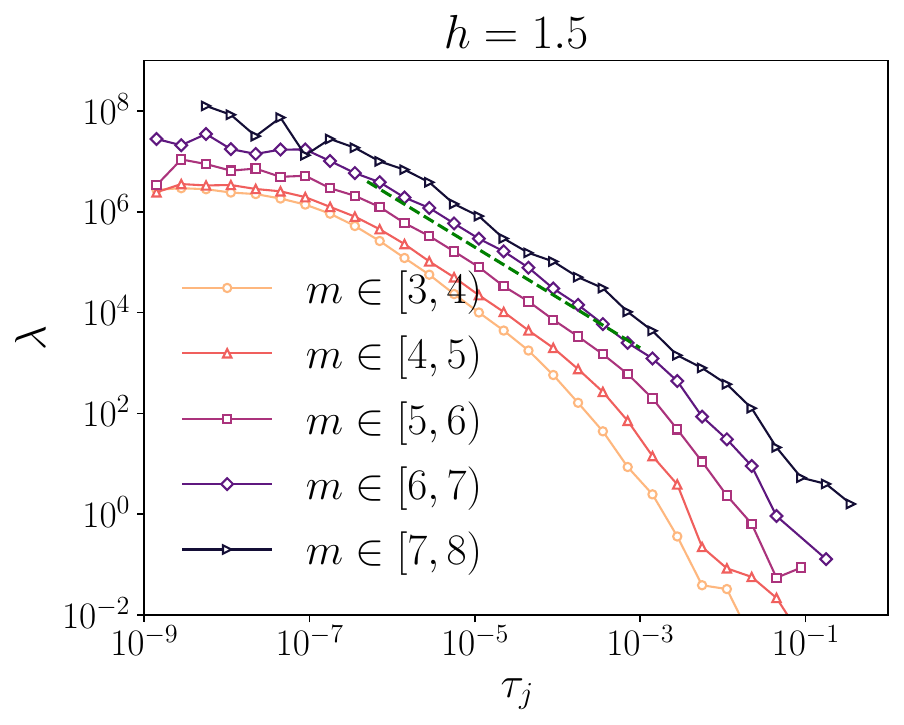}%
}

\subfloat[]{%
\includegraphics[width=0.25\textwidth]{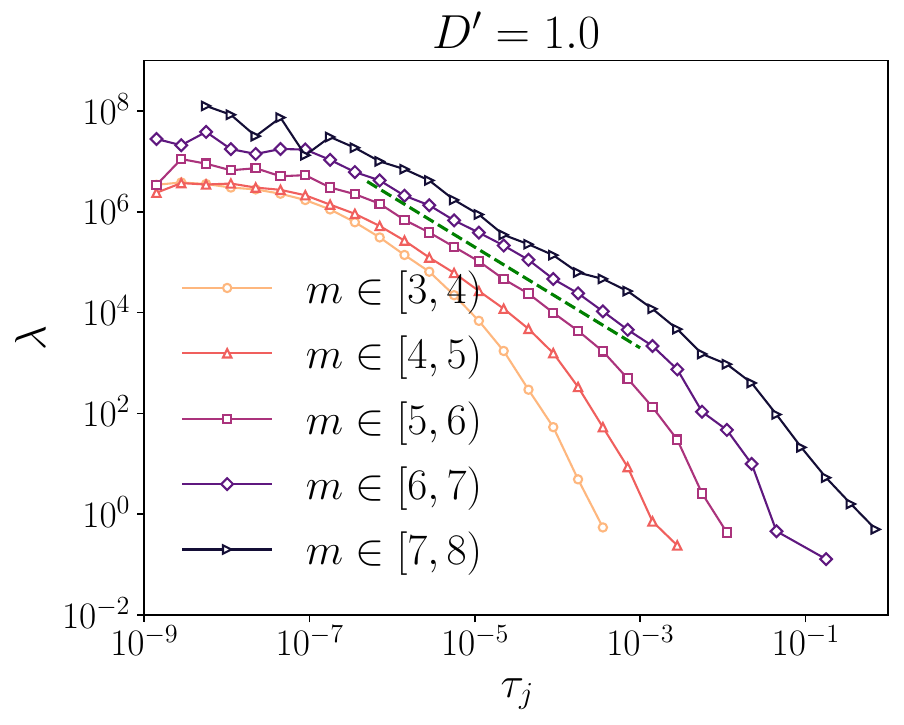}%
}
\subfloat[]{%
\includegraphics[width=0.25\textwidth]{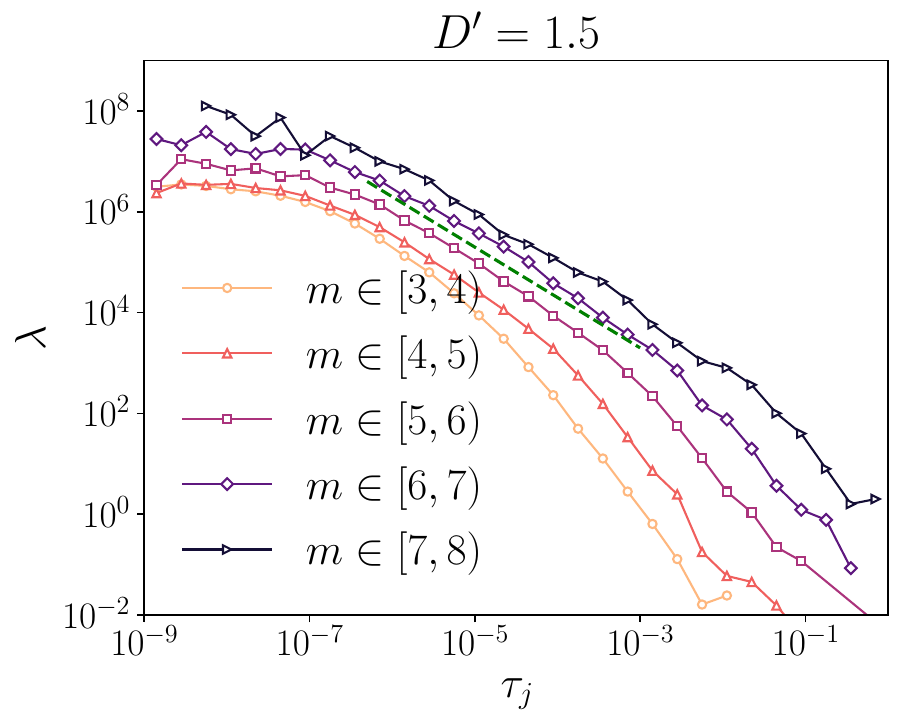}%
}
\subfloat[]{%
\includegraphics[width=0.25\textwidth]{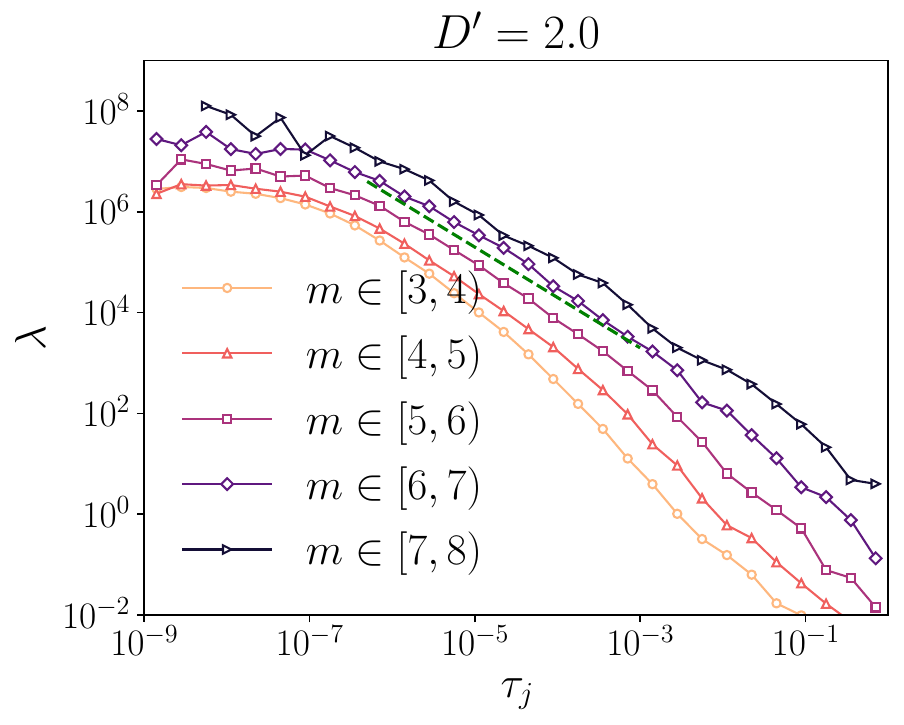}%
}
\subfloat[]{%
\includegraphics[width=0.25\textwidth]{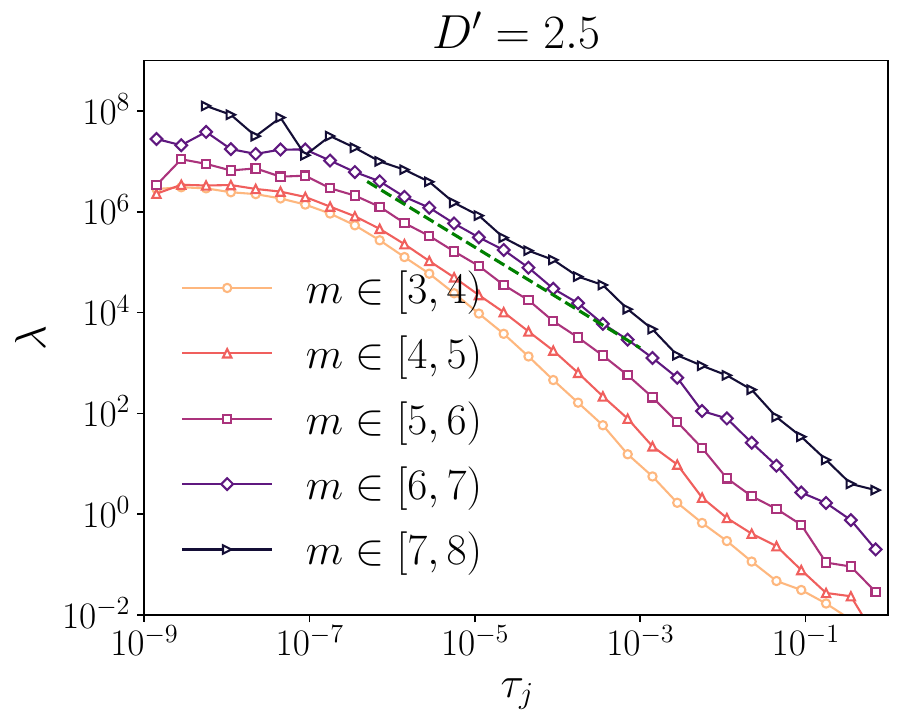}%
}

\subfloat[]{%
\includegraphics[width=0.25\textwidth]{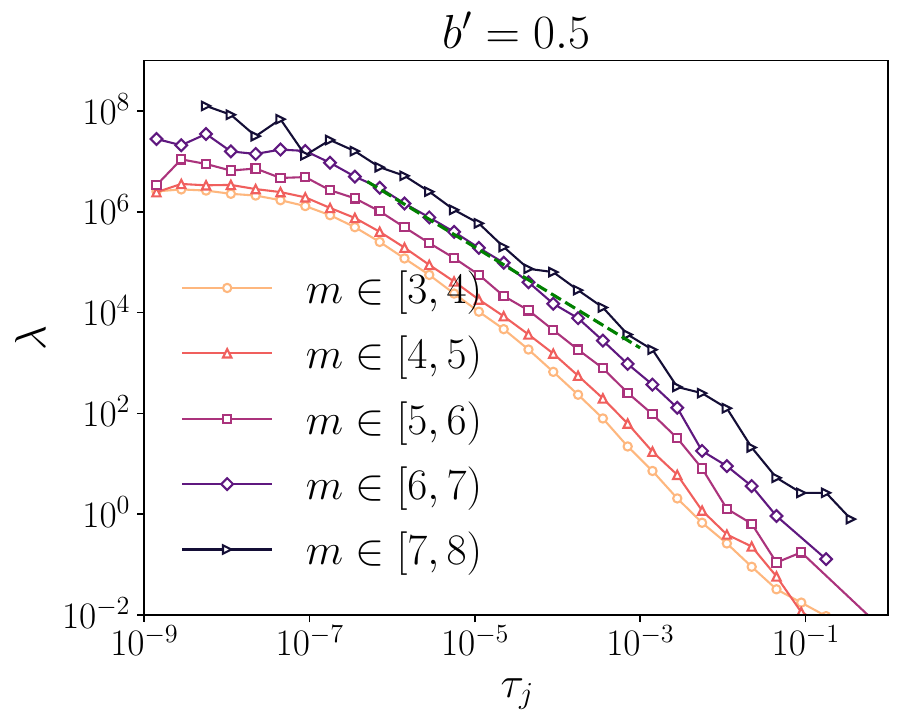}%
}
\subfloat[]{%
\includegraphics[width=0.25\textwidth]{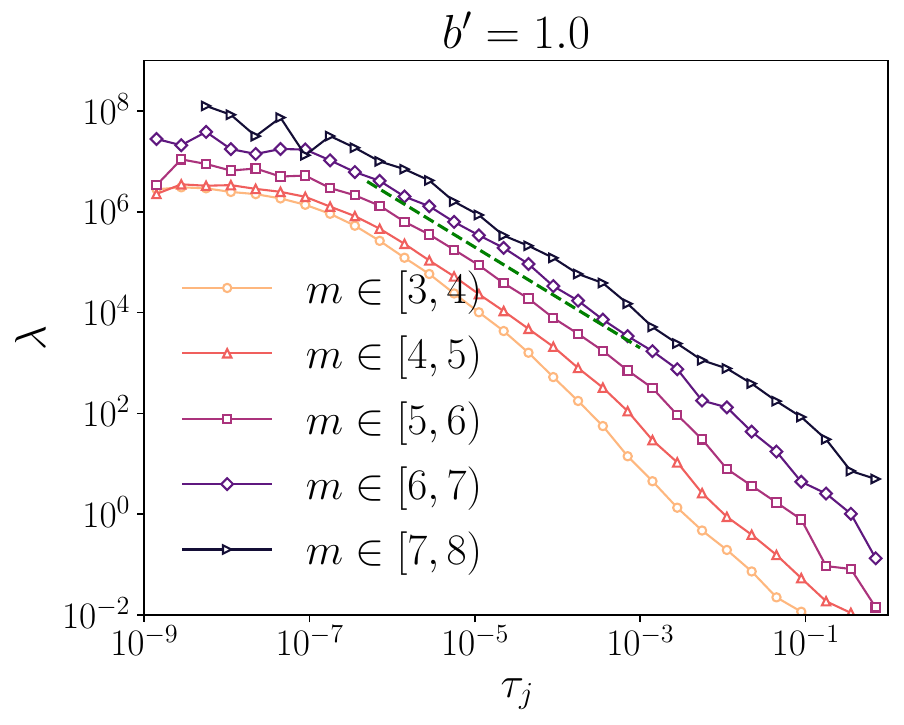}%
}
\subfloat[]{%
\includegraphics[width=0.25\textwidth]{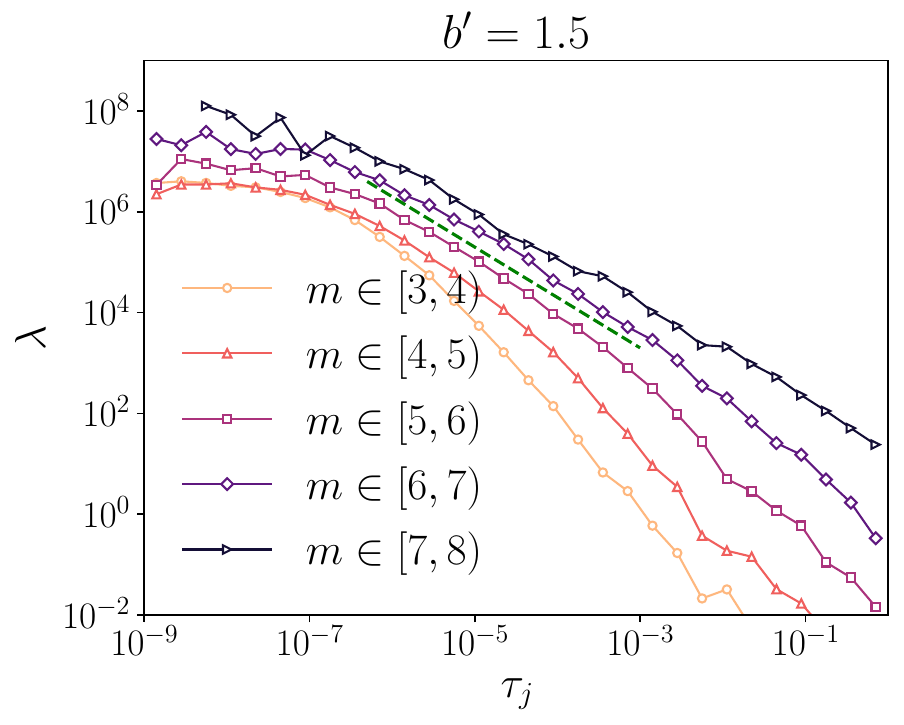}%
}
\caption{Rate of the number of aftershocks $\lambda (\tau_j)$ for events labeled as \textbf{aftershocks in the ETAS catalog}
 for different magnitudes $m$ varying (a)-(c) $h$, (d)-(g) $D'$ and (h)-(j) $b'$ values. We keep other parameters fixed at $h = 1$, $D' = 2$ and $b' = 1$. The green dashed line is a power law with exponent $-1$.}\label{supp:fig:Omori_ETAS_all_m}
\end{figure*}

%%%%---
\begin{figure*}[t!p]
\subfloat[]{%
\includegraphics[width=0.2\textwidth]{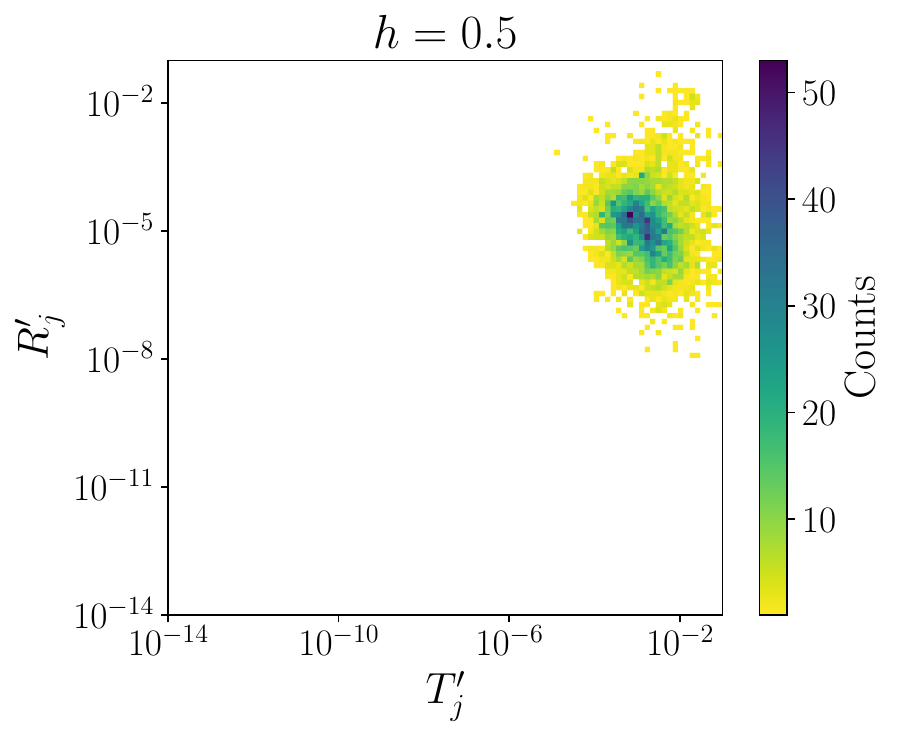}%
}
\subfloat[]{%
\includegraphics[width=0.2\textwidth]{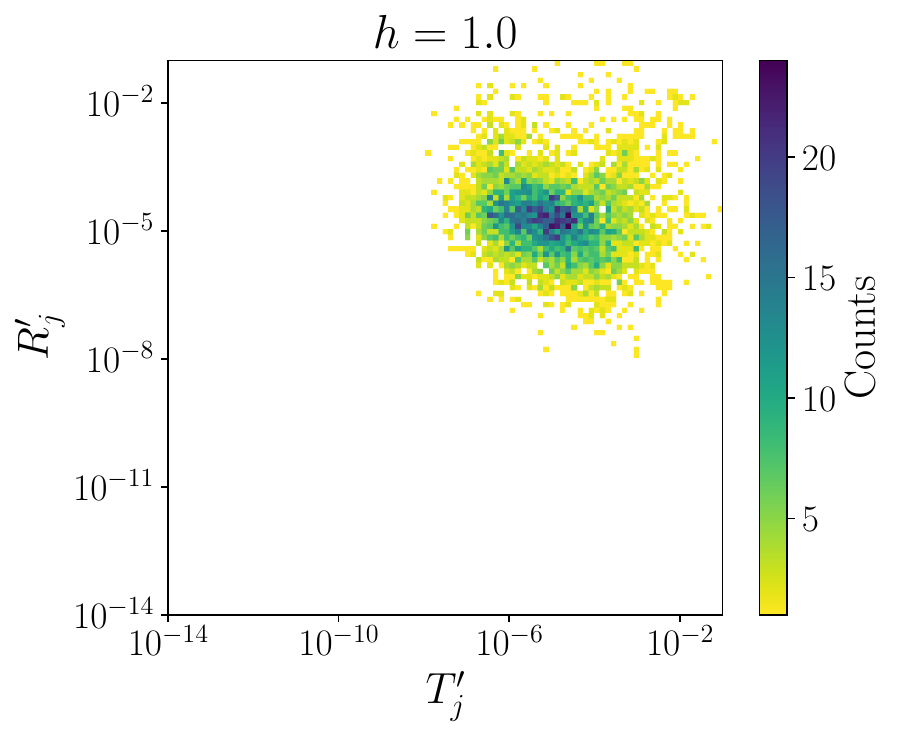}%
}
\subfloat[]{%
\includegraphics[width=0.2\textwidth]{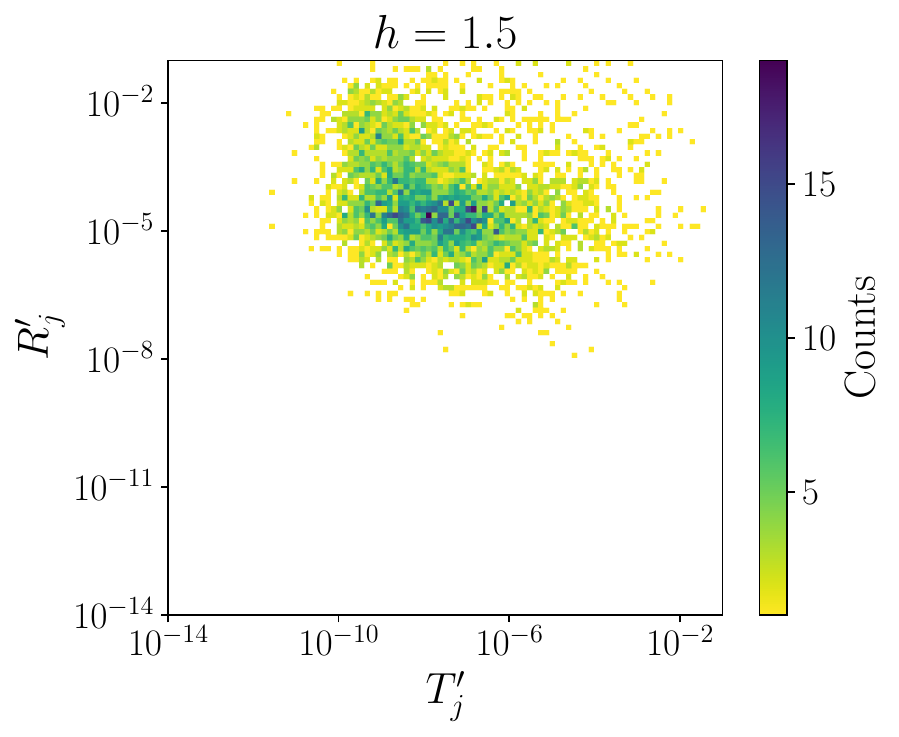}%
}

\subfloat[]{%
\includegraphics[width=0.2\textwidth]{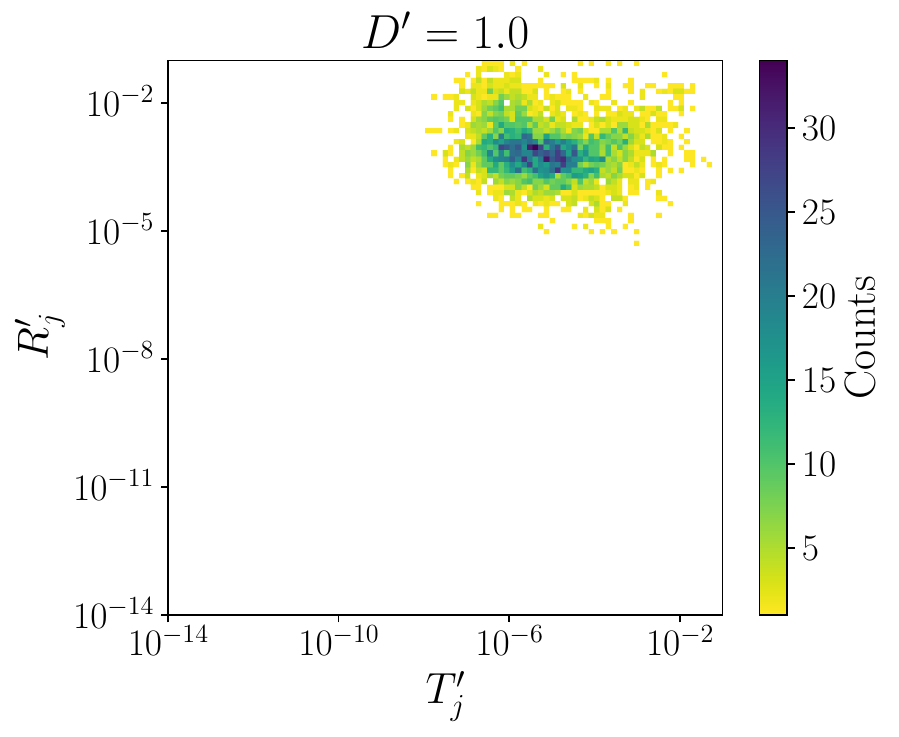}%
}
\subfloat[]{%
\includegraphics[width=0.2\textwidth]{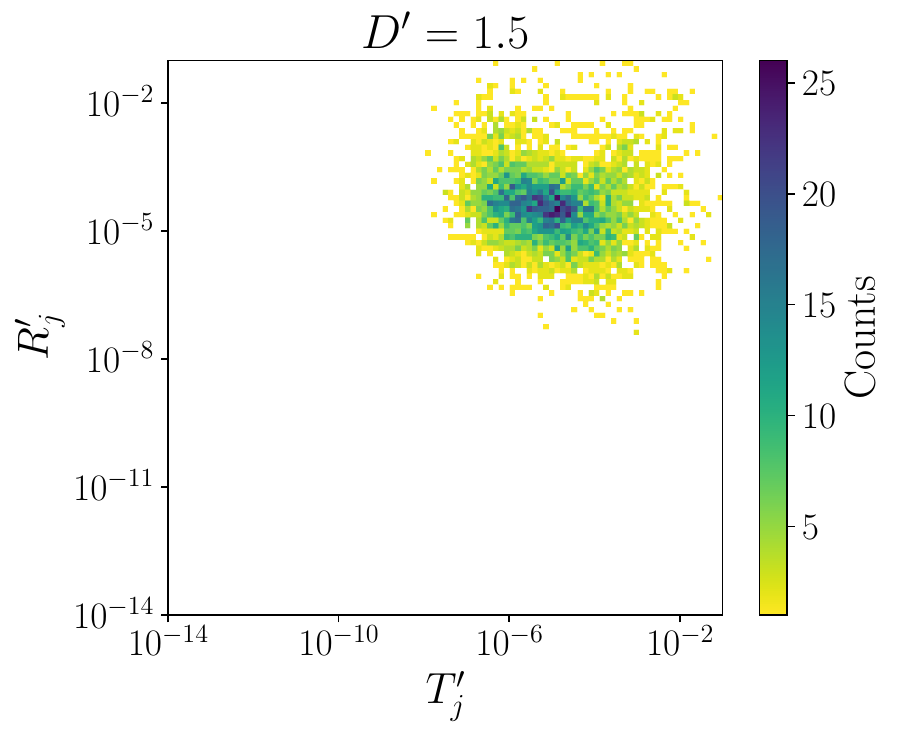}%
}
\subfloat[]{%
\includegraphics[width=0.2\textwidth]{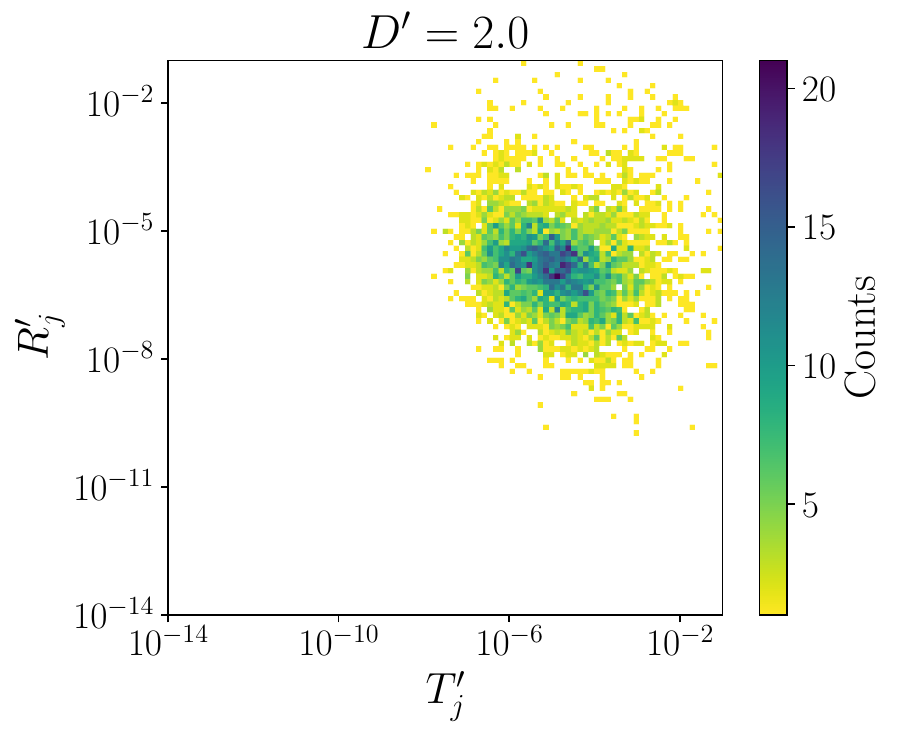}%
}
\subfloat[]{%
\includegraphics[width=0.2\textwidth]{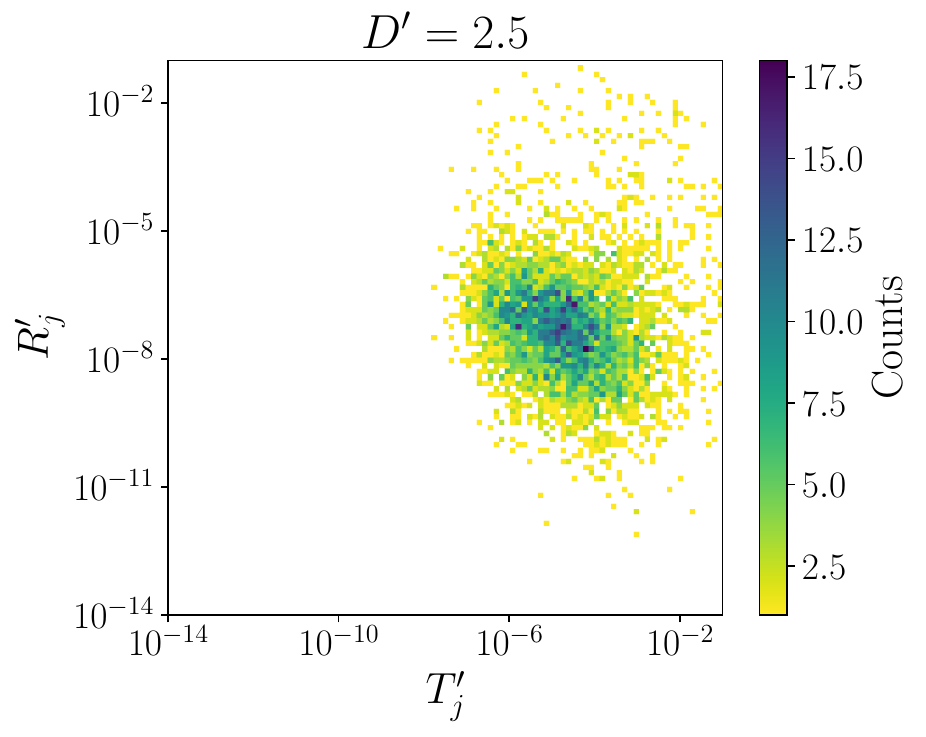}%
}

\subfloat[]{%
\includegraphics[width=0.2\textwidth]{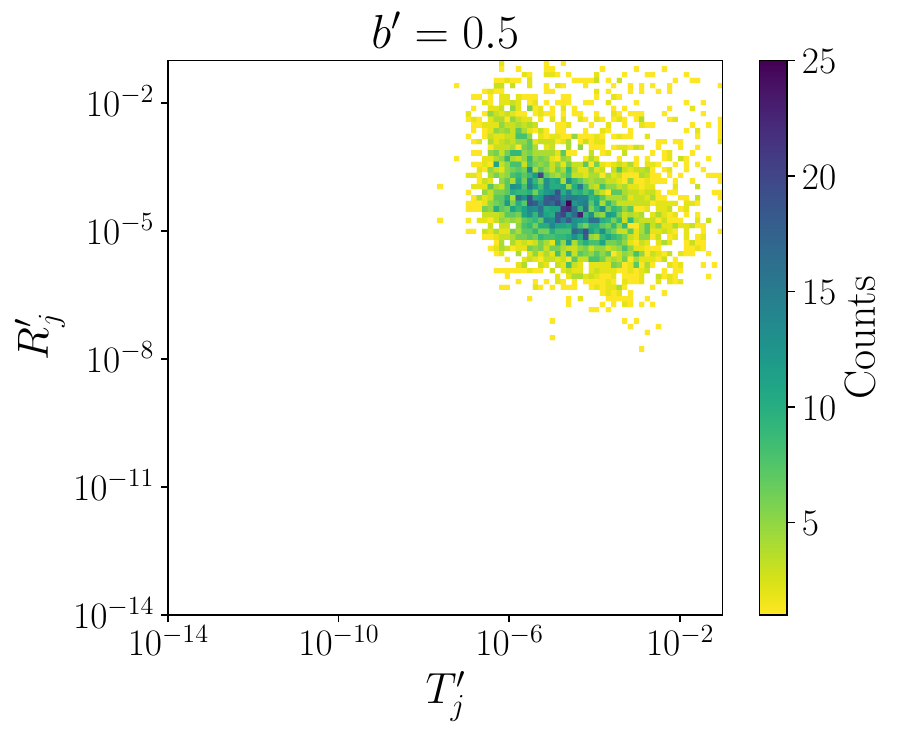}%
}
\subfloat[]{%
\includegraphics[width=0.2\textwidth]{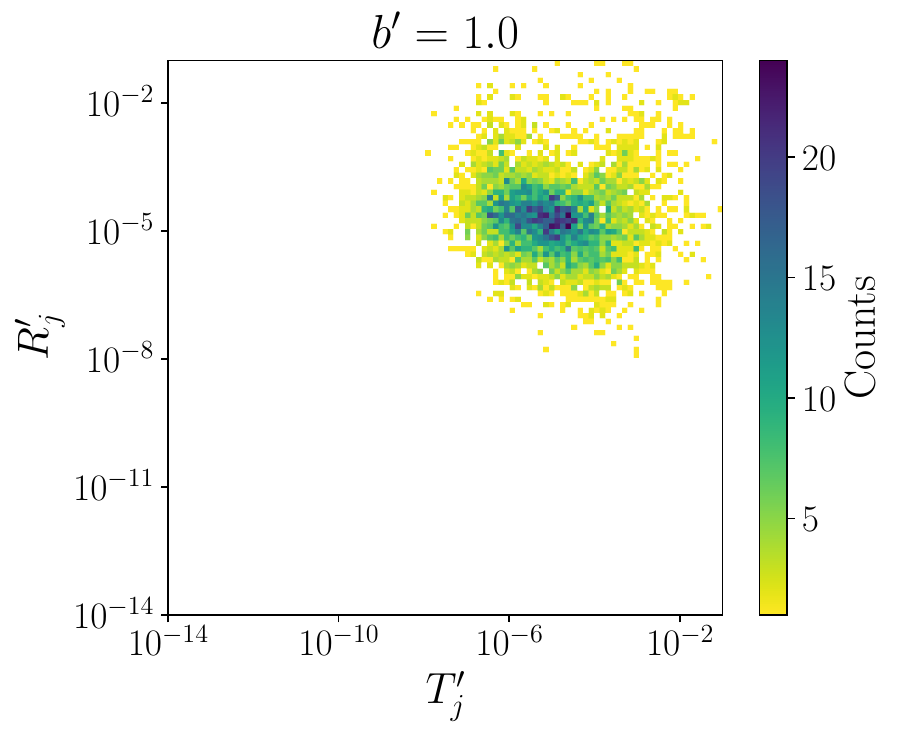}%
}
\subfloat[]{%
\includegraphics[width=0.2\textwidth]{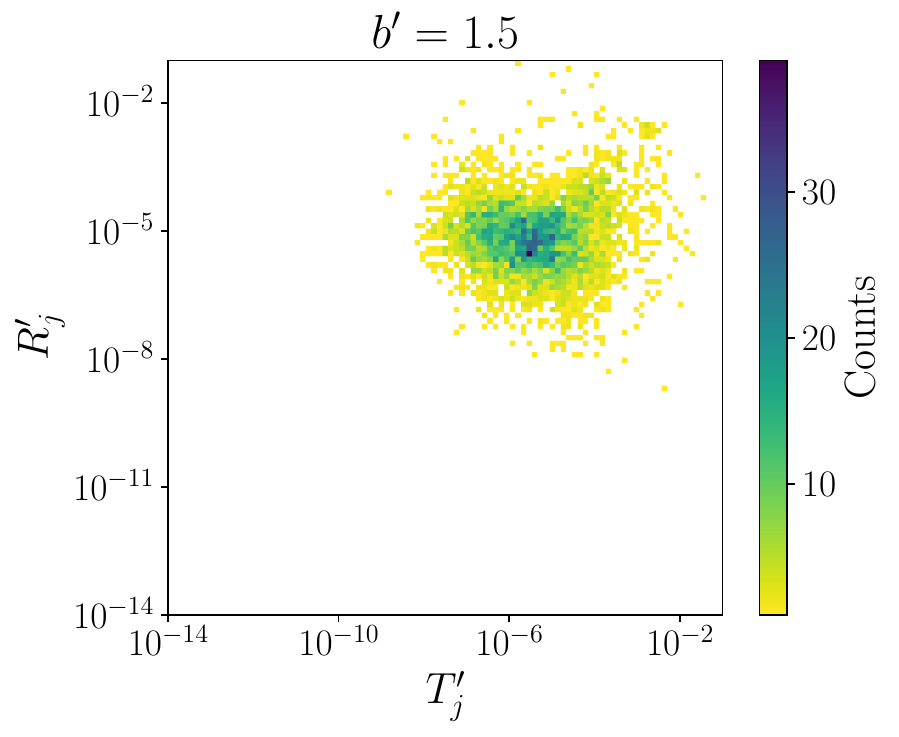}%
}
\caption{Joint distribution of the rescaled space $R'_j$ and time $T'_j$ for different $h$, $D'$ and $b'$ values in the \textbf{natural swarm catalog}. We keep other parameters fixed at $h = 1$, $D' = 1.6$ and $b' = 1$.} \label{supp:fig:rescaledDiagram_shelly}
\end{figure*}

%---
\begin{figure*}[t!p]
\subfloat[]{%
\includegraphics[width=0.25\textwidth]{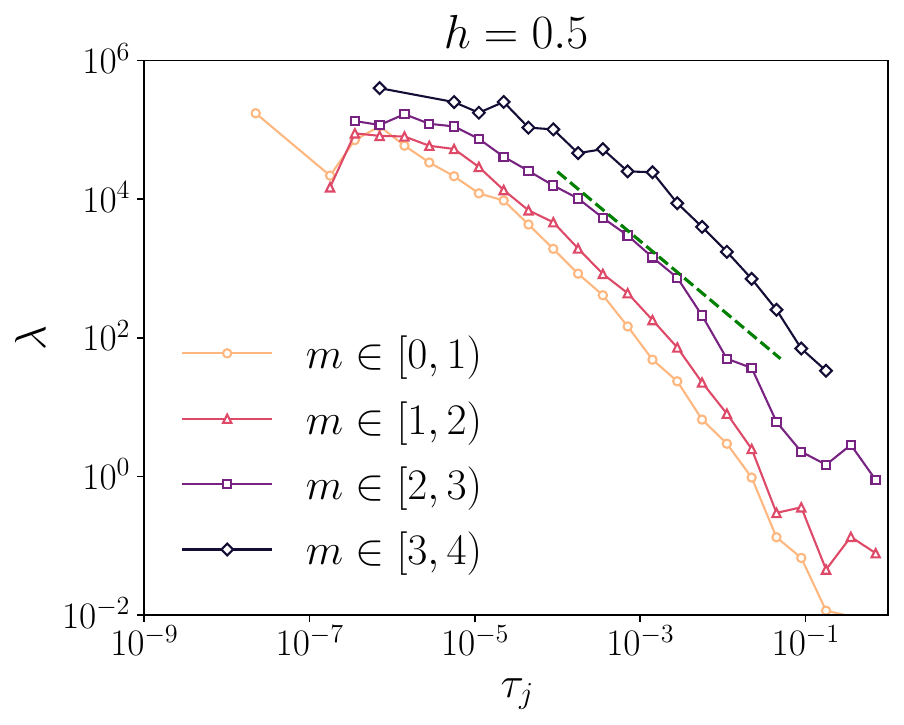}%
}
\subfloat[]{%
\includegraphics[width=0.25\textwidth]{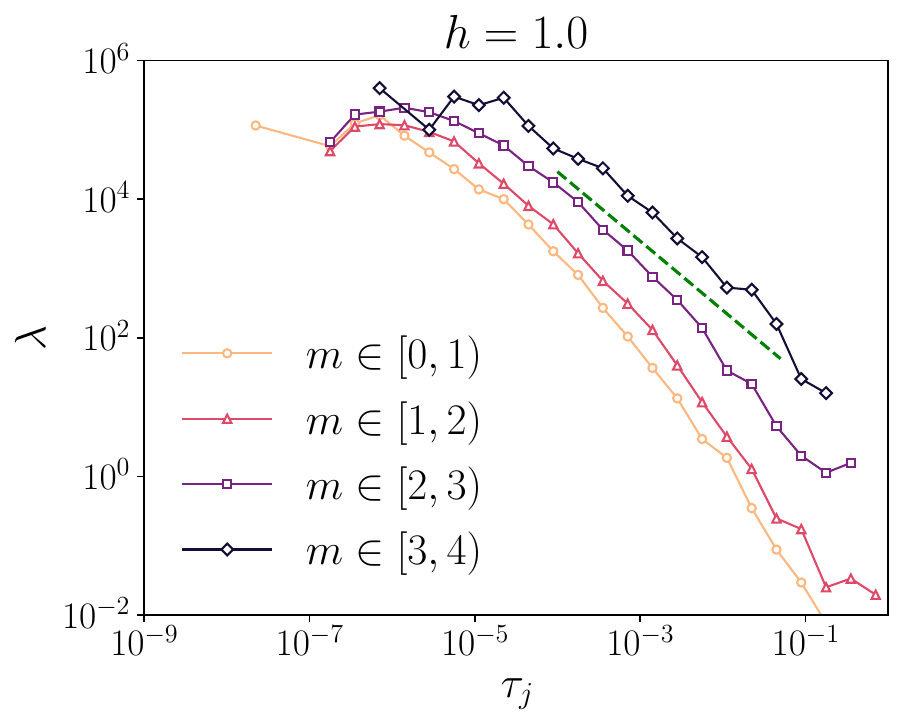}%
}
\subfloat[]{%
\includegraphics[width=0.25\textwidth]{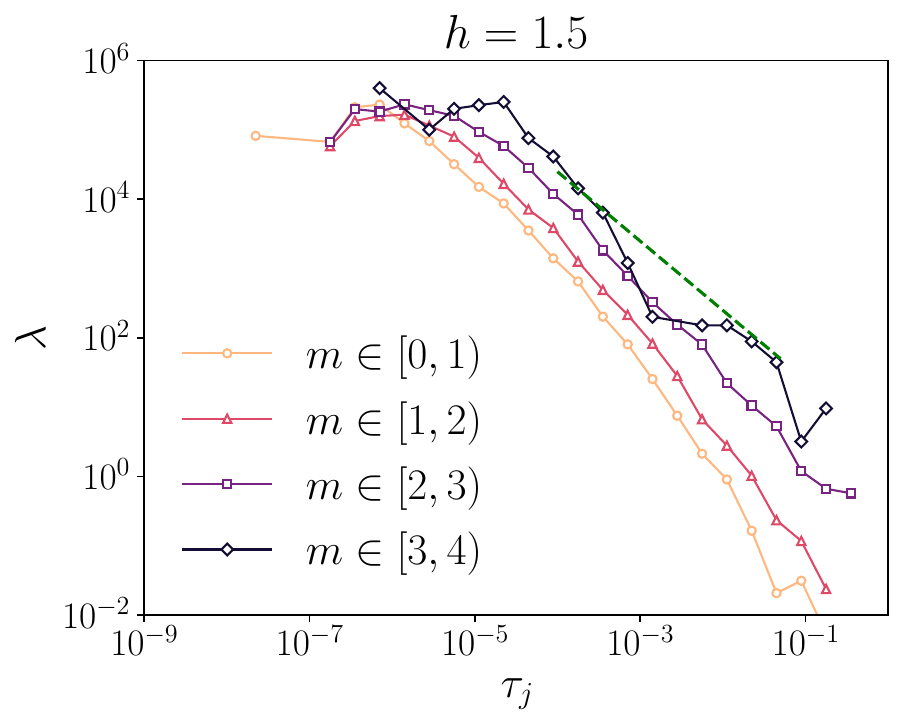}%
}

\subfloat[]{%
\includegraphics[width=0.25\textwidth]{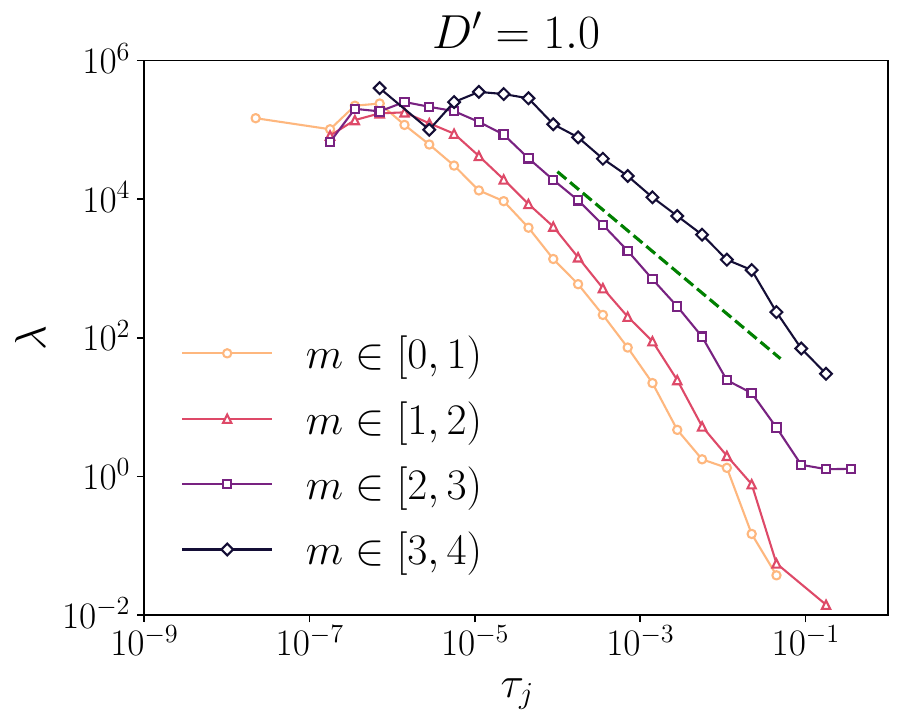}%
}
\subfloat[]{%
\includegraphics[width=0.25\textwidth]{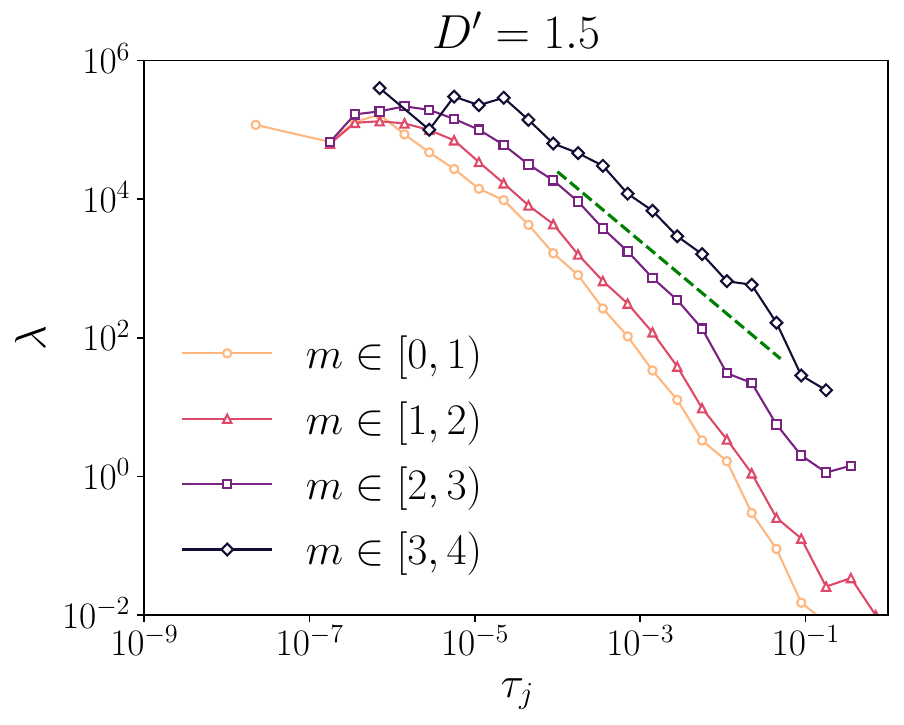}%
}
\subfloat[]{%
\includegraphics[width=0.25\textwidth]{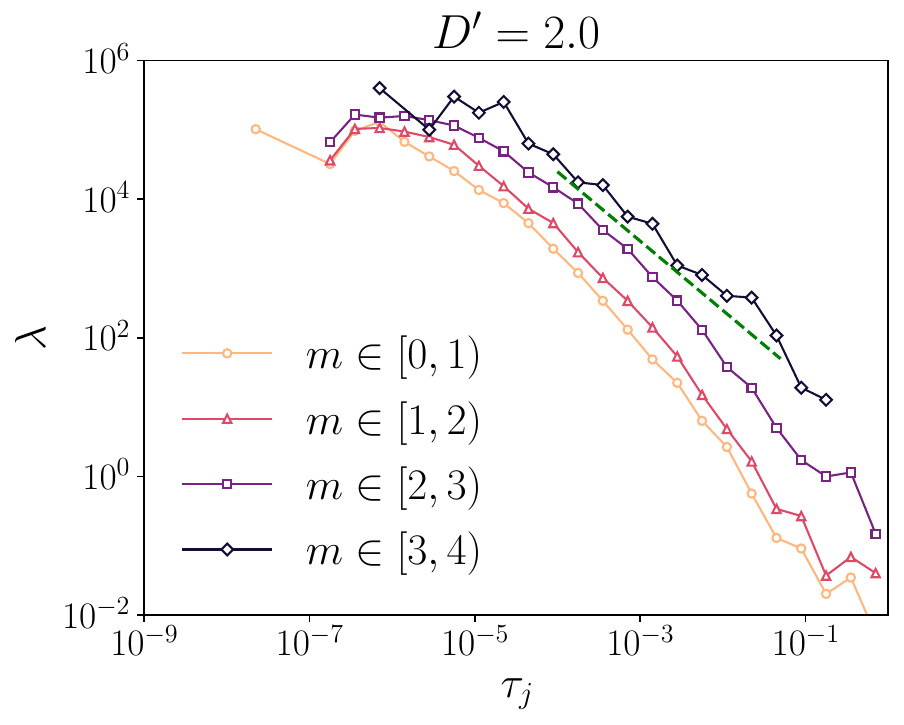}%
}
\subfloat[]{%
\includegraphics[width=0.25\textwidth]{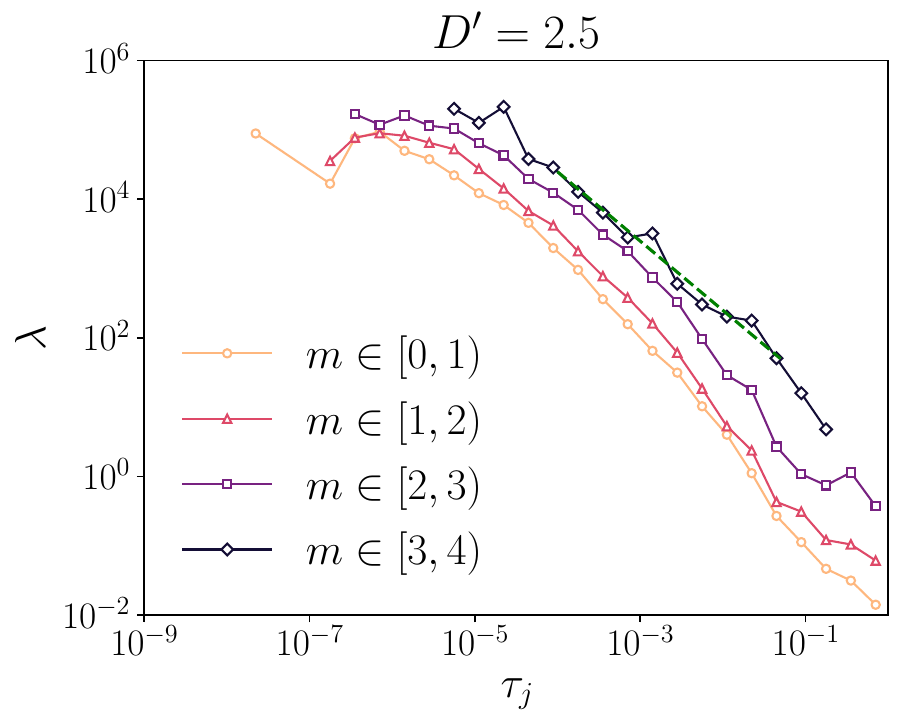}%
}

\subfloat[]{%
\includegraphics[width=0.25\textwidth]{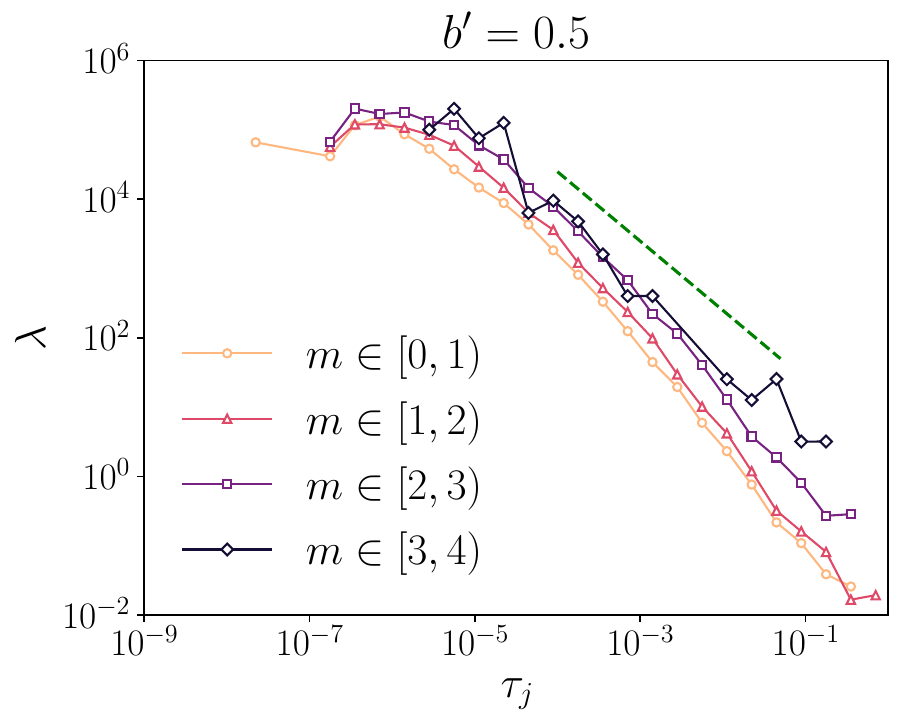}%
}
\subfloat[]{%
\includegraphics[width=0.25\textwidth]{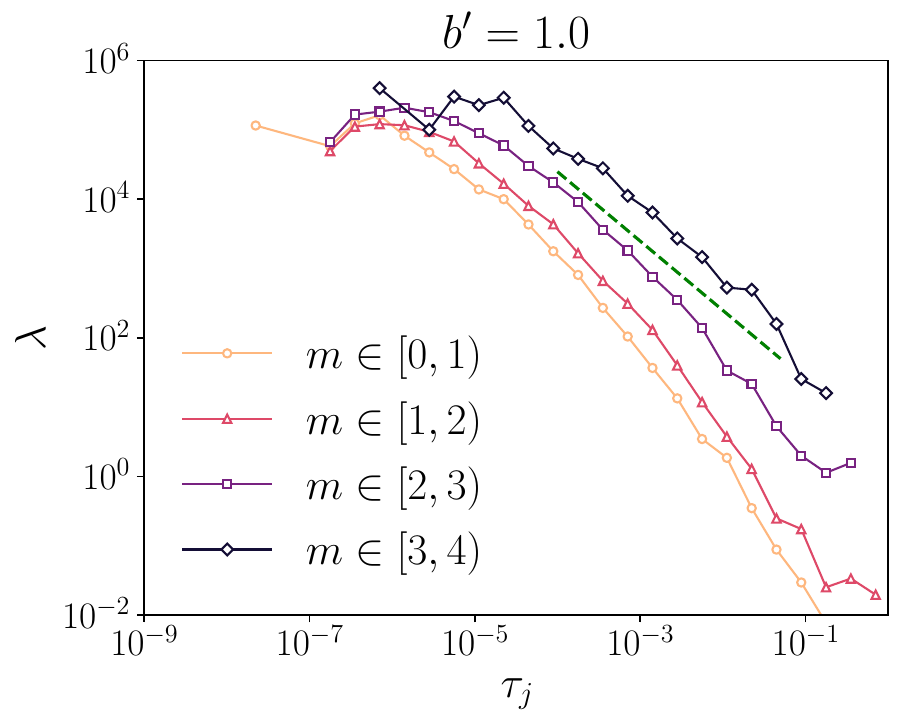}%
}
\subfloat[]{%
\includegraphics[width=0.25\textwidth]{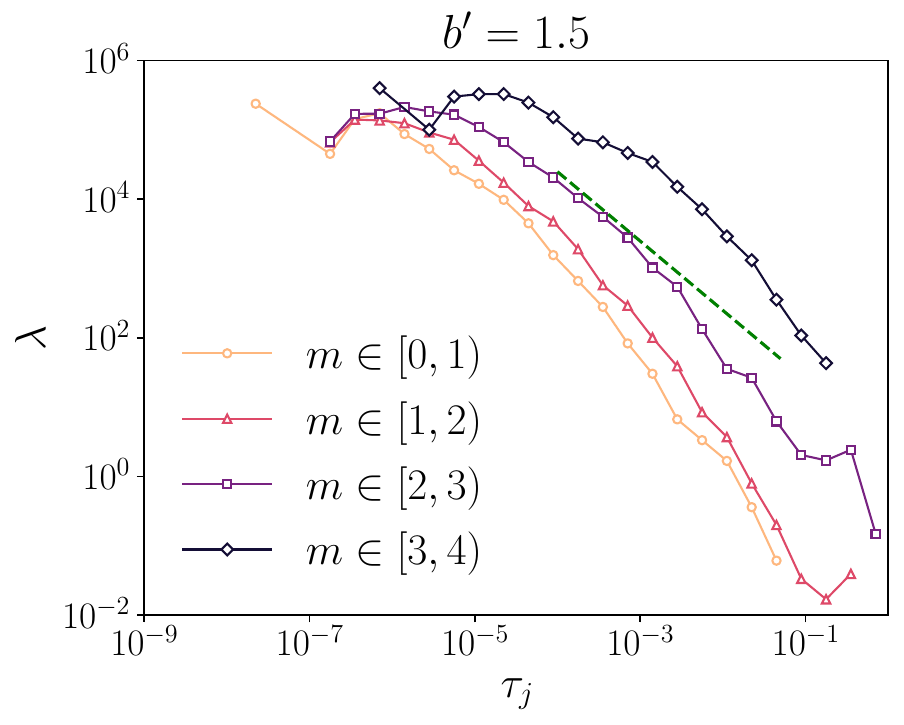}%
}
\caption{Rate of the number of aftershocks $\lambda (\tau_j)$ for the \textbf{natural swarm catalog} for different magnitudes $m$ varying (a)-(c) $h$, (d)-(g) $D'$ and (h)-(j) $b'$ values. We keep other parameters fixed at $h = 1$, $D' = 1.6$ and $b' = 1$. The green dashed line is a power law with exponent $-1$.}\label{supp:fig:Omori_Shelly_all_m}
\end{figure*}

\end{document}